\documentclass[11pt]{article}
\usepackage[T1]{fontenc}

\usepackage{listings}
\usepackage{datetime}
\usepackage{comment}					
\usepackage[Euler]{upgreek} 

\usepackage[usenames,dvipsnames]{xcolor}

\usepackage[parsep]{collref}	

\usepackage{amsmath, amssymb,amsthm}
\usepackage{stackrel}
\numberwithin{equation}{section}

\usepackage{bm,environ,mathrsfs,array,arydshln}
\usepackage{booktabs,float,slashed}
\usepackage{appendix}
\usepackage[mathcal]{euscript}
\usepackage{tensor} 						
\usepackage{mathabx}
\usepackage[vcentermath]{youngtab}
\usepackage{simpler-wick}

\usepackage{graphicx,epsfig,epic}
\usepackage[setpagesize=false,pagebackref=false, 
linktocpage, bookmarksopen=true, colorlinks=true, 
linkcolor=Maroon,citecolor=Maroon,urlcolor=Maroon]{hyperref}

\allowdisplaybreaks

\def\ba#1\ea{\begin{align}#1\end{align}}		        
\newcommand{\be}{\begin{equation}}
\newcommand{\ee}{\end{equation}}
\newcommand{\bea}{\begin{equation} \begin{aligned}} 
\newcommand{\eea}{\end{aligned} \end{equation}}

\newcommand{\la}{\label}

\newcommand{\lp}{\notag \\ & }

\DeclareMathOperator{\tr}{\text{tr}}
\DeclareMathOperator{\vol}{vol}

\def \AdS  {{\rm AdS}}
  \def \ZZ  {{\mathbb Z}}

\def \np { \newpage}
\def \ed {
 \small
\baselineskip 11pt
\bibliography{m11}
\small
\bibliographystyle{JHEP-v2.9}
\end{document}
}
\def \iffa  {\iffalse}
\def \te {\textstyle}
\newcommand{\rf}[1]{(\ref{#1})}
\def\ov{\over}
\def \ci {\cite}
\def \foot {\footnote}

\def\la{\label}\def \a {\alpha}
\def\foot{\footnote}

\def \tb {$T\bar T$ }
\def \s {\sigma} \def \del {\partial} 
\def \ha {\tfrac{1}{2}}
\def \DD {{\rm D}}

\def \OO {{\cal O}}

\def \no {\nonumber}

\newcommand{\dd}{\dot{d}}

\newcommand{\Ber}{\operatorname{Ber}}

\newcommand{\cL}{\mathcal L}

\newcommand{\cM}{\mathcal M}
\newcommand{\cO}{\mathcal O}
\newcommand{\Gb}{\bar\Gamma}
\newcommand{\Gk}{\Gamma_\kappa}

\newcommand{\cQ}{\mathcal Q}
\newcommand{\cJ}{\mathcal J}
\newcommand{\cI}{\mathcal I}

\newcommand{\al}[1]{\begin{align}{#1}\end{align}}

\def \fo {\tfrac{1}{4}} \def \beg {\be}
 
\def \DD {{\rm D}}
\def \ep {\epsilon}

\def \iffa {\iffalse} \def \aa {{\rm a }}  

\def \tr {{\rm tr}}
\def \vol   {{\rm vol}}

\def \OO {{\cal O}}

\def \ka {\kappa} 
\def \rT {{\rm T}}

\def \N   {{\cal N}}

\def \fo {{1\ov 4}}
\def \la {\label} \def \fo {{1\ov 4}}
\def \beg  {\begin{equation}}
\def \eeg {\end{equation}}

\def \rT {{\rm T}} 

\def \z  {\zeta}

\def \gs {g_{\rm s}}

\def \La  {\Lambda}

\def \lp {\ell_{\rm p}} 

\def \dd {d}
 \def \rT {{\rm T}}

\def \cL {{\cal L}}   
\def \X {{\rm X}}

\def \tb {\bar \theta}

\newcommand{\vev}[1]{\langle #1\rangle}
\newcommand{\CP}{\mathbb{CP}}

\newcommand{\PC}{P_C}
\newcommand{\PN}{P_N}
\newcommand{\Tmem}{\mathrm T_2}
\newcommand{\ii}{\mathrm i}

\newcommand{\CC}{\mathcal B}

\newcommand{\Dsl}{\slashed D}

\begin{document}
\begin{titlepage}
\begin{tabbing}
\hspace*{10.5cm} \=  \kill 
\>  Imperial--TP--2026--AAT--03 \\
\> 
\end{tabbing}



\vspace*{05 mm}
\begin{center}
{\Large\sc  \bf     
M2  brane  $S^3/\ZZ_k$ instanton partition function   in $\AdS_4\times S^7/\ZZ_k$:     \\[2mm]
  2-loop correction  }
\vspace*{07mm}

Arkady A. Tseytlin\footnote{Also at  ITMP of MSU  and Lebedev Inst.\ \ \ tseytlin@ic.ac.uk   }

\vspace*{4mm}
{\small
	
 Abdus Salam Centre for Theoretical Physics 
 \\ Imperial College London,  SW7 2AZ, U.K.
}
\vspace*{1.2cm}

\end{center}

\begin{abstract} 
As was shown in arXiv:2609.14497, the 2-loop $ {\rm T}^{-1}_2$  (the  first   inverse tension) correction
to the partition function of M2 brane  wrapping $\mathrm{AdS}_2\times S^1$ inside
$\mathrm{AdS}_4\times S^7/\mathbb{Z}_k$  vanishes. This is  in agreement with  the localization prediction for  $\frac{1}{2}$-BPS circular Wilson loop in ABJM theory interpreted in the grand-canonical ensemble   according to the conjecture of  arXiv:2505.21633. Here we study the ${\rm T}^{-1}_2$  correction in the case of the  M2 brane instanton wrapping $S^3/\mathbb Z_k\subset S^7/\mathbb Z_k$  dual to the leading non-perturbative  large $N$ contribution  to  the ABJM free energy on  3-sphere.  The   1-loop instanton M2  brane  partition function  was shown  in arXiv:2307.14112  to match the localization prediction. The 2-loop computation is complicated by the presence of bosonic and fermionic zero modes,
which   requires  to integrate over the collective coordinates.
The  collective-coordinate  Jacobian  provides a  non-trivial ${\rm T}^{-1}_2$  contribution   in addition to the  one of  the  quartic interaction  vertex in the M2 brane action. We find that for $k>2$ the 
 the quartic vertex contribution is cancelled by the Jacobian contribution 
  for the choice of the separation of the fermionic zero modes from the quantum fields 
under which these form separate supermultiplets. 
Assuming that the supersymmetric prescription for the paired zero modes, which defines the non-zero
1-loop result, extends to the interacting theory as a supersymmetric deformation of the path integral
with an invariant regulated measure, we show that the cancellation persists after the integration over
the collective coordinates: the deformation localizes the integral onto two configurations, and the
nonlinear measure selected by it reproduces the Jacobian contribution at the reference saddle, so that
the $\rT_2^{-1}$ correction to the instanton contribution vanishes, in agreement with the
grand-canonical prediction. 
 For   $k=1$  the quartic vertex  contribution  reduces to equation-of-motion terms as in the case  of  ${\rm AdS}_3 \subset  { \rm AdS}_7 \times S^4$  in  arXiv:2511.22306  which is 
related to $S^3 \subset {\rm AdS}_4 \times S^7$  case  by a formal analytic continuation.
\end{abstract}


\end{titlepage}
{\footnotesize
\makeatletter
\newcommand*{\toccontents}{\@starttoc{toc}}
\makeatother
\toccontents
}

\def \ff {{\rm f}}

\def \ree {{\rm e}}
\def \rF {{\rm F}} \def \rZ {{\rm Z}}
 \def \XX {{\cal X}}
\def \U {U}
\def \M {{\cal M}}  \def \JJ {{\rm \hat J}}
\def \wA  {{\mathcal T}_N}
\def \wJ  {{\mathcal  T}_C}
\def \yy   {{\rm y} }
 \def \ka {\kappa}
 \def \DH {{\rm D}}
\def \hh {{\rm h}}
\def \cU  {{\cal U}}
\def \EOM  {{\rm EOM}}
\def \CC {{\mathbb C}}
\def \k {\kappa}
\def \RRR  {\mathbb R}
\def \cI {{\rm I}}
\def \Gr {{\rm Gr}}
\def \l {\lambda}
\def \mod {{\rm mod}}
\def \rL  {{\rm L}}

\def \CJ  {{\mathscr  J}}
\def \aZ  {{\cal Z}}  \def \td {\tilde}
\def \tX    {\td \X}
\def \rJ {{\rm J}}
\def \ve {\varepsilon}
\def \La  {\Lambda}  \def \ma  {^{(a)}} \def \Z {\La} \def \N  {{\cal N}}
\def \cj {{\cal J}}
\def \aa {{\rm a }}
\def \rR {{\cal R}}
\def \ag {\alpha}
\def \cG  {\mathcal G}
\def \rK {{\rm K}}
\def \HJ {\upkappa}

 \def \JJJ    {{\cal J}}
\def \rN {{\rm N}} 
\def \hA {{\mathcal T}_N}
\def \JJ  {{\mathcal  T}_C}
\def \rb {{\rm b}}
\def \cZ  {{\cal Z}}\def \rq  {{\rm q}} 
\def \vva  {\upepsilon}

\def \vC {{\rm c}}
\def \mQ  {\widehat {\mathcal Q}}
\def \mL {\mathsf L}
\def \qq  {{\rm q}}
\def \ML {{\mathscr L}}
\def \cR {\mathcal R}   \def \cQ  {\mathcal Q}
\newcommand{\sL}{\mathsf L}
\def \WW {{\rm W}}
\def \rU {{\rm U}}

\setcounter{footnote}{0}
\section{Introduction and summary}\label{sec1}

 Quantum M2 branes \ci{Bergshoeff:1987cm,Bergshoeff:1987qx}  in 11d backgrounds provide an important tool for studying AdS/CFT duality 
 beyond the weak-coupling string theory  approximation. 
  While a   semiclassical quantization of  an M2  brane  near a 
  particular  saddle   is straightforward\ci{Duff:1987cs,Harvey:1999as,Giombi:2023vzu,Beccaria:2023ujc,Beccaria:2023sph,Drukker:2023bip,Giombi:2024itd,Beccaria:2025vdj,Beccaria:2025npl,Gautason:2025bft,Tseytlin:2026rxv,Kurlyand:2026yke}, 
    going beyond  the 1-loop approximation is, in general,     challenging 
   due, in particular, to    non-renormalizability of the M2 brane 3d world volume 
    theory  starting  from 2-loop order  \ci{Beccaria:2025xry}. 

     Still,  on    several recent  examples  it was  found that the 2-loop 
     correction to the  M2 brane   partition  function is free of logarithmic UV divergences and thus 
     well  defined  in   standard analytic regularization prescription  \ci{Beccaria:2025ahf,Tseytlin:2026ctl,Beccaria:2026zjj}.
     The present paper is  a  continuation of this   series of 2-loop   computations.  
     
   {\bf 1}.   Here we will 
   consider the  partition  function of  an M2  brane   in $\AdS_4\times S^7/\ZZ_k$  
   expanded  near  the Euclidean   instanton  3-surface $S^3/\ZZ_k \subset  S^7/\ZZ_k$.
      In the dual  ABJM theory~\cite{Aharony:2008ug}  it should  correspond to the leading one-instanton 
      part of   the
       non-perturbative  large-$N$  contribution  to the   ABJM free energy   
  on  a  3-sphere, 
  $\rF=-\log \rZ= \rF^{\rm p}+\rF^{\rm np}, $ \ $ \rF^{\rm np} = \rF^{\rm 1-inst}  + ...$, where 
  \ci{Drukker:2011zy,Hatsuda:2013gj}
   ($C=\frac{2}{\pi^2k}, \   B=\frac{k}{24}+\frac{1}{3k}$, $k>2$)   
  \al{
\la{01}
\rF^{\rm 1-inst}(N,k) &
 =  - \frac{1}{\sin^{2} \frac{2\pi}{k}   } 
 \frac{\text{Ai}[C^{-{1}/{3}}(N-B+\tfrac{4}{k})]}{\text{Ai}[C^{-{1}/{3}}(N-B(k))]} 
   =  \rF_1(N,k) \, \big[1  +   {\te   {\pi \ov  \sqrt {2k}  }  {k^2-40\ov 12k } {1 \ov \sqrt N} }+ ...       \big] \ , 
\\
 \rF_1(N,k)  &= 
 - \frac{1} {\sin^{2} \frac{2\pi}{k}   } 
 \, e^{-2\pi  \sqrt{2N\ov k }}  =  - \frac{1} {\sin^{2} \frac{2\pi}{k}  } \, 
 e^{-S_{0} } \ , \la{02} \\  \te 
    S_{0} &\te =  \rT_2 V \ , \qquad  \    V(S^3/\ZZ_k)= {2\pi^2\ov k} \ , \qquad \   \rT_2 = {1\ov \pi} \sqrt{ 2 N k} 
   \ . \la{03}
}
As was shown   in  \ci{Beccaria:2023ujc}  the leading   large $N$  term in the 1-instanton 
contribution in \rf{02}
   is   the same as  (minus)  the 1-loop  $S^3/\ZZ_k$   M2  brane   partition function.\foot{A
    complementary geometric  analysis of this 1-loop problem was given in \cite{Kurlyand:2026yke}.}
   The exponential part    comes  from the  classical action $S_{\rm cl} $ of an M2 brane  instanton with  the 
   $S^3/\ZZ_k$ world-volume geometry
   while the $ (\sin^2\frac{2\pi}{k})^{-1}  $ prefactor  represents  the 1-loop M2 brane
   correction.\foot{Such M2 brane  instanton wraps the 11d
 circle $S^1$ and a $\CP^1$ in $\CP^3$. It 
 represents the M-theory uplift of the    $\CP^1$ instanton  in  type IIA string theory on AdS$_4 \times \CP^3$ 
 \ci{Cagnazzo:2009zh}.  The IIA superstring   computation of the  leading large $k$ 
 term in this instanton prefactor $\frac{1}{\sin^{2} \frac{2\pi}{k} } \to {k^2\ov (2 \pi)^2}  = { 2 T_1 \ov \pi \gs^2} $ 
  was  presented  in  \ci{Gautason:2023igo}.
  } 
  
  Naively, one  could  expect  
   that the  subleading $ {   {\pi \ov  \sqrt {2k}  }  {k^2-40\ov 12k } {1 \ov \sqrt N} }$ term  in 
  \rf{01}  should   correspond to the 2-loop   M2  brane  correction as it scales  as  
   $\sim \rT_2^{-1}$, where $\rT_2$ is  the effective M2 brane tension. 
    However,  as was pointed out in 
  \cite{Gautason:2025per,Gautason:2025plx} (see also \cite{Gautason:2025bft,vanMuiden:2026nsp,Bobev:2026gir})
   the   M2  brane    partition function  should    correspond  not to the fixed $N$ (or ``canonical ensemble'') 
    expression in \rf{01}  but  to the grand-canonical  one where  the  chemical potential $\mu$ (conjugate to $N$)    and $k$  are   held fixed (cf.  $\rZ(N)= \int d \mu \, e^{J-\mu N}$).  
    This suggestion  implies   an  important sharpening of the AdS/CFT duality  rules   that  thus 
    requires    further  consistency  checks. 
    
    This ensemble conjecture  received 
    strong support in \ci{Beccaria:2026zjj}  where  the 
 2-loop correction to  the partition function of an M2   brane on $\AdS_2 \times S^1/\ZZ_k  \subset  \AdS_4 \times S^7/\ZZ_k$ 
 was   shown to be zero, in full  agreement with  the 1-loop exactness   of the grand-canonical ensemble  localization 
 prediction for the expectation value of the  $1\ov2$-BPS  Wilson loop in the ABJM theory.

    In the  case of free energy on 3-sphere  the grand-canonical analog of \rf{01} is  given by 
     \be 
     J^{\rm 1-inst}(\mu,k)   =    \frac{1} {\sin^{2} \frac{2\pi}{k}  }  e^{- S_0 } 
      \  , \te 
   \qquad \quad   S_0= {4\ov k} \mu  = V  \rT_2^{(\mu)} \ , \qquad \rT_2^{(\mu)}= {2\mu\ov \pi^2} \ .  \la{04}
       \ee
Thus  from the M2   brane  point of view   it   should be    again   1-loop   exact.

    As we find below,   the coefficient  of the $\rT_2^{-1}$ term  in  the M2 brane partition function in   $ \AdS_4 \times S^7/\ZZ_k$ 
 expanded near the 
     $S^3/\ZZ_k$   instanton saddle  vanishes, 
     providing another non-trivial   consistency check of the grand-canonical  ensemble
conjecture of \cite{Gautason:2025per,Gautason:2025plx}.
 We shall first establish the cancellation of  $\rT_2^{-1}$ term   at the origin of the collective-coordinate 
  integrand in section  \ref{sec5} and then in  section~\ref{n1} 
   extend this result to the integral over the
collective coordinates, assuming a particular   supersymmetric prescription.

 \
  
  {\bf 2}.
 The  computation of the 2-loop   contribution  to the  partition function of the M2 brane on 
 $S^3/\ZZ_k  \subset  \AdS_4 \times S^7/\ZZ_k$
 \be 
 Z=\int [D{\rm X}]\ e^{-S[{\rm X}] } \ , \la{229}
 \ee
   coming from the vertices in the expansion of the 
  M2  brane action $S$ discussed 
    below  is structurally 
  similar 
 to the one  for the M2  on  $\AdS_3   \subset  \AdS_7 \times S^4$ in 
 \ci{Beccaria:2025ahf}  and for the M2  on $\AdS_2 \times S^1/\ZZ_k  \subset  \AdS_4 \times S^7/\ZZ_k$   in     \ci{Beccaria:2026zjj}.
 In all of these three cases   the starting point is  the M2  brane action    \ci{Bergshoeff:1987cm,Bergshoeff:1987qx}  
 in the  corresponding 11d supercoset  background \ci{deWit:1998yu,Claus:1998fh}. 
 
 We use again  the static and adapted $\kappa$-symmetry gauges of
\ci{Seibold:2024oyr,Beccaria:2025xry,Beccaria:2025ahf,Beccaria:2026zjj}.
In these gauges the M2-brane action  expanded  in fluctuation  fields has no cubic interaction vertices: 
\be \la{005}  
S= S_0 + S_2 + \rT_2^{-1}  S_4 + \OO(\rT_2^{-2} ) \ , \ee
where   we  rescaled the fluctuations  by $\rT^{-1/2}_2$  to have   canonical normalization of the  quadratic term $S_2$. 

The $\rT_2^{-1}$  correction $\ff_2$  to the free energy   may be written as 
\be \la{05}
F=-\log Z = \rT_2  \ff_0   + \ff _1  + \rT_2^{-1} \ff_2  + ... \ .
\ee
It  receives, in particular, a   contribution from
 the expectation values  of the quartic (bosonic, mixed   and fermionic)  interaction terms in the classical action.
 
   The presence of the zero modes implies that \rf{229} should contain also the integral over the
collective coordinates, and that the collective-coordinate Jacobian may contribute to $\ff_2$ (see below).
Ignoring at first both the Jacobian and the collective-coordinate integral, the $\rT_2^{-1}$ term in \rf{005}
gives the contribution of the quartic vertices
  \be \la{06}
f^{(0)}_{2} = \langle  S_4 \rangle \ , \qquad \qquad \qquad 
S_4= S_{4b} + S_{2b,2f} + S_{4f} \ . \ee
 $ \langle  S_4 \rangle$  is   given by a sum of  products 
 of  8  bosonic  plus  8  fermionic  3d   fluctuation-field    Green's   functions and their first and second derivatives at coinciding points. In  3 dimensions   these Green's functions contain no log UV divergences, so 
  the resulting 2-loop correction  is  manifestly free of  these too and its finite part can be evaluated
  in the standard  analytic  regularization  prescription. 
  
  In both  cases  considered  in \ci{Beccaria:2025ahf}    and \ci{Beccaria:2026zjj}   
  the  2-loop   correction $f_2$ 
    was found to vanish. 
  The mechanisms  of this vanishing were, however,   different. 
  
  In the $\AdS_3   \subset  \AdS_7 \times S^4$ case  in \ci{Beccaria:2025ahf}
  it was  shown  that  each of  the  contributions $\langle S_{4b} \rangle, \ \langle S_{2b,2f}\rangle 
  $ and $\langle S_{4f}\rangle  $  in \rf{06}  were { separately} proportional to  the ``equations-of-motion'' (EOM)
   combinations of Green's functions  and their derivatives  at coinciding points 
   (e.g. $(-\nabla^2 + m^2) G(\s,\s')\big|_{\s\to\s'}$  in a scalar case).
    As a result,  the  three   contributions   vanished separately in an analytic (dimensional or $\z$-function)
     regularization  that sets $\delta(0)$   factors   to zero.
    
  This  appeared to  be a consequence of:
     (i)  the fact  that the  classical  surface   is embedded  
  geodesically  in  just 
   one   maximally symmetric-space 
   factor of the  11d  target space;
    (ii)  the  special structure of the quartic vertex in the bosonic
      (induced volume) part of the M2 brane action;
    (iii) an  effective supersymmetry  of the world-volume   action (a residue of the target space supersymmetry  in  the  $\kappa$-symmetry gauge)  that relates  the bosonic and fermionic quartic vertices. 
        
 In contrast, in  the  $\AdS_2 \times S^1/\ZZ_k    \subset  \AdS_4 \times S^7/\ZZ_k 
 $ case  in \ci{Beccaria:2026zjj}  the three   contributions to \rf{06} 
  were non-zero.
 However,  due to the underlying supersymmetry, 
 they  combined in a remarkable way   into a ``perfect square'', 
 $
\langle S_4 \rangle
\sim \mathcal X^2 , 
$
where $\mathcal X$  was   a linear  combination of the bosonic and fermionic Green's functions
at coincident points. Representing $\XX$  
as  a  sum over the $S^1$   Fourier  modes   and  accounting 
for  the contribution of the massless
(level-zero) fermion  contribution in a  supersymmetry-consistent 
 way  it was found   \ci{Beccaria:2026zjj} that  the 
 $\zeta$-function regularized value of ${\XX}$ vanishes,   so that    $f_2=0$. 

\
 
  {\bf 3.} The present case of   the  M2 brane on 
 $S^3/\ZZ_k  \subset  \AdS_4 \times S^7/\ZZ_k$  is   analogous  to the  one of 
 $\AdS_3   \subset  \AdS_7 \times S^4$  in \ci{Beccaria:2025ahf}. 
 First, we 
   note  that  the $\ZZ_k$ orbifolding  of $S^7$  is done on a single Hopf circle  
   coordinate  identified in the static gauge with  the third  world-volume   coordinate.
      As in \ci{Beccaria:2026zjj}, the 
 vertices  in the M2 brane  action do not depend on $k$,   and  thus, at least naively
 (ignoring at the moment   zero-mode  issues),  
 a    non-trivial $k$-dependence 
  should    enter  only   via the 3d   Green's  functions of the fluctuation  operators on $S^3/\ZZ_k$, 
      while the general   structure of  $\langle  S_4\rangle$  in terms of the 3d Green's functions  should   be 
     the same as in the    formal   $k=1$  case  of  $S^3  \subset  \AdS_4 \times S^7$.

        The second observation is  that the $k=1$ 
           case of an  M2  brane on $S^3  \subset  S^7$   in $ \AdS_4 \times S^7$  is related by a formal analytic continuation to the case of  an M2  brane on $\AdS_3  \subset  \AdS_7 $ in   $ \AdS_7  \times S^4$.\foot{This was mentioned already in 
         footnote 15 in   \ci{Beccaria:2025ahf} as a possible reason of the vanishing of 
         2-loop correction 
         in the $k=1$   case of   $S^3 \subset \AdS_4 \times S^7$.}
             Hence    the    structure of  $\langle  S_4\rangle$  in both cases   written in 
     terms of the corresponding Green's functions (on $\AdS_3$ and $S^3$  respectively) 
      should be  similar.  
      
       Then the result of \ci{Beccaria:2025ahf}  suggests  that   in the 
       $k=1$   case  of  $S^3\subset  \AdS_4 \times S^7$  the 2-loop contribution 
      $\langle  S_4\rangle$
         given   by the sum  of  $\langle S_{4b} \rangle, \ \langle S_{2b,2f}\rangle 
  $ and $\langle S_{4f}\rangle  $    each of  these terms  are 
        proportional to an  EOM  combination of the  corresponding Green's functions. 
      The   vanishing  of the 2-loop   correction   would then follow  after  using 
          the $\delta(0)=0$   regularization.

As we shall  see below, this  argument overlooks two important subtleties.

First, 
 for $k>2$  the definition  of the Green's functions on $S^3/\ZZ_k$  involves summing over images
in $S^1$   direction and  thus breaks the manifest global $S^3$  symmetry in the contractions of derivatives of the Green's functions,  leading to a somewhat different  structure of 
$\langle  S_4\rangle$
  compared to the  $k=1$ case   or  the $\AdS_3  \subset  \AdS_7\times S^4  $ case.
   Due to this anisotropy 
   $\langle S_{4b} \rangle, \ \langle S_{2b,2f}\rangle 
  $ and $\langle S_{4f}\rangle  $  each  will  not be separately proportional to the EOM terms.
  However, the effect of this  geometric  anisotropy
   cancels  in their  sum $\langle S_{4} \rangle$  (this should be due to  the underlying supersymmetry,
    cf. Appendix~\ref{api}).\foot{The criterion turns
out to be the isotropy group of the wrapped surface: for $k=1$ and $k=2$
 the isotropy is the full $SO(3)$, and each of the three terms 
$\vev{\cL_{4b}}$, $\vev{\cL_{2b,2f}}$, $\vev{\cL_{4f}}$ is separately of the 
EOM  type. For $k>2$ the isotropy  is  $U(1)$ and this  vanishing 
``sector-by-sector''   EOM property fails. 
 The bosonic and fermionic anisotropies turn out to
be effectively equal and cancel in the sum over sectors.}

   The  second   novel  issue  is  the presence of zero modes: 
    while  in the $\AdS_3 $  \ci{Beccaria:2025ahf}   and $\AdS_2 \times S^1/\ZZ_k$ \ci{Beccaria:2026zjj}  world volume  cases     the   fluctuation operators   had  no normalizable  zero-modes, 
   this is no longer  so in the  case  of  $S^3/\ZZ_k   \subset  S^7/\ZZ_k$.
  
  \
  
{\bf 4.}  
  The  kinetic   operators   of   8+8   fluctuation  fields on  $S^3/\ZZ_k $
   have  12 bosonic  and 12  fermionic zero modes   \cite{Beccaria:2023ujc}. The bosonic modes move the location  of the instanton in $\AdS_4$ and rotate the  corresponding  2-plane   in  the embedding   space $\mathbb C^4$, while the fermionic ones are generated by the 12   broken supersymmetries.  At 1-loop order 
these modes only affect the overall normalization of $\sin^{-2}$ factor in  \rf{nn2}.\foot{These modes appear at the
      $n=0$ (lowest)  $S^1$  level, i.e. they  are present already  in the corresponding  type IIA string  partition function 
      for $\CP^1\subset \CP^3$ instanton 
     \cite{Gautason:2023igo}. The contributions 
 of the equal number of 12+12  bosonic and fermionic modes  can  be regularized 
 and were  shown to  cancel   after introducing a squashing parameter in the  $\CP^3$ 
metric  \cite{Gautason:2023igo}. Then there are     two   possible string $\CP^1$ saddles  contributing equally 
 and   giving an extra  factor of 2  in the 1-loop   normalization
    \cite{Gautason:2023igo,Gautason:2025per}.
 As  suggested in  \ci{Beccaria:2023ujc},    
  it may be   possible to  derive this  result systematically without  squashing $\CP^3$ 
by  introducing  collective  coordinates for the bosonic and fermionic 
zero-modes and  computing the volume of the corresponding supercoset.}

 The presence of the zero modes has  several novel   consequences:
  
   (i) Green's functions are   projected ones,  i.e.  the  EOM terms  at coinciding points 
    contain  the  projector  contributions  that  survive  in the  $\delta(0)=0$  regularization, i.e.    the EOM terms 
   in $\langle S_{4} \rangle$  no longer vanish;
   
  (ii)  There are extra   anisotropic contributions to   $\langle S_{4} \rangle$   related to the   presence of the  zero modes;
  
  (iii) The zero modes of kinetic operators of fluctuation fields require the introduction of the associated collective   coordinates in the path integral;  the 
     corresponding Jacobian   (Berezinian) term $\JJJ$   in the    measure   happens   to 
     produce a  finite contribution to the  collective-coordinate integrand   starting at  the $\rT_2^{-1}$  order. 
     
     
     (iv) Finally, the remaining integral over the $12+12$ collective coordinates has to be defined and
computed. The result has two calculational parts and one assumption. The reference-saddle computation of
section \ref{sec5} gives $f_2(0;k)=0$ in the supersymmetry-adapted complement. Appendix \ref{apf}
computes the measure contribution selected by the paired zero-mode prescription, implemented as a
supersymmetric deformation of the path integral, at the two configurations onto which it localizes,
and finds that it reproduces the collective-coordinate Jacobian of section \ref{sec5}. Section \ref{n1}
then relates these local results to the full integral, assuming that the deformation is a symmetry of
the regulated measure at order $\rT_2^{-1}$. Under this assumption the $12+12$ zero modes, which are
paired by the preserved supercharge, are lifted as in \ci{Gautason:2023igo,Kurlyand:2026yke} at one loop,
and the integrated $\rT_2^{-1}$ correction vanishes.

  As we shall find below, different   effects   of the presence of zero modes effectively compensate each other 
  in the  $\rT_2^{-1}$  correction $f_2$  in \rf{05}. 
  Namely, the  non-zero value of $\langle S_{4} \rangle$  gets  cancelled by a   non-zero    contribution 
  of the collective coordinate Jacobian.

  Let $\mathrm X$ denote the gauge-fixed transverse field coordinates,
with the reference saddle at $\mathrm X=0$, and let
$\widetilde{\mathrm X}$ be the canonically normalized 3d quantum fields.
There are four real AdS bosons $x^r$, four real ``internal''  bosons $y^m$, and eight real
3d  fermions. The corresponding 
linear  decomposition is
\begin{equation}
 \mathrm X=\mathrm a^A\mathcal Z_A+\epsilon\widetilde{\mathrm X},
 \qquad \Lambda^A\widetilde{\mathrm X}=0,
 \qquad \Lambda^A\mathcal Z_B=\delta^A{}_B, \qquad  \ \   \epsilon\equiv \mathrm T_2^{-1/2} \ , 
 \label{002}
\end{equation}
where $\mathcal Z_A$ are the $12+12$ tangent zero modes  and $\mathrm a^A$ are the corresponding 
collective coordinates. A nonlinear analog of this  decomposition 
for a  family of saddles is  given in \eqref{928}.
Then the   M2 brane  path integral measure may be written as  
\begin{align}
 [D\mathrm X]& = d^{12|12}\mathrm a\,[D\widetilde{\mathrm X}]
       \,\delta(\Lambda\widetilde{\mathrm X})\,\operatorname{Ber}\mathcal M,
 \qquad \qquad \mathcal M^A{}_B=\Lambda^A\rK_B[\epsilon\widetilde{\mathrm X}],
 \label{001}\\
 &\log\operatorname{Ber}\mathcal M=\epsilon^2\mathcal J+O(\epsilon^3),
 \qquad \qquad\qquad  \mathcal J=\widetilde{\mathrm X}\,\mathcal W\,\widetilde{\mathrm X}.
 \label{010}
\end{align}
Here the generators $\rK_B$ include the compensating static-gauge and
$\kappa$-symmetry  transformations.  $\mathcal W$
is a  local  differential operator  (the notation 
$\widetilde{\mathrm X}\,\mathcal W\,\widetilde{\mathrm X}$ assumes    3d integral). 

Denoting the normalized  collective-coordinate measure as $d\mu_0$  
 the resulting   partition function \rf{229}  expanded in $\Tmem^{-1} $ may be written as 
\al{
 &Z=Z_1 \int d\mu_0(\mathrm a)
 \Big[1-\Tmem^{-1}f_2(\mathrm a;k)+\OO(\Tmem^{-2})\Big]  \ ,\qquad \qquad \qquad \int d\mu_0(\mathrm a) =1 \ ,  \la{n2}\\
 & \qquad \ \ \ \ \ \ \ 
   { Z}_1  =  e^{-\Tmem V}\,  ( \sin \tfrac{2\pi}{ k})^{-2} \ , 
 \label{nn2}
}
where we separated the classical  and 1-loop   contributions (cf. \rf{02},\rf{03}).\foot{Let us note that 
the integral over $\rm a$  in  \rf{n2},\rf{n3} is  formal: the Berezin integral over the 12 odd collective coordinates
of a constant density vanishes, while the integral over the non-compact $\AdS_4$ directions diverges.
The measure $d\mu_0$ is defined by the supersymmetric prescription for the paired zero modes
(see Appendix \ref{apps}) which is implemented in section \ref{n1} as a supersymmetric deformation of the path
integral. 
The normalization $\int d\mu_0=1$ in \rf{n2} is relative to the complete leading amplitude $Z_1$ taken
from the 1-loop computation of \ci{Beccaria:2023ujc}, whose prefactor $\gamma/[4\sin^2(2\pi/k)]$
with $\gamma=4$ includes the zero-mode and orientation counting discussed in \ci{Gautason:2023igo}.}
 $f_2(\mathrm a;k)$ denotes the $\Tmem^{-1}$ coefficient in the integrand  at fixed  values of collective
coordinates.  The final expression  for the $f_2$  coefficient in the free energy in \rf{05}   is then  given by 
\begin{equation}
 \ff_2(k)=\int d\mu_0(\mathrm a) \,  \,f_2(\mathrm a;k) \ . 
           \label{n3}
\end{equation}
  One  key point  is that the  contribution of the  Jacobian in  \rf{010}   corrects the  2-loop  expression  \rf{06}
 for   the  $\Tmem^{-1}$   correction computed at $\mathrm a=0$   to  
 \begin{equation}
 f_2(0;k)
 =\vev{S_4}-\vev{\cJ}
 =V\vev{\cL_4}-\vev{\cJ}\ , \qquad \ \ \ \   \qquad   V=2\pi^2 k^{-1}   \ .
 \label{011}
\end{equation}
Here we used the fact that  as   the world-volume is homogeneous,  the quartic vertex 
expectation value is proportional to the  3-volume $V$  ($S_4 =\int d^3 \s\sqrt g\, \cL_4$).

In \rf{011} and in sections~\ref{sec2}--\ref{sec5} below  the quantum
expectation values will  be always  evaluated at the origin 
$\mathrm a=0$ of the  collective-coordinate space (or at the  ``reference saddle").
We shall find that $f_2(0;k)=0$.

The relation  of the vanishing of $f_2(0;k)$ to    the vanishing of $\ff_2$ in \rf{05}, \rf{n3} 
will be  argued for   in section~\ref{n1}  (no  requirement  that
$f_2(\mathrm a;k)$  should be  zero on the full super-orbit $\mathrm a\ne 0$   will  be  needed).

$\vev{\cL_4}$ in \rf{011}  can be  expressed in terms of  quadratic combinations 
 of the coincident-point   values   of the bosonic and fermionic Green functions and their first and second  derivatives.
 After  using the $\delta(0)=0$ regularization in the EOM terms and accounting for the zero-mode   projector contributions 
  we will  find  below  that 
\begin{equation}
\vev{\cL_4}
=   -\tfrac23\,V^{-1}\big(28\,G_x+13\,G_y\big) + \tfrac53\,V^{-2}   \ . 
 \label{012}
\end{equation}
Here  $G_x$ and $G_y$ are  the  coincident values of the  Green functions of one real $x$ and $y$  bosons on 3-space  $S^3/\ZZ_k$ 
 that are non-trivial (transcendental) functions  of $k$. 
 
The expectation   value of  the quadratic term in the 
 collective-coordinate  Jacobian  in \rf{010} 
  happens  to be 
\begin{equation}
  \vev{\cJ}= -\tfrac23\, \big(28\,G_x+13\,G_y\big) +  \tfrac53\,V^{-1} \ . 
 \label{013}
\end{equation}
The    computation  of $  \vev{\cJ}$ uses   the  data from the   quadratic-fluctuation 
part $S_2$  of the M2 brane action together   with the symmetry transformations    and 
is done in a way  consistent 
with   the  residual   3d supersymmetry of  the action.\foot{The Gaussian contraction uses the quadratic fluctuation Green's  functions.
    The finite Jacobian is determined in addition by the target-space
    generators to quadratic order in the fields, including the
    compensation  transformations  restoring static and $\kappa$-symmetry 
     gauge. Linear world-volume
    supersymmetry is used to select the fermionic complement of the fermionic zero mode part 
     (see section \ref{sec5} and Appendix \ref{apd}).} 

Remarkably, the expression in \rf{013} 
 is the same  as \rf{012}   up to a factor of $V$. Thus    combining  it   with \rf{012}    according to   \rf{011}     gives 
 \be  f_2(0;k)=0 \ .
 \label{014}
\end{equation}
  The possibility of the cancellation between the  contributions of the 
 ordinary 2-loop vacuum diagrams
 \rf{012} and  the  local Jacobian \rf{010} associated with the $12+12$ collective-coordinate  separation  \rf{013}
   depends crucially   on the absence of the  quadratic $G\times G$   terms in  $\vev{\cL_4}$. Like 
    in the   $\AdS_3 \subset \AdS_7 \times S^4$   case in  \ci{Beccaria:2025ahf} the latter 
    is a consequence of the special  structure of the M2  brane action,  
    underlying supersymmetry and  an  analytic regularization where $\delta(0)=0$.\foot{As we shall see below,  the ``anisotropic'' terms
quadratic in the second derivative  constant  $G_{x,33}$ cancel between the  bosonic and fermionic sectors by
supersymmetry, and the remaining products of a propagator with an
EOM  combination reduce to  linear terms in 
Green's functions   once  one uses that $\delta(0)=0$.}

    The cancellation in \rf{014} has two parts of different status. The coefficients of
$G_x$ and $G_y$  functions  which carry all of the non-trivial dependence on $k$ 
 coincide in \rf{012}
and \rf{013} for any choice of the dual functionals $\Lambda^A$ in \rf{002}.
The rational
$V^{-1}= {1\ov 2 \pi^2} k $ terms, on the other hand, depend on how the  
fermionic zero modes
are separated from the quantum fields. 
For the  $\a$-parameter family of such separations
considered in Appendix \ref{apd}  one finds
\be \la{015}
 f_2(0;k;\alpha) = 4\,(1-\alpha^2)\,V^{-1}\ ,
\ee
where $\alpha=0$ corresponds to the ``orthogonal'' complement and $\alpha=\pm 1$ to the
complement  adapted to a preserved supercharge.
Thus \rf{014} is the vanishing of the reference-point ($\mathrm a=0$)  coefficient for 
the supersymmetry-adapted complement. 
The extension of this vanishing  to the integrated
coefficient in \rf{n3}  is  discussed  in section~\ref{n1}.

    Finally, let us note that a similar   2-loop   result  in the case of   type IIA string   $\CP^1$ instanton in 
    $\AdS_4 \times \CP^3$  \ci{Cagnazzo:2009zh,Gautason:2023igo}
     can be obtained  by taking the $k\to \infty $, $\rT_1= {\pi\ov 2k} \rT_2$=fixed 
       limit of the M2  brane  expression for the free energy, 
       i.e.   $
 \ff_{2,\mathrm{str}}
 =\frac{\pi}{2}\lim_{k\to\infty} (k^{-1} {\ff_2(k)})=0.
$
     In general, such    limit gives a properly regularized 
        expression for the type IIA   GS string partition function  (see   \ci{Giombi:2023vzu,Seibold:2024oyr,Beccaria:2026zjj}).

\


{\bf 5.}
The rest of the paper is organized as follows.  The quadratic part of the fluctuation action $S_2$   is  reviewed in section 2.
In section 3 we discuss  the Green's functions  for the  fluctuation  fields on $S^3/\ZZ_k$  and their  coincident-point limits,  
 their derivatives and the EOM  terms. We  comment on  geometric symmetries   that for $k>2$ 
 lead to an anisotropy  in the values of derivatives of Green's functions. We also explain the origin of the anisotropy using  the image sum representation for  the Green's functions. 

In section 4  we first present  the  expressions  for   each of the three  terms in the quartic interaction Lagrangian 
$\cL_4=\cL_{4b} + \cL_{2b,2f} + \cL_{4f}$  and then   compute their expectation values. Summing them up 
   leads to  the expression for $\vev{\cL_4}$  in \rf{012}.
 The contribution   of  the  collective-coordinate Jacobian $\cJ$ \rf{013}  is found   in section 5. 
 
 Section~\ref{n1} explains how the
same cancellation extends to the collective-coordinate integral \rf{n3} under
the  supersymmetric prescription. Some concluding remarks are made 
in section~\ref{sec6}.

 
 The spectral representation of Green's functions on $S^3/\ZZ_k$  and their derivatives  in the coincident-point limit are 
   presented  in Appendix  A.
  In Appendix B we  discuss the 12+12  zero modes.  In Appendix  C we review some general  facts about the collective 
   coordinates and the associated  Jacobian in    path integral measure.  In Appendix D  we  specialize   this 
     discussion to the present M2  brane  case. 
     
      In Appendix  E we  comment on  
     normalization of path integral  in the case of equal numbers of  bosonic and fermionic zero modes. 
     Appendix~\ref{apf} identifies the two supersymmetric configurations
selected by the deformation discussed  in section \ref{n1} and evaluates the local measure in the
paired zero-mode prescription, including the contribution associated with the
equation-of-motion term in the  supersymmetry algebra.

\section{Quadratic  part of fluctuation action}\label{sec2}

The  11d    $\AdS_4\times S^7/\ZZ_k$    background  
 and the   Euclidean  M2  brane action  are  the same as 
discussed   in  \ci{Beccaria:2023ujc,Beccaria:2026zjj}
\begin{align}
 \dd s_{11}^2&=R^2\big(\tfrac14\dd s^2_{\AdS_4}+\dd s^2_{S^7/\ZZ_k}\big),
 \qquad \qquad\te  F_4=dC_3= -{3\ov 8} \ii  R^3\vol(\AdS_4) \ , \la{1}\\
  S &=T_2\int\dd^3\sigma\sqrt h
 +\ii T_2\int C_3+\hbox{fermions}, \qquad \ \  T_2=\tfrac1{(2\pi)^2\lp^3},\qquad
 \Tmem\equiv R^3T_2 \ . 
 \label{2}
\end{align}
$S^7/\ZZ_k$   may be described  in terms of $\CC^4$ embedding coordinates as  
$\sum_{m=1}^4|Y_m|^2=1, \  Y_m \equiv  e^{2\pi\ii/k} Y_m $   and then 
the $S^3/\ZZ_k$  M2 brane  solution is  represented by $Y_3=Y_4=0$ and is point-like in $\AdS_4$.
The   world-volume metric     of the corresponding  3-surface $\Sigma_3 = S^3/\ZZ_k$ 
 is  ($i,j=1,2,3$)
\al{
& \dd s^2_{S^3/\ZZ_k } = g_{ij} d\s^i d \s^j = e^a\otimes  e^a=
 \frac{\dd z\dd\bar z}{(1+|z|^2)^2}
 +\big(k^{-1} \dd s+A\big)^2,
 \qquad s\equiv  s+2\pi\ , \la{3}\\
&
 e^1+\ii e^2=e^{2\ii s/k} 
 u^{-1} {\dd z},
 \qquad
 e^3=k^{-1} {\dd s}+A,
 \qquad
 A = \tfrac{\ii}{2}u^{-1} \,({z\,\dd\bar z-\bar z\,\dd z}), \ \ \ \  u\equiv  1 + |z|^2 \ ,
 \label{4}\\
&  \ree_1 -\ii \ree_2 = e^{-2\ii s/k} \big( 2 u \del_z   + \ii k \bar z \del_s \big) \ , \qquad \qquad \ree_3= k \del_s \ ,  \la{044}
}
where $\ree_a$  form  the dual vector-field basis ($a=1,2,3$ will denote  tangent-space indices).  
The pull-back of the bosonic $C_3$ form  vanishes on the classical solution  and 
 thus the  M2 brane  action is  
 \be S_{0}=\Tmem V \  , \qquad \qquad    \qquad   V\equiv \vol(S^3/\ZZ_k)\te  = {2\pi^2 \ov k } \la{ee4}\ .\ee
We shall parametrize  the  $\AdS_4$ and $S^7/Z_k $   metrics   as
\begin{equation}
 \tfrac14\dd s^2_{\AdS_4}=\frac{\dd x\cdot\dd x}{(1-x^2)^2},\qquad  \ \ 
 \qquad
 \dd s^2_{S^7/\ZZ_k}=(1-y^2)\dd s^2_{S^3/\ZZ_k }+ \nabla y\cdot  \nabla y
 +\frac{(y\cdot  \nabla y)^2}{1-y^2} \ . 
 \label{5}
\end{equation}
In the static gauge
the four  $x^r$   and four  $y^m$ will represent the 3d bosonic fluctuation fields   that depend on $z^1, z^2$ and 2$\pi$-periodic coordinate $s$. 
 $\nabla$ is the normal-bundle  differential.  $x^r$ are ``uncharged '' with respect to the   Hopf  circle $U(1)$ group, while $y^m$ are charged. 
  If we set 
 $\yy^1= y^1 + \ii y^2, \ \  \yy^2 = y^3 + \ii y^4$  then 
 \be 
  \dd s^2_{S^7/\ZZ_k}=(1-|\yy|^2)\dd s^2_{S^3/\ZZ_k }+ |\nabla \yy^\a|^2
 +\frac{({\rm Re}\,  \bar \yy^\a \nabla \yy^\a )^2}{1-|\yy|^2} \  , \qquad \ \ \ 
 \nabla \yy^\a= d \yy^\a + \ii k^{-1} \yy^\a d s  \ , \ \ \ \a=1,2 \ .  \la{6} 
 \ee
 Equivalently,   we may write
 \begin{align}\la{66}
 \nabla y^m &= \dd y^m + k^{-1}(J_y)^{mn}\,y^n\,\dd s \ , \qquad 
 J_y \equiv  \operatorname{diag}(J_2,J_2)\ ,
 \quad J_2=(\begin{smallmatrix}0&-1\\1&0\end{smallmatrix})\ , 
 \quad I_x \equiv \operatorname{diag}(-J_2,J_2)\ .
\end{align}
  For  future use  we defined also $I_x $, so that    
 $(J_y)_{12}=(J_y)_{34}=-1,
\ 
 (I_x )_{12}=-(I_x )_{34}=+1$.

The bosonic  part of the quadratic fluctuation Lagrangian   is   then 
\begin{equation}\la{7}
 \cL_{2b}=\tfrac12 \nabla_a  x^r \nabla_a  x^r +\tfrac12\big(   \nabla_a  y^m \nabla_a y^m  - 3 y^m y^m \big)
 \ . \ee
 Here  $\nabla_a  x^r = \ree_a x^r$  with $\ree_a$ in \rf{044}  (i.e. 
 $  \nabla_a  x^r \nabla_a  x^r = g^{ij}\del_i x^r\del_j x^r $)   while 
  $\nabla_a y^m $ in \rf{66}  contains, in particular,   a non-trivial  $U(1)$ connection in the 3rd    direction, i.e.
  \be (\nabla_1-\ii\nabla_2)\yy^\a=e^{-2\ii s/k}\big[2u\partial_z+\ii \bar z(k\partial_s+\ii) \big]\yy^\a,
\qquad \ \ \ 
 \nabla_3 \yy^\a = ( k \del_s    +  \ii)  \yy^\a \ .\la{707}  \ee 
 Expanding in Fourier modes in $\s^3=s$  we have $x^r$  modes  with  charge $\lambda= k n $  
 while  modes of  $\yy^\a$ in \rf{6}   have   charge $\lambda= k n +1 $  (see also   \ci{Beccaria:2023ujc}).

For the 11d Majorana spinor  coordinates  in  the M2  brane  action  we will use 
 $\tb=\theta^TC$, with  $C^T=-C$ and $C^{-1}\Gamma_A^TC=-\Gamma_A$. The
$\ka$-symmetry gauge   is  chosen as   
\begin{equation}
P_-\theta=\theta, \qquad \qquad  P_\pm=\tfrac12(1\pm\Gk), \qquad 
 \Gk=\ii\Gamma_{123},
  \label{y8}
\end{equation}
leaving 16  independent   spinor components. We also 
define the constant matrices
\al{
&\Gamma_*=\Gamma_{8\,9\,10\,11}\ , \qquad \qquad 
 \PC=\tfrac12(1-\Gamma_*),
 \qquad\qquad 
 \PN=\tfrac12(1+\Gamma_*),\la{9}\\
 &
  T_C=\rho_3 \wJ \    , \ \ \ \ \wJ= \tfrac{\ii}{2} (\Gamma_{45}+\Gamma_{67}),
 \qquad\qquad 
 T_N=\rho_3  \wA \ , \ \ \ \    \wA= \ii\Gamma_{89}\PN
 \label{1011}\ , \\
 & T_C^2=\PC,
 \qquad
 T_N^2=\PN,
 \qquad
 \tr\,  T_C=\tr\,  T_N=0,
 \qquad
 \tr\, \PC=\tr\, \PN=8\  .\la{1111}
}
Here 
 $\rho_3$  is  3rd component of $\rho_a$   which is   the pullback of the  11d  target space 
gamma matrices along the 3d world-volume. 
In the   gauge  \rf{y8} one has $\Gk\theta=-\theta$ and hence acting on $\theta$ we have 
$\rho_1\rho_2\rho_3=\ii$, so that
\begin{equation}
 \rho_a\rho_b=\delta_{ab}+\ii\epsilon_{abc}\rho_c \ . 
 \label{32}
\end{equation}
The labels  in $\PC$ and $\PN$  indicate that they select the charged ($C$)  and  uncharged  or neutral ($N$) 
 spinor  sectors. 
The quadratic  fermionic Lagrangian is
\al{
& \label{10}\qquad \ \ \   \cL_{2f}= \tfrac12\tb\Dsl\theta,\quad 
 \qquad
 \Dsl=\rho^aD_a=   \rho^a \nabla_a +3\ii\PN, \qquad \ \ \    \DH\equiv \ii\Dsl\ , \\
&
 D_a=\hat \nabla_a+\tfrac{1}{2} \ii \rho_a\Gamma_* =    \nabla_a +  {\ii} \rho_a  P_N   \ ,\qquad
 \qquad   \ \hat \nabla_a = \nabla_a + \tfrac{1}{2}\ii \rho_a \ , \ \ \la{11}\\
 &  \nabla_a = \ree^i_a \del_i=  \ree_a \ , \qquad \qquad 
 \rho_3D_3=\rho_3\ree_3+\ii\PN \ , \la{119}
}
where  $\hat \nabla_a $ is the 3d spinor  covariant derivative and 
   $\ree_a$ was given in \rf{044}. We also used that $\Gamma_*=2P_N-1$ in \rf{9}.\foot{On the unit radius 3-sphere in \rf{3}
$\nabla_{a}\ree_b=\tfrac12[\ree_a,\ree_b]=-\epsilon_{abc}\ree_c$, so that 
$\hat \nabla_a=\nabla_a + \tfrac14\omega_a{}^{bc}\rho_{bc}=\ree_a+\tfrac{\ii}{2}\rho_a.$}
The second part  of $D_a$ comes from the 4-flux term of the
11d supercovariant derivative.

Let $\gamma_*$ be the
eigenvalue of $\Gamma_*$, equal to $-1$ in the charged and $+1$ in the
uncharged sectors. Then on a Fourier mode in the 3rd direction with the corresponding charge $\lambda$ 
we have 
\begin{equation}
 \Dsl=\slashed\nabla^{(\l)}_{S^2}
 +\ii\big(\l\rho_3+\tfrac12+\tfrac32\gamma_*\big),
 \qquad\qquad 
 \rho_3D_3=\ii\big(\l\rho_3-\tfrac12+\tfrac12\gamma_*\big).
 \label{33}
\end{equation}
Let us   comment on the  zero modes for $k>2$ (see Appendix \ref{apa}). 
 They appear only at the $S^1$  Fourier  level $n=0$. 
The $\AdS_4$ operator in \rf{7}   has 4 constant  zero modes. 
Each  complex $\yy^\a$  kinetic field operator in \rf{7}  has two  linear  classical  solutions
 $z_1,z_2$, giving eight real zero modes. The fermionic 
  operator in \rf{10}  has 4 ``charged''  plus  8 ``neutral''  real  zero modes   corresponding to   12  broken supersymmetries.
  
  Below we   focus  on the case of $k>2$.  The cases $k=1,2$  differ in 
 the zero-mode content (see Appendix \ref{apa}) and in the residual  symmetry (isotropy) 
 of the correlators  (see Section \ref{sec3} and Appendix \ref{api}).

\section{Coincident-point   values of Green's  functions and their derivatives}\label{sec3}

The presence of the  zero modes  implies that we need to  define  the  projected Green's function $G'$
of the corresponding kinetic operator $\cO$. For example, for  the orthogonal
complement, $\cO G'=G'\cO=\cI-P_0$, where $P_0$ is the orthogonal
kernel projector  (cf. \rf{942}). 
Below we omit primes on $G$. 


For $k >2$   the $\ZZ_k$ orbifolding of $S^3$  implies that  Green's functions are given by  sums \rf{63}  over images on $S^1$, 
and that, in general,  breaks  the symmetry  between $a=1,2$ and $a=3$  directions in  their derivatives.

We shall  denote by $\vev{...}$   the free-theory expectation value  and by $[...]$ 
the  coincident-point limit (always taken after differentiation). 
Let us  define 
\begin{equation}
 \vev{x ^rx ^{r'}}=G_x \delta^{rr'},\qquad 
 \qquad
 \vev{y^my^{m'}}=G_y\delta^{mm'},
 \label{39}
\end{equation}
\begin{equation}
 \vev{x ^r\nabla_a x ^{r'}}=-G_{x ,a}(I_x )^{rr'},
 \qquad\qquad 
 \vev{y^m\nabla_ay^{m'}}=-G_{y,a}(J_y)^{mm'},
 \label{40}
\end{equation}
where $I_x $ and $J_y$ were introduced  in \rf{66}.  Note that $G_{x,a}$ and $G_{y,a}$  vanish for $a=1,2$ due to the residual symmetry. 
In fact, $G_{x,a}=0$ for all $a$ by symmetry, i.e. we have also 
\be \la{0022}  G_{x   ,3}=0\ . \ee
 For the $\AdS_4$  fluctuations  $SO(4)$  is an exact symmetry of the gauge-fixed
action. Since the bosonic zero modes are separated
by orthogonal projection, the $x^r$ propagator is $SO(4)$ invariant, while the
antisymmetric matrix $I_x$ in \rf{40} is invariant only under a subgroup
$U(2)\subset SO(4)$ (the same $U(2)$ that preserves $\wA$ in \rf{1011}).\foot{This follows also 
from  the representation in  \rf{63}. The Green's function
$G_x(\s,\s')$ depends    just on geodesic distance between  the two points   and  
 after the sum in \rf{63} 
 its $\ree_3= k \del_s$  derivative 
vanishes in the coincident limit. Alternatively,  this follows from the  representation in 
\rf{p2},\rf{pp3}  where the $n$ and $-n$  contributions   cancel each other.}
In some   equations   below we will still keep $G_{x,3}$  explicit  along with non-vanishing  $G_{y,3}$ 
 as a bookkeeping
device  for the $U(2)$-covariant terms of this type that appear in the
collective-coordinate Jacobian in section~\ref{sec5}.

Let us also define $G_{x ,ab}$  and $G_{y ,ab}$   by ($a=1,2,3$) 
\begin{equation}
 \vev{\nabla_a x ^r\nabla_b  x ^{r'}}
 =G_{x ,ab}\delta^{rr'}-G_{x ,3}\epsilon_{ab3}(I_x )^{rr'},
 \qquad
 \vev{\nabla_a y^m\nabla_b  y^{m'}}
 =G_{y,ab}\delta^{mm'}-G_{y,3}\epsilon_{ab3}(J_y)^{mm'} .
 \label{41}
\end{equation}
The equation-of-motion (EOM) combinations are (cf. \rf{7}) 
\al{
 & E_x =\big[\cO_x  G_x \big],\ \ \ \ \ 
 \qquad
 E_y=\big[\cO_yG_y\big] \ , \la{42}\\
 & \cO_{x,y} = -\nabla^2 _{x,y} + M^2_{x,y}  \ , \ \qquad    M^2_x =0 \ , \ \ \quad  M^2_y=-3 \ . 
 \label{442}
}
Then  summing  over  the  tangent   3d index  in \rf{41}  gives 
\begin{equation}
 G_{x ,aa}=E_x -M_x ^2G_x ,
 \qquad
 \qquad  G_{y,aa}=E_y-M_y^2G_y .
 \label{43}
\end{equation}
The  vertices  in the M2-brane  action depend on 
the     induced world-volume metric in the static gauge (cf. \rf{2},\rf{5})  
\al{
 &h_{ij}=\big(1-y^2\big)g_{ij}+\frac{\del_i  x\cdot\del_j  x}{(1-x^2)^2}
 +\nabla_i y\cdot\nabla_j y
 +\frac{(y\cdot\nabla_i  y)(y\cdot\nabla_j  y)}{1-y^2}=  g_{ij}  + H_{ij}  
 + ...\ , 
 \label{100}\\
 &\qquad \qquad  H_{ij} =\del_i  x\cdot\del_j  x+\nabla_i   y\cdot\nabla_j  y -y^2g_{ij}\ ,\la{101}}
 where $i,j$ are the 3d  world-volume indices  and $g_{ij}$ is the same as in \rf{3}. 
The expectation value   of $H_{ab}$ at coincident points is given by (cf. \rf{42})\foot{Note that $G_x$   and $G_y$ are defined 
as a   Green's function for a single  component of the corresponding field.} 
\begin{equation}
\hh_{ab}\equiv \tfrac{1}{4}  \big[\vev{ H_{ab}  }     \big]=G_{x ,ab}+G_{y,ab}-G_y\delta_{ab},
 \qquad\qquad 
 \hh_{aa}=E_x +E_y \ . 
 \label{44}
\end{equation}
For the fermions we  set 
\al{
& G_\theta=\big[\vev{\theta\tb }\big]\equiv  \vev{\theta(\s)\tb(\s')}\Big|_{\s\to \s'}\ ,
 \qquad
 G_{\theta,a}=\big[\vev{D_a \theta\tb }\big]   \equiv  D_{\s,a}\vev{\theta(\s)\tb(\s')}\Big|_{\s\to \s'}\ ,\la{50}  \\
 & \qquad \qquad \qquad 
  E_\theta=\rho^aG_{\theta,a} \ , 
 \label{504}
}
where $E_\theta$   is the corresponding EOM 
combination (cf. \rf{10},\rf{11}).

Expanding  in the gamma-matrix   basis in \rf{9},\rf{1011}   we define  the corresponding  constant  coefficients 
 $s_C, ..., q_N$  as  
\begin{align}
\qquad  G_\theta&=\ii\big(s_C\PC+s_N\PN+v_CT_C+v_NT_N\big)\ ,\la{522} \\
  \qquad \ E_\theta&=r_C\PC+r_N\PN+d_CT_C+d_NT_N\ , 
   \la{52} \\
    \rho_3G_{\theta,3}&=p_C\PC+p_N\PN+q_CT_C+q_NT_N \ , \ \ \ 
    \qquad   \rho_1G_{\theta,1} + \rho_2G_{\theta,2} =E_\theta-\rho_3G_{\theta,3} 
     \          
       ,  \la{5222}
\end{align}
where  the    labels $C$ and $N$ refer  to ``charged'' and ``neutral''  fermion  sectors. 
We shall  also   use 
\begin{equation}
 r\equiv r_C+r_N,
 \qquad \ \ 
 c_F\equiv p_C+p_N,
 \qquad\ \ 
 s_F\equiv s_C-s_N .
 \label{54}
\end{equation}
Let us  define  the ``anisotropy''  combinations (see \rf{44})
\al{
& \Delta\equiv \hh_{33}-\tfrac13  \hh_{aa}
 =\hh_{33}-\tfrac13\big(E_x   +E_y\big),
 \label{55}\\
& \Delta_F\equiv c_F-\tfrac13r,
 \qquad
 \delta_C\equiv q_C-\tfrac13d_C,
 \qquad
 \delta_N\equiv q_N-\tfrac13d_N .
 \label{56}
}

\subsubsection*{Comments on symmetries}

The allowed tensor and spinor structures of the coincident
Green's  functions are constrained by the isometries that leave
the coincidence point fixed. These transformations form the
isotropy group at that point. We assume that the Green
functions and their zero-mode constraints respect the
symmetries used below.

For $S^3/\ZZ_k$, we determine the continuous symmetries by
considering rotations of the covering sphere $S^3$ that
commute with the generator $g$  of the $\ZZ_k$-quotient identification  (``deck generator'') 
\begin{equation}
 g:\ (Y_1,Y_2)\longmapsto
 \big(e^{2\pi\ii/k}Y_1,\ e^{2\pi\ii/k}Y_2\big).
 \label{57}
\end{equation}
The relevant question is which of these symmetries fix a
chosen point and how they rotate its tangent directions.
For $k=1$ we have $g=I$,   the isometry is  $SO(4)$ and the isotropy  is
$SO(3)$.
For 
$k=2$  we have $g=-I$ acting on $\mathbb  C^2$ or $\mathbb R^4$, which is a central element in  $SO(4)$. The
isometry group of $\mathbb{RP}^3$ is
$SO(4)/\ZZ_2$, and the isotropy is again $SO(3)$. Thus the $\ZZ_2$  quotient changes the
global topology but not the local isotropy.

For $k>2$    we have $g$ as  a simultaneous rotation by $2\pi/k$ in two orthogonal planes
which  is not central. Writing
$SO(4)=(SU(2)_L\times SU(2)_R)/\ZZ_2$ with $g\in SU(2)_R$, the centralizer
is $(SU(2)_L\times U(1)_R)/\ZZ_2$,  since $SU(2)_L$ acts  transitively on
$S^3$. The connected point stabilizer is a diagonal $U(1)$, whose tangent
action rotates the two directions perpendicular to the Hopf fibre.\foot{The same distinction appears in the normal bundle. At $k=1$ the wrapped
surface is the great three-sphere $S^3=S^7\cap \RRR^4$, whose normal bundle is
the trivial bundle $S^3\times\RRR^4$ with the flat connection. The normal frame rotates along the fibre and $y^m$ appears
to carry  a unit charge; at $k=1$ that charge is  removable by
undoing the frame rotation, and thus it does not obstruct $SO(4)$ covariance. For
$k>2$ the quotient gives the normal bundle the flat but nontrivial holonomy
$e^{2\pi\ii/k}$. At $k=2$ the holonomy is
$-1$: this is nontrivial but central flat holonomy,
and hence does not reduce the connected isotropy.}


Since for  $k\le2$ the isotropy is $SO(3)$,  the only invariant symmetric tensor is
$\delta_{ab}$ and there is no invariant vector, so that 
\begin{equation}
k \le 2: \qquad \ \ \  G_{x   ,ab}=\tfrac13\delta_{ab} G_{x   ,cc}\ ,
 \qquad \ \ 
 G_{y,ab}=\tfrac13\delta_{ab} G_{y,cc}\ ,
 \qquad
 G_{x   ,a}=G_{y,a}=0 \ .
 \label{58}
\end{equation}
In the spinor sector, $\rho_a$ transforms as a vector while in \rf{9},\rf{1011}
$\PC,\PN$ are invariant  while  $T_C$ and $T_N$ are not
 and  thus cannot occur.  The  corresponding coefficients in \rf{52},\rf{522}    should  then  vanish\foot{The fermionic zero-mode projector is then also an $SO(3)$-invariant matrix
 $\sim (\alpha\PC+\beta\PN)$  consistently
with \rf{59}.}
\begin{equation}
 v_C=v_N=0,
 \qquad
 d_C=d_N=0,
 \qquad
 q_C=q_N=0 .
 \label{59}
\end{equation}
The only $SO(3)$-covariant form for a vector-valued spinor matrix
is $G_{\theta,a}=\rho_a {\rm G}$ with $\rm G $  being  invariant  and hence
\begin{equation}
 \rho_3G_{\theta,3}=\tfrac13E_\theta,
 \qquad
 p_C=\tfrac13r_C,
 \qquad
 p_N=\tfrac13r_N,
 \qquad
 c_F=\tfrac13r .
 \label{60}
\end{equation}
Combining \rf{58}--\rf{60} with \rf{55}--\rf{56} gives 
\begin{equation}
k \le 2: \qquad \ \ \   \Delta=\Delta_F=\delta_C=\delta_N=0
\ .
 \label{62}
\end{equation}
For $k>2$ the isotropy is $U(1)$, i.e. the invariant symmetric tensors are 
$\delta_{ab}$ and $\delta_{a3}\delta_{b3}$, an invariant vector along $\ree_3$
is allowed, and $\rho_3$ is invariant, being the generator of the rotation
itself. Hence in \rf{44} (the   indices are tangent frame ones)
\begin{equation}
 \hh_{ab}=\operatorname{diag}\big(\hh_\perp,\hh_\perp,\hh_{33}\big) \ . 
 \label{61}
\end{equation}
In    \rf{522}--\rf{5222}  all 12  coefficients 
are independent if $k > 2$.

\subsubsection*{Image   sum representation and  origin of  anisotropy}

To recall, in  static gauge the $\ZZ_k$ quotient   may be 
imposed a posteriori, as an identification of  the 3rd world-volume coordinate (see also \ci{Beccaria:2026zjj}). The quartic 
vertices below  \rf{69}--\rf{72}  are local and contain no $k$ dependence. How then  can the structure of $\vev{\cL_4}$
 expressed  in terms of Green's functions depend on  whether $k\le 2 $  or $k >2$?
The point   is  that the Green's function values  are effectively  non-local  and sensitive 
 to direction of  derivatives.

The  propagator  $G(\s,\s')$ is the  inverse of the kinetic operator on the whole manifold, so its  coincidence limit 
 is a global object. On the quotient it is the image sum
\begin{equation}
 G_{S^3/\ZZ_k}(\s,\s')=\sum_{p=0}^{k-1}\omega^p\,G_{S^3}\big(\s,g^p\s' \big),
 \label{63}
\end{equation}
where $\omega$ is the holonomy carried by the field: $\omega=1$ for
$x^r$, $\omega=e^{2\pi\ii/k}$ for the complex combinations $\yy^\a$  of $y^m$, and
for $\theta$ the  holonomy   $\omega$  is the corresponding rotation of the spinor bundle.\foot{Eq.\rf{63} refers to the non-zero-mode parts; the kernels will be  discussed below.}


For a scalar $x^r$ with trivial holonomy, the image sum is invariant
under the normalizer of the deck group.
For charged fields and spinors,
the transformation must also lift consistently to the
corresponding bundles. The connected symmetry transformations
centralize $g$, giving the same connected-isotropy argument
as above.

The images of  point $\s'$ are  $g^p  \s' $, which for the Hopf  generator   lie on
the Hopf fibre through $\s'$. On the covering sphere that fibre is a great
circle, hence a geodesic of length $2\pi$, and $g^p$ displaces $\s'$ along it
by  $2\pi p/k$. For a scalar $G_{S^3}(\s,\s')=G\big(d(\s,\s')\big)$,   where $d(\s,\s')$   is the geodesic length so that
\begin{equation}
 \nabla_{\s,a}   \nabla_{\s', b} G_{S^3}\big(\s,g^p  \s'\big)
 =A_p\delta_{ab}+B_p  u_au_b+C_p \epsilon_{ab3},\ \ \ \ 
 \qquad u_a=\delta_{a3},
 \label{64}
\end{equation}
where $u_a$ is the unit tangent vector  to the connecting geodesic.  
 The $p=0$ term is the regularized coincidence limit on the round $S^3$ 
 and is isotropic  (the stabilizer of a point there is 
$SO(3)$).  An anisotropy of $\hh_{ab}$  and  possible nonzero $G_{x   ,3}$ or
$G_{y,3}$  come from the terms with $p\ne0$.   

For $k>2$ the general rule is that the $p$-th image contributes a tensor,  invariant
under the stabilizer of the {pair} $(\s,g^p\s)$. For central $g$ the
image is fixed by all of $SO(3)_\s$ and the contribution is isotropic; for
non-central $g$ the image is at a  finite distance along the fibre
and only $U(1)$ survives.\foot{ For  $k=2$ there is an
image term, the antipode  $-\s$ of $\s$. But every great circle through $\s$ passes
through $-\s$: the connecting geodesic is not unique and no direction is
singled out. Equivalently,  $-\s$ is fixed by the whole stabilizer $SO(3)_\s$,
so the bitensor at $(\s,-\s)$ is $SO(3)$-invariant and its coincidence limit
is isotropic; in \rf{64} one has $B_1=C_1=0$.}

The intuition that  a  posteriori  $\ZZ_k$ identification cannot change  local properties  
applies to short-distance divergences, the $\delta(0)$ terms,   etc.
 These come from the short-distance expansion of the $p=0$ term in the image  sum
 and are $k$-independent (per unit
volume $V= {2\pi^2\ov k}$). 
This  logic fails  for finite global quantities  like  a coincident
propagator: 
the image terms are  at finite separation and contribute finite amounts. 

The same applies  to the fermionic sector. 
The deck
element acts on the spinor bundle by a rotation about $\ree_3$ combined with
the internal phase rotation, so its generator is built from $\rho_3$,
$\wJ$ and $\wA$ in \rf{1011}. The image terms of the spinor propagator are
polynomials in these three matrices, and after use of $\rho_3^2=1$,
$\wJ^2=\PC$, $\wA^2=\PN$ they produce exactly the four
structures $\PC,\PN,T_C,T_N$  in \rf{522},\rf{52}. The two
that carry $\rho_3$  disappear  for $ k \le 2$ when the sum over images
stops selecting a direction, which  leads to  \rf{59}.


\section{Quartic interaction vertices and their expectation values }\label{sec4}

The structure of the interaction terms in the  M2 brane   action is   determined   in the same way as in \ci{Beccaria:2026zjj}
where more details can be found. 

There are no cubic vertices in the static and $\k$-symmetry  gauge, 
so the first inverse-tension
correction 
 comes from the expectation value of the quartic terms in the action \rf{06}   or in the  Lagrangian, 
\be \la{666}
\cL_4=\cL_{4b}+\cL_{2b,2f}+\cL_{4f}\ . \ee
Here
\al{
& \cL_{4b}=x   ^2(\nabla x   )^2+\tfrac12(y\nabla y)^2
 +\tfrac18C^2-\tfrac14C_{ij}C^{ij}-\tfrac14y^2C+\tfrac38(y^2)^2
 -\tfrac1{4\sqrt g }\epsilon^{ijk}\epsilon_{rstu}
 x   ^r  \del_i x   ^s   \del_j  x   ^t  \del_k x   ^u , \la{69}\\
& \qquad \ \ \ C_{ij} \equiv  \nabla_i x^r \nabla_j x^r + \nabla_i y^m  \nabla_j  y^m, \ \ \ \ \   \ \  \  C=C_{ij} g^{ij} \ .
 \la{699}
}
 $i, j=1,2,3$ are world-volume indices contracted with $g_{ij}$ in \rf{3}    and $\nabla_i x^r=\del_i x^r$.
 
The term quadratic in both bosons  and fermions is found to be 
\begin{align}
 \cL_{2b,2f}={}&\big(\tfrac14Hg^{ij}-\tfrac12H^{ij}\big)B_{ij}
 +\tfrac12g^{ij}\Big[ \tb Q_iD_j\theta
 +\tb R_iU_j\theta
 +\tb\rho_iV_j\theta\notag\\
 &-\tfrac12\tb\Gamma_{(1)}R_iD_j\theta
 -\tfrac12\tb\Gamma_{(1)}\rho_iU_j\theta
 -\tfrac12\tb\Gamma_{(2)}\rho_iD_j\theta \Big],
 \label{70}\\
 & B_{ij}\equiv \tb\rho_iD_j\theta, \qquad  \qquad H= g^{ij}H_{ij} \ , 
\end{align}
where 
$H_{ij}$ is the quadratic part of the induced metric in \rf{101},  
 $R_i,U_i$ are the terms linear and $Q_i,V_i$  quadratic in the
bosons in the expansions of the pulled-back gamma matrix and Killing
derivative respectively.
  $\Gamma_{(1)},\Gamma_{(2)}$ are the first two corrections to
the   matrix    (see  Appendix C in  \ci{Beccaria:2026zjj} for details)
\be \Gamma_{\rm B}=\tfrac{\ii}{3!\sqrt{h}}\,
\epsilon^{ijk}E_i^AE_j^BE_k^C\,\Gamma_{ABC}
=\Gk+\Gamma_{(1)}+\Gamma_{(2)}+...   \ . \la{kaa}   \ee
For completeness, the bosonic insertions in \eqref{70} may be written
explicitly in the orthonormal tangent frame. With
$\bar\Gamma=\ii\Gamma_*$ as in \eqref{72}, they are
\al{
 R_a&=\nabla_a x^r\Gamma_r+\nabla_a y^m\Gamma_m,
 \qquad Q_a=-\tfrac12 y^2\rho_a,\qquad 
 U_a=-\nabla_a x^r\Gamma_r\bar\Gamma
       -\tfrac12 y^m\Gamma_{am}
       +\tfrac12\nabla_a y^m\Gamma_m\bar\Gamma, \la{3399}\\
 V_a&=-x^r\nabla_a x^s\Gamma_{rs}
       +\tfrac14y^m\nabla_a y^n\Gamma_{mn}
       -\tfrac14y^2\rho_a\bar\Gamma.
\label{903}
}
Here $r,s$ and $m,n$ label the 4 AdS and 4  internal transverse
directions, respectively. Writing $R_a=R_a^A\Gamma_A$ and similarly
for $Q_a$ and $\rho_a$,  we   have for  the terms in \rf{kaa}\foot{Note that 
$\{\Gamma_\kappa,\Gamma_{(1)}\}=0$ and
$\{\Gamma_\kappa,\Gamma_{(2)}\}+\Gamma_{(1)}^2=0$ follow from
$\Gamma_{\rm B}^2=1$.}
\begin{equation}
\begin{aligned}
 \Gamma_{(1)}=\tfrac{\ii}{2}\epsilon^{abc}
                 R_a^A\rho_b^B\rho_c^C\Gamma_{ABC},\qquad \qquad 
 \Gamma_{(2)}=-\tfrac12 H\Gamma_\kappa
       +\tfrac{\ii}{2}\epsilon^{abc}
        \big(Q_a^A\rho_b^B\rho_c^C+R_a^AR_b^B\rho_c^C\big)\Gamma_{ABC}.
\end{aligned}
\label{904}
\end{equation}
   The four-fermion vertex is given by 
\al{
 &\cL_{4f}=\tfrac1{96}\tb\rho^i\mathscr M_4D_i\theta
 +\tfrac1{16}\big[\big(B_i{}^i\big)^2-B_{ij}B^{ji}\big],
 \label{71}
\\
 \mathscr M_4=&\Gb\Gamma_a\theta\,\tb\Gamma_a
 +\big(\Gamma_{rs}\theta\,\tb\Gamma_{rs}
 -\tfrac12\Gamma_{ab}\theta\,\tb\Gamma_{ab}
 -\tfrac12\Gamma_{mn}\theta\,\tb\Gamma_{mn}\big)\Gb,
 \qquad \Gb=\ii\Gamma_* .
 \label{72}
}
Here  $a,b$ are tangent indices and $r,s$  and   $m,n$ are again  the  4+4  sets of the transverse  indices.

To compute the  2-loop correction we are to apply Wick contractions to \rf{69},\rf{70} and \rf{71}.

\subsubsection*{Bosonic sector }

Let us assume that $k>2$. 
By \rf{61} the normalized expectation value $\hh_{ab}$ of
$H_{ab}=C_{ab}-y^2\delta_{ab}$ in \rf{44} is axially symmetric, so that 
\begin{equation}
 \hh_{aa}=2\hh_\perp+\hh_{33},
 \qquad\qquad 
 \hh_{ab}\hh_{ab}=2\hh_\perp^2+\hh_{33}^2 .
 \label{73}
\end{equation}
As a result, 
\begin{equation}
 \vev{\cL_{4b}}
 =-6\hh_{33}^2+4\big(E_x   +E_y\big)\hh_{33}
 +16G_x    E_x   -4G_yE_y-E_x   ^2-E_y^2+24G_{x,3}^2 \ ,
 \label{74}
\end{equation}
where we used  the definitions in \rf{40},\rf{42} and \rf{442},\rf{44}. 
The terms quadratic in $G_{y,3}$ cancel out. 
Of the two contributions to $24G_{x   ,3}^2$, one comes from the volume part of the M2 brane 
action and one from the 3-form $C_3$  part.

Using then \rf{55}   we  get 
\begin{equation}
  \vev{\cL_{4b}} =\tfrac23\big(E_x   +E_y\big)^2
 +16G_x    E_x   -4G_yE_y-E_x   ^2-E_y^2
 +24G_{x   ,3}^2-6\Delta^2 .
 \label{76}
\end{equation}
Note that only the last   two terms  are not proportional to the EOM 
  combinations  $E_x   $ or $E_y$.

In the regularization where $\delta(0)=0$\foot{Equivalently, one  may use spectral $\zeta$-function  regularization:
 for a Laplace-type operator   $\cO$ on a smooth closed
3-manifold the local heat expansion has no constant term, so
$\zeta_\cO(0)=-\dim\ker\cO$ 
and the coincident-point value of the    full identity operator   
is equal to   zero.}
 the single-field  combinations 
  $E_x$ and $E_y$  in \rf{42}   are given just by the 
zero-mode projector contributions 
\begin{equation}
 E=-\big[P_0\big](\s,\s),
 \qquad
 \big[P_0\big](\s,\s)=\frac{\#\ \text{modes}}{\#\ \text{components}} V^{-1} \ , \qquad 
 V\equiv  \vol{(S^3/\ZZ_k)} = 2 \pi^2 k^{-1} \ .
 \label{123}
\end{equation}
Explicitly,  here we get 
\begin{equation}
 E_x   =- V^{-1} , \qquad 
 \qquad
 E_y=-2 V^{-1}  \ . 
 \label{79}
\end{equation}
 Using  that    $G_{x   ,3}=0$ in \rf{0022},  we may write      \rf{76} as  
\begin{equation}
      \vev{\cL_{4b}} \Big|_{k>2}
 ={\rm EOM} -6\Delta^2 \ , 
 \label{80}
\end{equation}
where EOM stands for terms  proportional to $E_x$ or $E_y$.
Note  the anisotropy $\Delta$ in \rf{55}  is independent of the
zero-mode prescription  (the projectors entering only the trace through
\rf{44} and \rf{79}), i.e.  it is produced
entirely by the image terms of the propagators. 
This is thus  a consequence of the reduced isotropy, and would  be present 
  in a
theory with the same holonomy and no zero modes.

In the case of $k\le2$  
using \rf{58},\rf{62},\rf{0022}  we find that the last two  terms in  \rf{76} vanish and thus 
\begin{equation}
   \vev{\cL_{4b}} \Big|_{k\le2}=\tfrac23\big(E_x   +E_y\big)^2
 +16G_x    E_x   -4G_yE_y-E_x   ^2-E_y^2 \ , 
 \label{77}
\end{equation}
which  is   expressed  only in terms of  the  EOM  combinations. 
This is the same  result as was  found 
 in the case of  $\AdS_3\subset\AdS_7\times S^4$ in \ci{Beccaria:2025ahf}  related to the present  one  of 
 $S^3\subset\AdS_4\times S^7$  by  the analytic continuation. 
 This property is   related to the fact   that  in both cases 
the 3d   M2 brane  surface is totally geodesic in a maximally symmetric
space and the isotropy is the full $SO(3)$.
 In  the $\AdS_7\times S^4$  case with $\AdS_3$ world-volume metric there
are no normalizable zero modes, so  that $E_x   =E_y=0$ and the bosonic  contribution  \rf{77} vanishes on its own. 

Here the  EOM  combinations equal minus the zero-mode projector
densities (setting to zero the  $\delta(0)$  terms).  For  $k\le2$  the zero-mode content  is  4 modes  in  $x^r$ sector   and 
$4\times 4$   modes in $y^m$   sector (see section \ref{sec2})  and thus 
$ E_x   =-V^{-1} , \ \ 
 E_y=- 4 V^{-1} $ so that\foot{Note that for  $k=1$ the value of $G_y$ 
requires a continuation past the negative mode reflecting instability of $S^3$  surface  in $S^7$.    Eq.\rf{62} is, however, 
 unaffected,
since it does not use the value of  $G_y$.}
 \be   \vev{\cL_{4b}} \Big|_{k\le2}
 =-\tfrac1{3}V^{-2}- 16 V^{-1} \big(G_x   -G_y\big).
 \label{78}
\end{equation}

\subsubsection*{Mixed sector}

Let us  give the expressions   for the separate contractions of the seven terms in \eqref{70}.
The quantities $G_x,G_y$ and their derivative coefficients refer to one
real component. Thus the normalized tensor used below is
$\mathrm h_{ab}=\tfrac14\langle H_{ab}\rangle$ as in \eqref{44}.
For the  fermionic constants  in \rf{50}--\rf{5222}  we have   
\begin{equation}
\begin{aligned}
 G_{\theta,a}&=\tfrac12\rho_a\big[
 (r_C-p_C)P_C+(r_N-p_N)P_N
 +(d_C-q_C)T_C+(d_N-q_N)T_N\big],\qquad a=1,2,\\
 G_{\theta,3}&=\rho_3\big(p_CP_C+p_NP_N+q_CT_C+q_NT_N\big).
\end{aligned}
\label{905}
\end{equation}
The coefficients are those of \eqref{522}--\eqref{54}. 
For a numerical even spinor matrix $M$, the  contraction  rules are
$
 \langle\bar\theta M \theta\rangle=-\operatorname{tr}(M G_\theta),$ \ 
 $
 \langle\bar\theta M D_a\theta\rangle
       =-\operatorname{tr}(M G_{\theta,a}),
$
where the trace is over the sixteen  independent  spinor components.
Let us 
denote by $\mathcal V_I$, $I=1,\ldots,7$, the seven terms of
\eqref{70}, in their displayed order, including their numerical
prefactors. Contracting the bosons first, forming the Clifford
products before projection, and then using the above rules   gives 
\al{
& \langle\mathcal V_1\rangle
   =24\mathrm h_{33}c_F-8(E_x+E_y)c_F-8\mathrm h_{33}r,\qquad \qquad 
 \langle\mathcal V_2\rangle=8G_y r, \qquad \label{908} \\
 & \langle\mathcal V_3\rangle
   =8(2E_x-E_y-3G_y)s_F+8G_{y,3}v_C,\qquad \qquad 
 \langle\mathcal V_4\rangle
   =12G_y s_F-4G_{y,3}v_C-16G_{x,3}v_N, 
\no 
\\
&
 \langle\mathcal V_5\rangle
   =-12\mathrm h_{33}c_F+4(E_x+E_y)c_F
       +4\big(\mathrm h_{33}-E_x-E_y-2G_y\big)r +8G_{y,3}(q_C-d_C)+8G_{x,3}(d_N-q_N), \no \\
& \langle\mathcal V_6\rangle
   =-4(2E_x-E_y-3G_y)s_F-4G_{y,3}v_C-16G_{x,3}v_N,\qquad\no \\
 &\langle\mathcal V_7\rangle
   =4(E_x+E_y+3G_y)r+8G_{y,3}d_C-8G_{x,3}d_N.
\label{910}
}
The $G_{y,3}v_C$,  $G_{y,3}d_C$,  $G_{x,3}d_N$ 
 and $G_y s_F$  terms  cancel in the sum  of $\mathcal V_I$  and we get 
\begin{align}
 \vev{\cL_{2b,2f}}=\sum_{I=1}^7 \vev{ \mathcal V_I} = {}&12\,\hh_{33}\,c_F+4r\big(3G_y-\hh_{33}\big)
 -4\big(E_x   +E_y\big)c_F\notag\\
 &+4\big(2E_x   -E_y\big)s_F+8G_{y,3}q_C
 -32G_{x   ,3}v_N-8G_{x   ,3}q_N , 
 \label{84}
\end{align}
where the constants $c_F, r, ...$  appeared in \rf{52},\rf{522},\rf{54}.
Inserting $\hh_{33}=\tfrac13(E_x   +E_y)+\Delta$  from \rf{55}
and $c_F=\tfrac13r+\Delta_F$  from \rf{56}   we get 
\begin{align}
 \vev{\cL_{2b,2f}}={}&12G_y r-\tfrac43\big(E_x   +E_y\big)r
 +4\big(2E_x   -E_y\big)s_F
 +\tfrac83G_{y,3}d_C-\tfrac83G_{x   ,3}d_N\notag\\
 &+12\Delta_F\Delta+8G_{y,3}\delta_C
 -8G_{x   ,3}\delta_N-32G_{x   ,3}v_N .
 \label{85}
\end{align}
Here the   first   line  is proportional to the  EOM terms  (with $E_\theta$  given in \rf{52}).
For 
$k\le2$  using  \rf{59}, \rf{60} and \rf{62}  we get 
\begin{equation}
 \vev{\cL_{2b,2f}}\Big|_{k\le2}=12rG_y-\tfrac43\big(E_x   +E_y\big)r
 +4\big(2E_x   -E_y\big)s_F .
 \label{86}
\end{equation}
Since $r=r_C + r_N$ in \rf{54} is part of $E_\theta$ in \rf{52} 
 this  would vanish in a theory without zero modes, but in the present case  \rf{86}  will 
  be determined   by  the zero-mode  projector contributions.


For $k>2$  the EOM terms  in \rf{85}   contain
\begin{equation}
 r_C=d_C=-\tfrac{1}{2}V^{-1},
 \qquad\qquad 
 r_N=d_N=-V^{-1},
 \label{87}
\end{equation}
corresponding to 4 charged and  8 uncharged fermionic zero modes.

The remaining coefficients start from the orthogonal per-level
inverse \eqref{aa88}, with sums \eqref{bc14}--\eqref{bc16}.
 Altogether
\al{
 &q_C=-\tfrac{1}{2}V^{-1},
 \qquad
 q_N=0,
 \qquad
 v_N=\tfrac{1}{2}V^{-1},
 \qquad
 v_C=-2G_x   +G_y-\tfrac3{8}V^{-1},
 \label{88}\\
 &p_C=G_{x   ,33}+\tfrac{1}{2}V^{-1},
 \qquad
 p_N=G_{x   ,33},
 \qquad
 G_{y,3}=2G_x   -G_y+\tfrac1{4}V^{-1} ,
 \qquad
 G_{x   ,3}=0 .
 \label{89}
}
Hence from \rf{56}
\begin{equation}
 \delta_C=-\tfrac1{3}V^{-1} ,
 \qquad
 \delta_N=+\tfrac1{3}V^{-1}, 
 \qquad
 \delta_C+\delta_N=0,
 \qquad
 G_{y,3}+v_C=- \tfrac1{8}V^{-1}, 
 \label{90}
\end{equation}
and thus  
\begin{equation}
 \vev{\cL_{2b,2f}}\Big|_{k>2}={\rm EOM} + 12\Delta_F\Delta +8G_{y,3}\delta_C \ , 
 \label{91}
\end{equation}
where EOM stands   for the terms in the  first line in \rf{85}. 


\subsubsection*{Fermionic sector}

We use the same Majorana Wick contraction rules as in  \ci{Beccaria:2026zjj}. Applied to
\eqref{71} and \eqref{72}, they give $\langle\mathcal L_{4f}\rangle$
in terms of the constants in \eqref{52}--\eqref{54}. 
Let   us  separate the mass-matrix term of \eqref{71} from the two
$B_{ab}$ invariants.

It is   useful to  introduce the  formal  4-component derivative  notation $D_\mu$    with $\mu=\circ,1,2,3$  so that 
$D_\circ=1$ and $D_a$  for $a=1,2,3$ are  given by \rf{11}.  Then we  define  as in \rf{50} 
the coincident-point combinations ($\overline{D_a\theta}=(D_a\theta)^TC$)
\begin{equation}
 \mathcal S_{\mu\nu}=
\langle D_\mu\theta(\sigma)
              \overline{D_\nu\theta}(\sigma')\rangle\Big|_{\sigma=\sigma'},
 \qquad \mu,\nu\in\{\circ,1,2,3\},
 \qquad
 \mathcal S_{\circ\circ}=G_\theta,
 \quad \mathcal S_{a\circ}=G_{\theta,a}.
\label{911}
\end{equation}
 Under the conjugation $M^\vC\equiv C^{-1}M^TC$, the 
Majorana antisymmetry gives
$
 \mathcal S_{\nu\mu}=\mathcal S_{\mu\nu}^{\vC},
 \  G_\theta^\vC=G_\theta,
 \  \rho_a^\vC=-\rho_a.
$
In particular, the right-point derivative is not in general the
negative of the left-point derivative.

For two   constant  even spinor matrices $M_1,M_2 $  the  Wick contraction rule is
\begin{equation}
\begin{aligned}
 \mathcal U_{\mu\nu}(M_1,M_2)
 &\equiv\langle(\bar\theta M_1 D_\mu\theta)
                    (\bar\theta M_2 D_\nu\theta)\rangle\\
 &=\operatorname{tr}(M_1\mathcal S_{\mu\circ})
       \operatorname{tr}(M_2\mathcal S_{\nu\circ})
   -\operatorname{tr}(M_1\mathcal S_{\mu\circ}M_2\mathcal S_{\nu\circ})-\operatorname{tr}(M_1^\vC G_\theta M_2\mathcal S_{\nu\mu}).
\end{aligned}
\label{913}
\end{equation}
Then from   \eqref{72}  we get 
\begin{equation}
\begin{aligned}
 \mathcal F{_{\mathscr M}}
 \equiv\tfrac1{96}\langle\bar\theta\rho^a\mathscr M_4D_a\theta\rangle
 =&\tfrac1{96}\sum_a\Big[
 \sum_b\mathcal U_{\circ a}(\rho_a\bar\Gamma\Gamma_b,\Gamma_b)
 +\sum_{r,s}\mathcal U_{\circ a}
                   (\rho_a\Gamma_{rs},\Gamma_{rs}\bar\Gamma)\\
 &\qquad -\tfrac12\sum_{b,c}\mathcal U_{\circ a}
                   (\rho_a\Gamma_{bc},\Gamma_{bc}\bar\Gamma)-\tfrac12\sum_{m,n}\mathcal U_{\circ a}
                   (\rho_a\Gamma_{mn},\Gamma_{mn}\bar\Gamma)\Big].
\end{aligned}
\label{914}
\end{equation}
For the $BB$ part  of \rf{71}  we define
$\rb_{ab}=\langle B_{ab}\rangle=-\operatorname{tr}(\rho_aG_{\theta,b})$.
The ``disconnected'' $\tr\, \tr$  part  and the two ``exchange'' $\tr$  parts  in \rf{913}   of the  $BB$ term  in \rf{71}   we get 
\al{
&\tfrac1{16}\vev{\big[\big(B_i{}^i\big)^2-B_{ij}B^{ji}\big]}\equiv  \mathcal F_{\rm disc}  +  \mathcal F_1 +  \mathcal F_2\ , \qquad \ \ 
 \mathcal F_{\rm disc}
  =\tfrac1{16}\big[(\operatorname{tr}E_\theta)^2
                   -\sum_{a,b}\rb_{ab}\rb_{ba}\big],\la{978} \\
& \mathcal F_1
  =\tfrac1{16}\big[-\operatorname{tr}(E_\theta^2)
        +\sum_{a,b}\operatorname{tr}
          (\rho_aG_{\theta,b}\rho_bG_{\theta,a})\big],\quad
\qquad \ \ \  \mathcal F_2
  =\tfrac1{16}\sum_{a,b}\operatorname{tr}
        \big[\rho_aG_\theta\rho_b
                       (\mathcal S_{ba}-\mathcal S_{ab})\big]. \la{979}
}
Using that  the  spinor derivative in \rf{10},\rf{11}  is 
$D_a=\mathrm e_a+\Omega_a$, $\Omega_a=\ii\rho_aP_N$, and that 
$[\mathrm e_a,\mathrm e_b]=-2\epsilon_{abc}\mathrm e_c$ we find
\begin{equation}
\begin{aligned}
 \mathcal S_{ba}-\mathcal S_{ab}
   ={}&\mathrm e_aG_{\theta,b}-\mathrm e_bG_{\theta,a}
      +[\Omega_a,G_{\theta,b}]-[\Omega_b,G_{\theta,a}]
      +2\epsilon_{abc}G_{\theta,c}.
\end{aligned}
\label{916}
\end{equation}
For the present homogeneous  metric case  the  Green's functions  in \eqref{905} 
are constant, so the first two terms vanish.

Evaluating the traces in \rf{914},\rf{978},\rf{979} using \eqref{522}, \eqref{52} and
\eqref{905}  gives
\begin{equation}
\begin{aligned}
 \mathcal F_{\mathscr M}
   &=2r_Cs_C-2r_Ns_N+6v_Cq_C+6v_Nq_N
       -\tfrac83d_Cv_C-\tfrac43d_Nv_N,\qquad  \mathcal F_{\rm disc}=-6c_F^2+4rc_F+2r^2,\\
&
 \mathcal F_1 =2q_C(q_C-d_C)+2q_N(q_N-d_N),\qquad 
 \mathcal F_2=-2r_Cs_C+2r_Ns_N+2v_Cq_C+2v_Nq_N.
\end{aligned}
\label{917}
\end{equation}
 As a result, 
\begin{align}
 \langle\mathcal L_{4f}\rangle=
\mathcal F_{\mathscr M}+\mathcal F_{\rm disc}+\mathcal F_1+\mathcal F_2={}&-6c_F^2+4rc_F+2r^2+8v_Cq_C+8v_Nq_N\notag\\
 &-\tfrac83d_Cv_C-\tfrac43d_Nv_N
 +2q_C(q_C-d_C)+2q_N(q_N-d_N).
 \label{81}
\end{align}
This may be rewritten as (see \rf{55},\rf{56})
\begin{align}
 \vev{\cL_{4f}}=&\tfrac83r^2+\tfrac43v_Nd_N
 -\tfrac23\delta_Cd_C-\tfrac23\delta_Nd_N
 -\tfrac49d_C^2-\tfrac49d_N^2\notag\\
 &-6\Delta_F^2+8v_C\delta_C+8v_N\delta_N
 +2\delta_C^2+2\delta_N^2 \no\\
 = & {\rm EOM} -6\Delta_F^2+8v_C\delta_C+8v_N\delta_N
 +2\delta_C^2+2\delta_N^2  \ . 
 \label{83}
\end{align}
Here  $r=r_C+ r_N $, $d_C$ or $d_N$  appear in the  $E_\theta$  term in \rf{52}
and thus  vanish  if $\delta(0)=0$  up to  the  contribution of the  zero-mode projectors.  

The   non-EOM  ``residue''   term in \rf{83}   is absent for $k\le 2$   where   we get 
using  \rf{59}, \rf{60} and \rf{62} 
\begin{equation}
 \vev{\cL_{4f}}\Big|_{k\le2}=\tfrac83r^2  \ , \la{884}\ee
 with $r$ being again a  part of the  $E_\theta$ term.

\subsubsection*{Total   quartic vertex   contribution}

The sum of the above expressions in \rf{76},\rf{85},\rf{83}   may be written as 
(using that  $G_{x   ,3}=0$)
\al{ \vev{\cL_{4}} = \te \EOM -6\big(\Delta-\Delta_F\big)^2
 +8\big(G_{y,3}+v_C\big)\delta_C+8v_N\delta_N
 +2\delta_C^2+2\delta_N^2 \ . \la{92}
 }
 Here in  the  non-EOM residue    the 
  geometric anisotropy  parts cancel out. Indeed, \rf{b5} 
 gives $G_{y,33}=G_{x   ,33}+G_y$, so by \rf{44}
$
 \hh_{33}=G_{x   ,33}+G_{y,33}-G_y=2G_{x   ,33},
 $
while \rf{89} and \rf{54} give $c_F=2G_{x   ,33}+\ha V^{-1}$. With
$E_x   +E_y=-3V^{-1} $ and $r=-{3\ov 2} V^{-1}$ from \rf{79} and \rf{87},  we get for  $k>2$   (see Appendix \ref{api})
\begin{equation}
 \Delta=\Delta_F=2G_{x   ,33}+V^{-1}\ , \qquad  \hh_{33}-c_F=\tfrac13\big(E_x   +E_y-r\big)=-\tfrac{1}{2}V^{-1}.
   \label{94}
\end{equation}
The   contribution $-\ha V^{-1}$   is  the difference of the two EOM terms
with the geometric anisotropy contributions cancelling between the bosonic and fermionic parts;
equivalently,  this  is manifested in 
 the cancellation of
the $\hh_{33}^2$   terms  in the sum of  contributions of   the three sectors.

From   \rf{94} the non-EOM residue  of \rf{92} 
is due to the $\delta_C$ and $\delta_N$ terms in \rf{56}.  Using 
\rf{88} and \rf{90} gives   it as\foot{Let us note that the separation of \rf{92} into ``EOM terms''  and a residue is not
unique: the relations \rf{87}--\rf{90} between $r_{C,N}$, $d_{C,N}$, $q_{C,N}$ and $c_F$
allow one to move rational $V^{-2}$ terms between the two. The value \rf{96} refers to
the particular form of the EOM terms in \rf{83} and \rf{85}; only the total in  \rf{902}  below 
is unambiguous.  }
\be \la{96}\te 
8\big(G_{y,3}+v_C\big)\delta_C+8v_N\delta_N
 +2\delta_C^2+2\delta_N^2=
 \frac1{3}V^{-2}+\frac4{3}V^{-2}+\frac4{9}V^{-2}
 =\frac{19}{9} V^{-2} \ .
\ee
Unlike $\Delta$   this  is a   ``non-geometric''  anisotropy   being  due to the projecting out of the 
 fermionic zero modes.\foot{Both 
$\delta_C$ and $\delta_N$ are differences between a directional derivative
and one third of an EOM  term,  and  both are produced   by the
removal of the fermionic zero-mode  eigenvectors. Thus, if  one were not isolating the zero-mode 
 contributions  $q_C,q_N,d_C,d_N$  would vanish and
one would get 
 $\delta_C=\delta_N=0$   and no residue part left. }

For $k>2$  the  expressions in \rf{76},\rf{85},\rf{83}  may be  written also in the form 
\al{
 \vev{\cL_{4b}}&=-16\,V^{-1}G_x+8\,V^{-1}G_y-5\,V^{-2}
 -24\,G_{x,33}^2-24\,V^{-1}G_{x,33}\ ,
 \label{499}\\
 \vev{\cL_{2b,2f}}&=-8\,V^{-1}G_x-14\,V^{-1}G_y+5\,V^{-2}
 +48\,G_{x,33}^2+48\,V^{-1}G_{x,33}\ ,
 \label{599}\\
 \vev{\cL_{4f}}&=\tfrac{16}3\,V^{-1}G_x-\tfrac83\,V^{-1}G_y+\tfrac53\,V^{-2}
 -24\,G_{x,33}^2-24\,V^{-1}G_{x,33}\ .
 \label{899}
}
The $G_{x,33}$ terms here   stand for  the anisotropy terms 
 $-6\Delta^2$, $+12\Delta_F\Delta$ and
$-6\Delta_F^2$ respectively  (see \rf{94}).  
These terms   cancel in \rf{92},  i.e. in  the sum \rf{666}  of the three  terms in \rf{499}--\rf{899}  and thus we get 
\al{ 
\vev{\cL_4} = & \vev{\cL_{4b}} +  \vev{\cL_{2b,2f}} +  \vev{\cL_{4f}} \no \\  
= & -\tfrac23\,V^{-1}\big(28\,G_x+13\,G_y\big)+  \tfrac53\,V^{-2} =    {\rm EOM}   +\tfrac{19}9\,V^{-2} \ .
\la{902}
}
As in \rf{92}  here  we split the result into the EOM  part   and a residue.
The EOM part  consists  of  products 
of   
$E_x$, $E_y$  and $r$, $d_C$, $d_N$  terms in  $E_\theta$   which are rational multiples of $V^{-1}= {k\ov 2 \pi^2}$.
At  the same time, the 
propagators  $G_x$ and $G_y$ are non-trivial   transcendental functions of $k$.    

To summarize, 
for $k\le2$  the isotropy is $SO(3)$, all 
anisotropy contributions  vanish identically by symmetry, and each of the three sectors in $\vev{\cL_4} $
is separately proportional to the EOM  combinations,  in parallel  to   what    happened  in the  
 $\AdS_3\subset\AdS_7\times S^4$  case in \ci{Beccaria:2025ahf} which is related by a formal analytic continuation
 to the $k=1$ case. 

For $k>2$ the isotropy is only $U(1)$   and  for   this reason 
  there are anisotropic non-EOM   terms in each sector. 
In the  sum in \rf{92} 
these   geometric anisotropies cancel because of  $\Delta_F=\Delta$.   The only
surviving  non-EOM residue  is the zero-mode part \rf{96}. 

 Once again,  if all the zero-modes were absent   then the residue term \rf{96}   
 as well as EOM terms would  not be there, so the non-vanishing of \rf{902} is  essentially 
  due to the presence of the zero modes. 
  
 The non-vanishing   contribution \rf{902} of the quartic vertex 
  will be 
 compensated   by  another  consequence  of the presence of  zero modes -- the 
collective-coordinate Jacobian  contribution  discussed   below. 


\section{Contribution of  collective-coordinate Jacobian}\label{sec5}

Let us now determine the  path integral  measure contribution associated with
separating collective coordinates from constrained quantum
fluctuations. Throughout this section $k>2$, and the collective
coordinates are held fixed at the  given $S^3/\ZZ_k$ instanton   saddle  point  
 $\mathrm a=0$ in the collective  coordinate  space. 
The result is
 the local coefficient  in  \rf{011}. The remaining integration over $\mathrm a$ 
 using the  supersymmetric treatment of the paired
zero modes, is considered in section~\ref{n1}.

Let $\mathrm a=(a,\zeta)$ denote the 12 even and 12  odd collective
coordinates, and let $\widetilde{\mathrm X}=(x,y,\theta)$ denote the
canonically normalized quantum fields.  Let $\mathcal Z_A$ 
be  the tangent zero modes and $\Lambda^A$  their
dual functionals, i.e. 
\begin{equation}\Lambda^A\mathcal Z_B=\delta^A{}_B , \qquad \ \ 
 \Lambda^A\widetilde{\mathrm X}=0\ .  \label{927}
\end{equation}
The linear decomposition into zero modes and transverse
fields specifies the tangent spaces, but the collective-coordinate change
of variables  involves  the non-linear family of saddles $\X_0({\rm a})$. Assuming the reference saddle  is at  ${\rm a}=0$, with $\X_0(0)=0$
 and rescaling the gauge-fixed  
quantum fields by $\epsilon = \rT^{-1/2}_2$
 we   set  (cf. Appendix \ref{apco}) 
\begin{equation}
\begin{aligned}
 \mathrm X&=\mathscr U(\mathrm a)
                 [\epsilon\widetilde{\mathrm X}],\qquad  \ \     \X_0({\rm a}) =  \mathscr U(\mathrm a)[0], \ \  \qquad 
 \mathcal C^A(\mathrm a;\mathrm X)
       =\Lambda^A\big[\mathscr U(\mathrm a)^{-1}[\mathrm X]\big],\\
 &\qquad \qquad \ \ \mathcal M^A{}_B
       =-\frac{\partial\mathcal C^A}{\partial\mathrm a^B}
                   \Big|_{\mathrm X\,\mathrm{fixed},\,\mathrm a=0}
        =\Lambda^A \rK_B[\epsilon\widetilde{\mathrm X}].
\end{aligned}
 \label{928}
\end{equation}
The transformation $\mathscr U$  acts on the entire field
configuration   and  includes the compensation transformations  restoring the 
static and $\kappa$-symmetry gauge.\foot{$\mathscr U(\mathrm a)[\,\cdot\,]$ denotes the nonlinear
action of the target-space symmetries associated with the
collective coordinates. 
In particular, $\mathscr U(0)[\mathrm X]=\mathrm X$, while
$\mathscr U(\mathrm a)[0]=\mathrm X_{0}(\mathrm a)$
is the corresponding member of the family of classical
solutions. 
Only the
infinitesimal generators of this map at $\mathrm a=0$,
expanded through quadratic order in the fluctuations,
are needed for the present Jacobian calculation.}
 Its infinitesimal generators $\rK_B$ obey
$\rK_B[0]=\mathcal Z_B$, so that using  \rf{927}  gives 
$\mathcal M|_{\widetilde{\mathrm X}=0}=1$.

Factoring out the field-independent 
normalizations, the path integral  measure contains
\begin{equation}
 [D\mathrm X]\ =\
 d^{12}a\,d^{12}\zeta\,[D\widetilde{\mathrm X}]\,
 \delta(\Lambda\widetilde{\mathrm X})\,
 \operatorname{Ber}\mathcal M\ .
 \label{929}
\end{equation}
Expanding  in $\epsilon$ we  have, in general, 
\begin{equation}
\begin{aligned}
 \mathcal M&=1+\epsilon\mathcal M_1+\epsilon^2\mathcal M_2+
 \OO(\epsilon^3),\qquad\ \ \ \ 
 \log\operatorname{Ber}\mathcal M
 =\epsilon\operatorname{STr}\mathcal M_1
 +\epsilon^2\operatorname{STr}
       \big(\mathcal M_2-\tfrac12\mathcal M_1^2\big)
 +\OO(\epsilon^3).
\end{aligned}
\label{930}
\end{equation}
As  found in   Appendix~\ref{apd}, in the present case  $\mathcal M_1$   and thus 
 $\operatorname{STr}\mathcal M_1 $
and   $\operatorname{STr}\mathcal M^2_1$ 
 vanish. 
 Therefore, 
\begin{equation}
 \log\operatorname{Ber}\mathcal M
       =\epsilon^2\mathcal J+\OO(\epsilon^3),\qquad\qquad 
 \mathcal J=\operatorname{STr}\mathcal M_2\ . 
 \label{931}
\end{equation}
The  resulting expression for the 2-loop  coefficient  at the reference point $\rm a=0$ is then   as in \rf{011}
\be\la{976}
f_2(0;k)=\vev{S_4}-\vev{\mathcal J} \ . 
\ee
The M2  brane  action   has  residual 3d supersymmetry  after fixing the $\kappa$-symmetry gauge. 
A preserved
supersymmetry, acting linearly on fluctuations, specifies a convenient
complement of the fermion zero mode  field direction. The broken target-space supersymmetries generate
the odd parts  of $\mathcal M$ in \rf{928}. Computing those requires their
quadratic field dependence and the gauge compensation transformations;  these   are not 
determined  just by the quadratic part of the action. 


Let us introduce the  constant operator ($V= 2\pi^2 k^{-1}$) 
\begin{equation}
\mathcal P\equiv V P_{0,F}(\sigma,\sigma)
=\tfrac12(P_C+T_C)+P_N+T_N,\qquad 
\qquad \operatorname{tr}\mathcal P=12 .
 \label{932}
\end{equation}
Here $P_{0,F}(\sigma,\sigma')$ is the integral kernel of the
fermionic zero-mode projector $\mathcal Z_A\Lambda^A$.
Its coincident value is proportional to $V^{-1}$, and thus 
$\mathcal P$ is a  dimensionless spinor matrix.\foot{Note that $\tr\, \mathcal  P=12$ is the number of the fermionic
      zero modes. 
      The identity $P_{0,F}^2=P_{0,F}$ involves integration over an
intermediate point; it does not imply $\mathcal P^2=\mathcal P$.   }

As discussed  in Appendix~\ref{apd}, the
supertrace in \rf{931} reduces to the following functional  quadratic in the quantum fields (cf. \rf{010})  
\begin{equation}
\begin{aligned}
 \mathcal J=V^{-1} \int d^3\sigma\sqrt g\,\Big[&
 -4x^2-\tfrac52y^2+y^m(J_y)^{mn}\nabla_3y^n
             +4x^r(I_x)^{rs}\nabla_3x^s\\
 &\te -\frac{i}{3}\bar\theta (T_C- 2 T_N) \theta
       -\frac14\bar\theta\rho^a\mathcal P D_a\theta\Big].
\end{aligned}
\label{933}
\end{equation}
Here  we suppress
tildes on the quantum fields, 
$x^2=x^rx^r$, $y^2=y^my^m$, and $I_x,J_y$ are the matrices in
\eqref{66}. $T_C$ and $T_N$ were defined in \rf{1011} and $D_a$ is the same as in \rf{11}. 

The expectation  value of \rf{933}  is then   just linear in Green's functions and their derivatives at coincident points. In the  conventions  of    section \ref{sec3} we have 
\begin{equation}
\begin{aligned}
 &\langle x^2\rangle=4G_x,\qquad 
 \langle y^2\rangle=4G_y,\qquad 
 \langle x^r(I_x)^{rs}\nabla_3x^s\rangle =-4G_{x,3},\qquad 
 \langle y^m(J_y)^{mn}\nabla_3y^n\rangle&=-4G_{y,3},\\
& \qquad \qquad \langle\bar\theta M\theta\rangle
   =-\operatorname{tr}(M G_\theta),\qquad \qquad 
 \langle\bar\theta M D_a\theta\rangle
   =-\operatorname{tr}(M G_{\theta,a}), 
\end{aligned}
\label{934}
\end{equation}
where $M$ stands for any  constant spinor matrix. 
The trace is over 16 physical spinor components. Using the directional decomposition \eqref{905} and the definitions
\eqref{52}, \eqref{5222} we have
\begin{equation}
\sum_{a=1}^3G_{\theta,a}\rho^a
 =r_CP_C+r_NP_N+(2q_C-d_C)T_C+(2q_N-d_N)T_N.
 \label{935}
\end{equation}
  Note that compared to
$E_\theta=\rho^aG_{\theta,a}$ in \rf{504},\rf{5222}   here $\rho_a$ multiplies on the right. The difference is
$2(q_C-d_C)T_C+2(q_N-d_N)T_N$.
The derivative  acts at
the first endpoint. 
As a result, we find from \rf{933}
\begin{equation}
\begin{aligned}
 \langle\mathcal J\rangle={}&
 -16G_x-10G_y-4G_{y,3}-16G_{x,3}\\
&-\tfrac83v_C+\tfrac{16}{3}v_N   +r_C+2r_N-d_C-2d_N+2q_C+4q_N.
\end{aligned}
\label{936}
\end{equation}
To compute this   explicitly we use 
the 
relations  from \eqref{87}--\eqref{89}.
Note  that the EOM terms   from $E_\theta$ here  cancel  since 
$ r_C+2r_N-d_C-2d_N=0$.
 
As a result, 
\begin{equation}
 \langle\mathcal J\rangle
       =-\tfrac23\big(28G_x+13G_y\big)+\tfrac5{3} V^{-1}  \ . 
 \label{939}
\end{equation}
This  is the same as $\vev{S_4}= V \vev{\cL_4}$ in \rf{902} 
and thus  reference-point coefficient  $f_2$ in \rf{976}  vanishes, as was already   announced in \rf{011}--\rf{014}. 
This cancellation happens  for any   $G_x$ and $G_y$   which  are transcendental functions
of $k>2$ (cf. Appendix \ref{api}). Section~\ref{n1} relates this local result to the vanishing  of the 
 integrated $f_2$  coefficient.


Let us   add    a few comments on the dependence on the fermionic zero-mode complement.
The dual functionals in \rf{927} are fixed by the conditions
$\Lambda^A\mathcal Z_B=\delta^A{}_B$ only up to functionals vanishing on the zero modes.
For the bosons and for the charged fermionic zero modes we used the orthogonal choice.
As discussed in Appendix \ref{apd},  for  the  8 neutral fermionic zero modes there is a natural one-parameter family of
complements labelled by $\alpha$, with coincident zero-mode  projector density  (cf. \rf{932})
\be \la{016}
 \mathcal P_\alpha=\tfrac12(\PC+T_C)+\PN+\alpha\, T_N\ ,
\ee
where $\alpha=0$ is the orthogonal complement and $\alpha=1$ is the supersymmetric  choice used above.
Both $\vev{\cL_4}$ and $\vev{\cJ}$ depend on $\alpha$ through $v_N=\tfrac12\alpha V^{-1}$
and $d_N=-\alpha V^{-1}$  and one finds (see Appendix~\ref{apd})
\al{
 V\vev{\cL_4}_\a =-\tfrac23\big(28G_x+13G_y\big)
  & +\big(1+\tfrac23\alpha^2\big)V^{-1}\ , \quad 
 \vev{\cJ}_\a =-\tfrac23\big(28G_x+13G_y\big)
   -\big(3-\tfrac{14}3\alpha^2\big)V^{-1}\ , \la{5555} \\
 &f_2(0;k;\alpha)=V\vev{\cL_4}_\a -\vev{\cJ}_\a
   =4\,(1-\alpha^2)\,V^{-1}\ . \la{018}
}
Three features of \rf{018} are worth noting:

(i) The coefficients of $G_x$ and $G_y$ do not depend on $\alpha$, so the cancellation
of the whole transcendental $k$-dependence between the quartic vertex and the Jacobian
does not depend on the choice of the complement.

(ii) $f_2$ is even in $\alpha$. The replacement $T_N\to-T_N$ is implemented by the
rotation by $\pi$ in the $(9,10)$ plane, which belongs to the $SO(4)$ of $\AdS_4$
rotations 
and leaves $\Gamma_*$, $\PC$, $\PN$, $T_C$ and
$\Gk$ invariant (it also maps $I_x\to-I_x$). The complements with $\alpha$ and $-\alpha$
are thus related by a symmetry, and $\alpha=\pm1$ are adapted to two different
preserved supercharges.

(iii) $f_2$ vanishes precisely at $\alpha=\pm1$, i.e. for the complements for which the
residual pair and the quantum pair in \rf{950} transform separately under the
corresponding preserved supercharge 
(see Appendix~\ref{apd}).


The $\alpha$-dependence in \rf{018} concerns the coefficient at the
reference point, not the result after the complete  collective-coordinate  integral. Changing
$\Lambda^A$ mixes the odd collective coordinates with the quantum
fermions, through a shift of the form
$\zeta\to\zeta+\rT_2^{-1/2}\Lambda'(\theta)$. It can therefore
redistribute the first inverse-tension terms among the Grassmann
components of the    density on the
collective-coordinate super-orbit. Independence of the integrated
result requires that the constraints, measure and integration
prescription be transformed together.\foot{The paired prescription in section~\ref{n1} selects the adapted local
complement using  the same preserved supercharge. Its nonlinear
measure contribution is calculated in Appendix~\ref{apf}: the sum
$\mathcal J_{\mathcal Q}+\mathcal J_{\rm EOM}$ has the same expectation
value as $\mathcal J$, see \rf{n11}. These are two evaluations of the
local measure in the respective descriptions, not additive
corrections. 
This equality is the local input to the cancellation of the  integrated
 coefficient in  \rf{n14}.}

\section{Integration over the collective coordinates}\label{n1}

The computation of section \ref{sec5} treats the $12+12$ zero modes by the nonlinear change of variables
\rf{928}, which makes the exact flatness of the zero-mode directions manifest and reduces the problem to
the constrained fluctuation theory at the reference point, at the price of introducing the Jacobian $\Ber\cM$ and the corresponding $\Tmem^{-1}$
 insertion $\cJ$.
 
  In this section we evaluate the same integral in a different and more standard way:
the zero modes are kept  in a linear expansion of the fields, and are given
a quadratic weight by a supersymmetric deformation of the action. This is  analogous  to the way the paired zero-mode
prescription of Appendix \ref{apps} is implemented in \ci{Gautason:2023igo,Kurlyand:2026yke} at the
one-loop level. Extending it  to order $\Tmem^{-1}$ requires the supersymmetric completion of the
deformation in the interacting theory. The resulting interaction terms replace the Jacobian insertion of
section \ref{sec5}. We shall verify that the two treatments agree.

The  idea is to use a localization-type  argument.\foot{A possibility of  applying localization to M2  
brane path integral 
was discussed  by J. van Muiden (unpublished, see also  earlier in \ci{Gautason:2025per}). We thank him   for related  discussions.}
  It  requires both deformation
independence of the path  integral and a calculation of a local
measure selected by the deformation.
The argument consists of the following steps. 

(i) The paired zero-mode prescription is implemented
as a deformation of the path integral by a $\mQ$-exact term, where $\mQ$ is the full form of the  preserved supercharge
of \rf{948}; the deformed integral is independent of the deformation parameter $t$ provided $\mQ$ is a symmetry of
the regulated measure and closes on a bosonic symmetry $\cR$, see \rf{n4}.
This deformation lifts the $12+12$ paired zero modes. At the quadratic level it is of the same type as the
regularization of \ci{Kurlyand:2026yke} and implements the paired prescription motivated by
\ci{Gautason:2023igo}; its new content is at order $\Tmem^{-1}$.

 (ii) For $t\to\infty$ the integral
over the collective coordinates localizes onto the zeros of the odd vector field $\qq$ generated by $\cQ$ on the
family of saddles; there are two such zeros, see \rf{n5}. 

(iii) Near each zero the integral over the $12+12$
collective coordinates is a Gaussian one, which reproduces the unit factor per boson-fermion pair discussed in
Appendix \ref{apps}.

 (iv) 
 The deformation must be exact with respect to the nonlinear supersymmetry of the interacting
theory; its completion beyond the quadratic order couples the lifted zero modes to the quantum fields.
Integrating out the lifted modes in the presence of these completion terms produces the measure
insertions $\cJ_\cQ$ (from the nonlinear dependence of the fermionic partner $\mathcal F=\cQ\mathcal B$
of the bosonic projections on the quantum fields) and $\cJ_{\rm EOM}$ (from the equation-of-motion term
in $\cQ^2$); they are computed in Appendix \ref{apf}.

(v) $\vev{\cJ_\cQ+\cJ_{\rm EOM}}$ equals $\vev{\cJ}$ of
section \ref{sec5}, so that the $\rT_2^{-1}$ coefficient at each  localized configuration   equals the reference-point value
$f_2(0;k)=0$ of \rf{014}, and hence the integrated coefficient $\ff_2$ in \rf{n3} vanishes.

As in section~\ref{sec5}, we assume $k>2$ and denote the 12 even and 12 odd collective coordinates by
$\mathrm a=(a^I,\zeta^I)$. The 12 even directions are tangent to the family of isometric saddles
$\AdS_4\times\Gr_\CC(2,4)$, $\Gr_\CC(2,4)=U(4)/[U(2)\times U(2)]$: the Euclidean AdS position and the
complex two-plane $\Pi\subset\CC^4$ (complex for the common-phase structure of the coordinates $Y_a$)
wrapped by the instanton, $\Sigma_\Pi=(S^7\cap\Pi)/\ZZ_k$. The 12 odd directions are generated by the
12 broken supersymmetries. 

The integral over $\mathrm a$ was given in \rf{n2}. It is over 
the orbit of the reference instanton under the {broken} symmetries; it is not the
moduli space of configurations preserving the supercharge $\cQ$ of \rf{948} used below. None of the
$12+12$ zero modes preserves $\cQ$: the generator $\cR=\cQ^2$ acts on all of them with non-zero weights
(see Appendix \ref{apf}), so that they form 12 $\cQ$-doublets, and $\cQ$ is preserved only at two isolated
points of the family, see  \rf{n5}.\foot{In the terminology of section 4.1 of \ci{Kurlyand:2026yke}, the $12+12$
zero modes are ``non-cohomological'', i.e.\ paired zero modes of the quadratic operator which are not in
the kernel of $\mathcal R=\cQ^2$, while the ``BPS moduli'' of $\cQ$ (zero modes in $\ker \mathcal R$) are absent for $k>2$,
in agreement with the removal of the corresponding modes by the $\ZZ_k$ projection found there.   For
$k\le2$ the $\ZZ_k$ projection no longer removes the modes in $\ker \mathcal R$ and the counting of zero modes
becomes $20+16$ (see Appendix \ref{apa}). The additional $8$ bosonic and $4$ fermionic zero modes should be
the BPS moduli and obstruction modes of \ci{Kurlyand:2026yke} (the $n=0$ holomorphic normal deformations
and the $n=-2$ co-kernel modes there, cf.\ the footnote after \rf{n42} in Appendix \ref{apf}). In this case the
instanton is not isolated and the argument below does not apply as it stands.}

Along the even directions the fluctuation problem is transported by the isometries, provided
the fields, the constraints and the regulator are transported together, so that
$f_2(a,0;k)=f_2(0;k)$ for any choice of the complement. Along the odd directions this is not so:
the complement adapted to one preserved supercharge is not adapted to the other eleven
(the analog of \rf{950} for them contains the terms mixing the odd collective coordinates with the quantum
fermions), and $f_2(\mathrm a;k)$ depends on $\zeta$ in a complement-dependent way, cf.\ the discussion
at the end of section \ref{sec5}. The non-vanishing of the integral $\int d\mu_0$ in \rf{n2} thus does not mean that
only the highest Grassmann component of the density is non-zero, and we shall neither need nor claim
that $f_2(\mathrm a;k)$ vanishes point-wise on the super-orbit.


\subsubsection*{Supersymmetric deformation and the paired zero-mode prescription}

We shall assume that the prescription for the paired zero modes extends to the
interacting path integral and its regulated measure for one preserved
supercharge.
 Let ${\mQ}$ denote the full supercharge, including the static-gauge  and $\kappa$-gauge compensating transformations.
We reserve the notation  $\mathcal Q$ for the reduced physical-component transformation
 used in the local calculation below and in Appendix~\ref{apf}.
 Here $ {\mQ}$  is supposed  to act off shell on an extended  set 
 of fields  which,   for notational simplicity,  we shall still denote as $\rm X$.
 
 Let  us assume that    the measure and  the 
action $S$ are  invariant under $\mQ$, i.e. 
${\mQ}  S=0$.
Suppose  that    ${\mQ}^2 =  \mathcal R$   where    $\mathcal R$  
 is a  compact bosonic symmetry  and choose  a special 
  odd  functional $\Psi[\mathrm X]$  which is  invariant under   $\mathcal R$, 
 i.e.  ${\mQ}^{\,2}\Psi=0$.
 
 The required deformation-independence   identity  then follows  from the above properties  ($t>0$) 
\al{  \frac{dZ(t)}{dt}
   =-\int[D\mathrm X]\,\, {\mQ}
       \big(\Psi\,  e^{-S-t\,{\mQ}\Psi}\big)=0 , \qquad \qquad \ \ \ 
 Z(t)\equiv \int[D\mathrm X]\,
       e^{-S[\mathrm X]-\, t\, {\mQ}\Psi[\mathrm X]}\ .
 \label{n4}
}
Here $t>0$. Formally $Z(0)$ is the M2 brane partition function \rf{229}, but, as noted after
\rf{n3}, it is not defined as it stands: the integral over the odd collective coordinates vanishes and
the one over the $\AdS_4$ position diverges, so that $Z(0)$ is of the form $0\cdot\infty$. The
supersymmetric prescription for the paired zero modes of Appendix \ref{apps} is the definition
\be \la{n41}
Z\equiv Z(t)\ , \qquad t>0 \ ,
\ee
which, by \rf{n4}, does not depend on $t$ or on the choice of $\Psi$ (the deformations for two choices
differ by the $\mQ$-exact term $t\,\mQ(\Psi_1-\Psi_0)$, and the same argument applies along an
interpolation between them). Its value may thus be computed in the limit $t\to\infty$, where the integral
localizes, see below. We shall assume that \rf{n4} holds at the leading and at the first
inverse-tension order.\foot{We require ${\mQ}^{\,2}\Psi=0$, or the corresponding integrated identity,
to hold to the required order.  For the reduced component transformation $\cQ$
in the $\kappa$-symmetry gauge, 
 $\cQ^2=\cR$ holds  on the fermions only
up to the equation-of-motion terms, see \rf{n24},\rf{n25}. Their effect on the localized integral,
$\cJ_{\rm EOM}$ in \rf{n27}, is shown in Appendix \ref{apf} to have vanishing expectation value at order
$\rT_2^{-1}$, which is the form in which the condition $\mQ^2\Psi=0$ is used here. The integration cycle
must remove the boundary term at the non-compact $\AdS_4$ part. We assume the same analytic regularization
with $\delta(0)=0$ as in the preceding sections, in which the Jacobian of the local field transformation
$\mathrm X\to\mathrm X+\nu\mQ\mathrm X$ is equal to one (Appendix \ref{apco}).}

The invariance of the regulated measure under $\mQ$ is the assumption on which the argument rests, as in
any localization computation.  
What supports
this  assumption is the following. At the quadratic level it is established for the present system in
\ci{Kurlyand:2026yke}: after the introduction of auxiliary fields (the field $h_0$ of \rf{948}) the
linearized supersymmetry closes off shell and the quadratic fluctuation action is $\cQ$-exact. The
transformation $\mathrm X\to\mathrm X+\nu\mQ\mathrm X$ is local, and its Jacobian is one in the analytic
regularization with $\delta(0)=0$ used in sections 2--5 (Appendix \ref{apco}). This rule removes the local
contact terms but not the finite zero-mode projector terms, which are retained in the local computation
of Appendix \ref{apf}. 

The   complication that the  gauge-fixed algebra  closes only 
 on-shell,  i.e. the equation-of-motion term in $\cQ^2$, is analysed  below and shown to be
harmless at order $\Tmem^{-1}$.
 At the quadratic level the deformation \rf{n4} is equivalent to the
regularizations used previously to lift the paired zero modes: the shift of the mass parameter of the
quadratic action in \ci{Kurlyand:2026yke}, which is itself a $\cQ$-exact deformation since that action is
$\cQ$-exact. The mass deformation of the ABJ(M) background in \ci{Gautason:2023igo} is instead a physical
deformation, whose undeformed limit motivates the same paired prescription (see also remark (i) after
\rf{m13} below). The deformation is formulated at the level of the full path integral because at
order $\Tmem^{-1}$ the lifting is no longer a quadratic operation: the dependence of the lifted zero-mode
coordinates on the quantum fields produces the measure insertions \rf{n10}--\rf{n11} below.

\subsubsection*{The preserved supercharge and its two zeros}

The preserved supercharge pairs all 12  even and 12  odd tangent
directions at the reference point. Its 
square   $\mathcal R$    is a rotation of $\AdS_4$   with equal  weights 
in two real planes, together with the relative phase rotation of the planes 
$\Pi_0$ and $\Pi_0^\perp$ in $\mathbb C^4$ (the latter is generated by $\operatorname{diag}(1,1,-1,-1)$).

 Let $\qq$ denote the
induced odd vector field on the family of saddles, namely,  $\qq {\rm a}^A= \cQ {\rm a}^A\big|_{\widetilde{ \rm X}=0}$, 
$\  {\rm a}^A = (a^I, \zeta^I)$.

Its zeros are the embeddings preserving the chosen supercharge, rather than all the points fixed by
$\cR=\qq^2$ \ci{Schwarz:1995dg}. For the $\AdS_4$ part, the zero is the fixed point of the Killing spinor
of $\cQ$, used in \ci{Gautason:2025per} to fix the position of the M2 brane in $\AdS_4$. The quadratic
worldvolume supersymmetry around the instanton and the structure of its zero modes are described in
sections 3.3 and 4.1 of \ci{Kurlyand:2026yke}.


The decomposition of the Killing spinors given in Appendix~\ref{apf}
shows that there are two such configurations
\begin{equation}
 p_+=(0,\Pi_0),\qquad p_-=(0,\Pi_0^\perp),
 \qquad \Pi_0=\operatorname{span}(e_1,e_2)\subset\mathbb C^4 .
 \label{n5}
\end{equation}
Here $\Pi_0$
is the reference complex two-plane  ($e_i$  being the standard orthonormal basis).
The membrane wraps $(S^7\cap\Pi_0)/\mathbb Z_k\simeq S^3/\mathbb Z_k$.\foot{The  two configurations 
correspond   to M2  branes wrapping  the two $S^3$'s   defined  in embedding coordinates  by 
$Y_3=Y_4=0$ and as $Y_1=Y_2=0$. 
These are the two configurations found in the deformed problem of \ci{Gautason:2023igo} and the two
isolated saddles $\widehat\Sigma_\pm$ of \ci{Kurlyand:2026yke}. The equality of their leading
contributions is the factor $Z_{\rm zero\text{-}modes}=2$ of \ci{Gautason:2023igo}. The two points are
isolated within the family: all $12+12$ tangent directions at $p_\pm$ are rotated by $\cR$ with
non-zero weights, so that no tangent displacement preserves $\cQ$ (Appendix \ref{apf}).}

For example, let $\vartheta_1,\vartheta_2\in[0,{\pi\ov 2}]$ be the principal
angles of 2-plane $\Pi$ relative to $\Pi_0$, and let $\varrho<1$ be the radial
coordinate of the instanton centre in  AdS$_4$. A positive bosonic
part of the geometric deformation is then 
\begin{equation}
 v(\varrho,\Pi)=\frac{4\varrho^2}{1-\varrho^2}
   +2\big[\sin^2(\vartheta_1-\vartheta_2)
          +\sin^2(\vartheta_1+\vartheta_2)\big] .
 \label{n32}
\end{equation}
It vanishes precisely at \rf{n5}.\foot{The mixed planes, spanned by one line
in $\Pi_0$ and one in $\Pi_0^\perp$, form a component
$\mathbb{CP}^1\times\mathbb{CP}^1$ fixed by the internal circle. They do
not preserve the odd symmetry: their internal contribution to \rf{n32}
is $4$, not zero.}
Here \rf{n32} is the squared norm of the component of the Killing spinor of $\cQ$ along
the 12 supersymmetries broken by the displaced saddle $\mathrm X_0(\mathrm a)$. 

In terms of
the odd vector field $\qq$ introduced above it is the square norm of the body of $\qq\,\zeta$: at the point
$\rm a$ of the family of saddles the fixed supercharge $\cQ$ decomposes into a part preserved
by $\mathrm X_0(\mathrm a)$ and a part along the 12 broken supercharges, with bosonic coefficients
$\phi^I(\mathrm a)$, and
\be \la{n44}
 \qq\,\zeta^I=\phi^I(\mathrm a)+\OO(\zeta^2)\ , \qquad\qquad
 v(\varrho,\Pi)=\sum_{I=1}^{12}\big|\phi^I(\mathrm a)\big|^2 \ ,
\ee
where the sum is over an orthonormal frame of the broken supercharges, in the normalization \rf{945} of
the fermionic zero modes, so that $v=0$ exactly where $\cQ$ is preserved. The $\AdS_4$ term in \rf{n32} follows from the
dependence of the Killing spinor on the position: with $\varrho=|x|$ the ball coordinate of
\rf{5} and $\varrho=\tanh u$, the broken part at geodesic distance $u$ from the centre has norm
squared $4\sinh^2u$. The internal term is derived in Appendix \ref{apf}, see \rf{n37}.
Near $p_\pm$ both terms are quadratic in the collective coordinates with the same coefficient, i.e. 
$
 v\simeq 4\varrho^2+4\big(\vartheta_1^2+\vartheta_2^2\big) ,
$
in agreement with the weights $\pm2$ of $\cR$ on all even directions. The identification with
$|K\mathcal B|^2=4a^2$ in \rf{n7},\rf{mm8} below fixes the relative normalization of the internal
collective coordinates and of the broken Killing-spinor components, cf. \rf{n48}.
A localizing functional whose bosonic part is
\rf{n32} on the whole family of saddles may be constructed from the broken component, e.g.
$\Xi=\sum_I\zeta^I\phi^I(\mathrm a)$ (in the normalization \rf{945} of the fermionic zero modes),
for which $\qq\,\Xi=v+\OO(\zeta^2)$. Near $p_\pm$ it reduces to the local form $\Psi_p$ of \rf{n6}.

We shall assume that the
full-field $\rm X$ deformation in \rf{n4} extends this geometric choice  and  agrees
with the paired local form below near both points, 
 and has no other contributing configurations in the instanton sector.

\subsubsection*{Local form of the deformation and the paired Gaussian}

Let $\Lambda_B^I$ and $\Lambda_F^I$ be the bosonic and fermionic  zero-mode duals  defined  in 
Appendix~\ref{apd}  ($I=1,\ldots,12$).\foot{We shall use  the same  label $I$ for the   bosonic  and fermionic  duals, 
the identification should be clear  from the  setting.}
 In a local frame at either
configuration let  us define functionals of the full   fields by
\al{
& \mathcal B^I[\mathrm X]=\Lambda_B^I\mathrm X_B,
 \qquad\qquad 
 \mathcal F^I[\mathrm X]={\mathcal Q}\mathcal B^I[\mathrm X] \ , \la{mn6}\\
 &
 \Psi_p\equiv (K\mathcal B)^T\mathcal F =  (K\mathcal B)^T {\mathcal Q}\mathcal B  = 
 (K  \Lambda_B \mathrm X_B)^T {\mathcal Q}\,  \Lambda_B \mathrm X_B
 \ . 
  \label{n6}
}
Here $\mathrm X_B$ denotes the full bosonic field.  
The functionals $\mathcal B^I$ are even and their superpartners $\mathcal F^I$ are odd.
$\Psi_p$ is the local component representative of the
full-field deformation near one of the two    supersymmetric
configurations $p$  in \eqref{n5}.
Here $\mathcal Q$ denotes
the reduced-component  supersymmetry  transformation in the compact phase convention
specified in Appendix \ref{apf}. It must not be confused with the extended  operator
$\widehat{\mathcal Q}$ (whose algebra should close off shell) in \eqref{n4}. 

The numerical matrix $K$ (not to be confused with  the generators $\mathrm K_B$ in \rf{928})
represents
the bosonic symmetry on the twelve projections,
$\mathcal R\mathcal B=K\mathcal B$, and satisfies
$K^T=-K$ and $K^2=-4\cdot \mathbf1$.

At quadratic order in the fields the supersymmetry variation is 
\begin{equation}
 ({\mathcal Q}\Psi_p)_2
       =|K\mathcal B|^2-\mathcal F_1^TK\mathcal F_1 ,
 \label{n7}
\end{equation}
where $\mathcal F_1$ is the linear part of $\mathcal F$.
Then   at large $t$ the dominant  contribution in \rf{n4} comes from the Gaussian factor 
\be \la{m5}  \exp\!\big[-t(\mathcal Q\Psi_p)_2\big]
 =\exp\!\big[-t|K\mathcal B|^2
                   +t\mathcal F_1^TK\mathcal F_1\big] \ . \ee
  Let us introduce local linear
integration coordinates near a fixed  saddle  parametrized by  $(a^I,\zeta^I)$ as ($\epsilon= \rT^{-1/2}_2$)
\begin{equation}
 \mathrm X_B=\mathcal Z_Ba+\epsilon\widetilde{\mathrm X}_B,\ \ \ 
 \qquad
 \mathrm X_F=\mathcal Z_F\zeta+\epsilon\theta,\ \  \ 
 \qquad\ \ \ 
 \Lambda_B\widetilde{\mathrm X}_B=0, \ \ \ \Lambda_F\theta=0 .
 \label{n8}
\end{equation}
 $\mathcal Z_B$ and $\mathcal Z_F$  stand for  the bosonic and fermionic
zero-mode profiles. 

 $(a^I,\zeta^I)$ in \rf{n8} are the coefficients of the zero modes in a linear mode
expansion of the fields, as in any perturbative computation. They are not the nonlinear collective
coordinates $\mathrm a$ of \rf{928}, and the change of variables \rf{n8} has a field-independent Jacobian.
We nevertheless keep the same notation, since at the reference point the two sets of variables agree to
leading order.

  Then 
\begin{equation}
 \Lambda^I_B({\mathcal Q}_1{\mathrm X_F})_B
       =\mathsf L^I_J\Lambda^J_F X_F,\qquad \ \ \  \mathsf L^I{}_J=\Lambda_B^I
({\mathcal Q}_1\mathcal Z_{F,J})_B \ , \qquad 
 \qquad\qquad  \det\mathsf L\ne0 \ .
 \label{n9}
\end{equation}
This is the functional version of the separation in \rf{950}. 
Eq.\ \rf{n9} states that the supercharge $\cQ$ does not map the constrained quantum fermions into
any of the 12 bosonic collective coordinates, $\Lambda_B^I(\cQ_1\theta)_B=0$ for $\Lambda_F\theta=0$.
For the internal duals this is the content of \rf{950} at $\a=1$. For the four $\AdS_4$ duals
one has $\Lambda_r(\cQ_1\theta)_B=-V^{-1}\int d^3\sigma\sqrt g\ \bar\varepsilon\Gamma_r\theta$, which
vanishes since the constant spinors $\Gamma_r\varepsilon$ have zero pairing with the constant modes
of the constrained $\theta$ (these lie in $\ha(\PC-T_C)$, cf.\ Appendix \ref{apa}); we have checked this
directly. The matrix $\mathsf L$ is invertible because $\cQ$ pairs all 12 even with all 12 odd tangent
directions.

Together with the constraints and the normalization of the zero-mode
duals, this gives 
\begin{equation}
 \mathcal B=a,\qquad \qquad \mathcal F_1=\mathsf L\zeta \ , 
 \label{mm8}
\end{equation}
i.e. the adapted projection  removes  the 
quantum-field contribution from $\mathcal F_1$.
The  localizing weight \rf{m5}  in the path integral  is then the exponent of a  quadratic form, i.e. 
$ \exp\!\big[-4t\,a^Ta
        +t\,\zeta^T\mathsf L^TK\mathsf L\zeta\big]. 
$
 The even  and odd Gaussian integrations  then 
contribute the   $t^{\mp 6}$  factors   so that   their 
product is finite and independent of $t$, as in the paired normalization
discussed in Appendix~\ref{apps}.\foot{
In more detail,  for odd  variables 
the  exponential is a finite polynomial  with its 
highest term $ \frac{t^6}{6!}
 \big(\mathcal F_1^TK\mathcal F_1\big)^6
 \ \sim \ t^6\,\mathcal F_1^1\dots\mathcal F_1^{12}.
$}
In the limit $t\to \infty$  the bosonic integrand is localized at $a=0$, i.e. 
 the leading localized integration functional is proportional to (cf. \rf{mm8})
\begin{equation}
 \delta^{(12)}(a)\,
 \mathcal F_1^1\dots\mathcal F_1^{12}  \ \  \sim \ \   \delta^{(12)}(a)\,
 \zeta^1\dots\zeta^{12} 
 \label{m13}
\end{equation}
The quadratic localizing weight  in \rf{m5} thus  suppresses the twelve
bosonic zero-mode coefficients and supplies the leading product of their
twelve fermionic partners.

The bare odd integral $\int d^{12}\zeta\,1$ vanishes, and equal numbers of bosonic and fermionic zero
modes do not change this. The non-zero result comes from the paired Gaussian: the top component of the odd
exponential supplies the twelve odd factors, the bosonic exponential the regulator of the twelve
even ones, and the powers $t^{\pm6}$ cancel. The factor \rf{m13} is thus a distributional statement for
the leading localizing weight under the integral. It does not mean that the original measure had only
a highest odd component, and it fixes neither the absolute value nor the orientation of the local
normalization, which are inherited from $Z_1$. Its use for evaluating the  instanton
contribution rests on the deformation independence \rf{n4}.

Three remarks are in order.

(i)  In the variables \rf{n8} the quadratic part of the deformation,
\be \la{n53}
t\,\big(\mQ\Psi\big)_2=t\big(4\,a^Ta-\,\zeta^T\sL^TK\sL\,\zeta\big)\ ,
\ee
is a mass term for the $12+12$ zero modes and nothing else: in the deformed quadratic action these
modes are no longer zero modes, their propagators are  proportional to $t^{-1}$ and $\sim  t^{-1}(\sL^TK\sL)^{-1}$
respectively, 
and all other modes are unaffected. This is a $\cQ$-exact ``regularization'' of the quadratic action
which leaves the linearized supersymmetry transformations unchanged.

It is  of the same type as the shift of
the mass parameter of the quadratic action used in \ci{Kurlyand:2026yke} to lift the paired zero modes
(that shift is also $\cQ$-exact, since the quadratic action is $\cQ$-exact). The only difference is that
\rf{n53} is supported on the zero-mode subspace, via  the projections $\mathcal B=\Lambda_B\mathrm X_B$
and $\mathcal F=\cQ\mathcal B$.\foot{The regularizations
differ in what is deformed. The shift of the mass parameter in \ci{Kurlyand:2026yke} and the
deformation \rf{n53} are $\cQ$-exact and leave the supersymmetry transformations unchanged, so that
the result is independent of the deformation parameter. The background deformation of
\ci{Gautason:2025per} changes the action, the transformations and the $R$-charges, so that the result
depends on the deformation parameter and the undeformed one is recovered in the limit.} 
Both lift the 12 paired doublets and both reproduce the one-loop result \rf{nn2}; we do not need to
identify the two deformed quadratic operators or their nonlinear completions.
Then  $t\to\infty$ is the limit in which the regularized zero-mode propagators vanish and
the Gaussian exponential insertion  gives   \rf{m13}.

(ii) Eq. \rf{n8} is a linear split of the fields around the reference saddle, not the nonlinear family
of saddles $\mathrm X_0(\mathrm a)$ of \rf{928}. Along the linear tangent directions the action is not
flat beyond the quadratic order: in terms of \rf{n8} it contains terms of the form $\Tmem a^4$, $a^2\tX^2$,
$\Tmem^{1/2}a^3\tX$, $\Tmem^{-1/2}a\tX^3$ and similar ones involving $\zeta$. These do not contribute
because $a,\zeta=\OO(t^{-1/2})$ in the localized Gaussian, i.e.\ the limit $t\to\infty$ in \rf{n41}
is to be taken at fixed $\Tmem$ before expanding in $\Tmem^{-1}$:  in the regularized-propagator
language, each  leg costs a factor $t^{-1/2}$.

(iii) The deformation is $t$-independent only if it is exact with respect to the supersymmetry $\mQ$ of the
interacting theory, which is nonlinear in the fields. The quadratic weight \rf{n53} is therefore not the
complete deformation. Writing $\mathcal F=\mathcal F_1+\mathcal F_3+\dots$ with
$\mathcal F_1=\cQ_1\mathcal B$, $\mathcal F_3=\cQ_3\mathcal B$, and denoting by
$\mathcal D_3=(\cQ_1\cQ_3+\cQ_3\cQ_1)\mathcal B$ the cubic part of $(\cQ^2-\cR)\mathcal B$
(the subscripts indicate the degree in the fields), the graded product rule gives
\be \la{n54}
t\,\cQ\Psi_p=t\big[|K\mathcal B|^2-\mathcal F_1^TK\mathcal F_1\big]
+t\big[-2\,\mathcal F_1^TK\mathcal F_3+(K\mathcal B)^T\mathcal D_3\big]+\OO(t\,\mathrm X^6)\ ,
\ee
where $\mathcal F_3^TK\mathcal F_1=\mathcal F_1^TK\mathcal F_3$ was used ($\mathcal F_{1,3}$ are odd and $K$
is antisymmetric). The two terms in the second bracket couple the lifted modes to the quantum fields and are
fixed by supersymmetry once the quadratic deformation \rf{n53} is chosen: the first comes from the nonlinear
part of $\cQ$ acting on the projections, $\mathcal F=\sL\zeta+\ep^2\sL A_2\zeta+\dots$, the second from the
equation-of-motion term in $\cQ^2$. At order $\Tmem^{-1}$ these terms are what distinguishes the computation
of this section from a naive one with regularized propagators and the undeformed vertices.

\iffa 
(iii) The deformation is $t$-independent only if it is exact with respect to the supersymmetry $\mQ$ of the
{interacting} theory,  which is nonlinear in the fields. Its complete form is therefore not
\rf{n53}  but \rf{54} below.
For the particular component functional in \eqref{n6}, define
\be 
 \mathcal F_1=\mathcal Q_1\mathcal B,\qquad
 \mathcal F_3=\mathcal Q_3\mathcal B,\qquad
 \mathcal D_3=(\mathcal Q_1\mathcal Q_3+
                 \mathcal Q_3\mathcal Q_1)\mathcal B.
\ee
The subscripts denote degree in the unrescaled physical fields.
The graded product rule then gives
\begin{equation}
 \begin{aligned}
 t\mathcal Q\Psi_p
 ={}&t\big[|K\mathcal B|^2-\mathcal F_1^TK\mathcal F_1\big]+t\big[-2\mathcal F_1^TK\mathcal F_3
                 +(K\mathcal B)^T\mathcal D_3\big]
 +\OO(t\,\mathrm X^6).
 \end{aligned}
 \label{n54}
\end{equation}
This is the local component expansion whose identification with
the reduced full-field deformation is part of the prescription
in \eqref{n4}.
\fi

\iffa In \rf{n54}  the second term, which couples the lifted modes to the quantum fields, is fixed by supersymmetry
once the quadratic deformation is chosen. It  comes from the nonlinear part of $\cQ$ acting on the
projections, $\mathcal F=\cQ\mathcal B=\sL\zeta+\ep^2\sL A_2\zeta+\dots$, and from the equation-of-motion
term in $\cQ^2$.  
At order $\Tmem^{-1}$ these  terms required by supersymmetry 
 are what distinguishes the computation of this section from a
naive one with regularized propagators and the undeformed vertices. \fi

Integrating out the lifted modes
in their presence produces the insertions $\cJ_\cQ$ and $\cJ_{\rm EOM}$ of \rf{n10}--\rf{n11} below,
which take over the role of the Jacobian insertion $\cJ$ of section \ref{sec5}. In the linear
parametrization there is no Jacobian, so the agreement \rf{n11} between $\vev{\cJ_\cQ+\cJ_{\rm EOM}}$ and
$\vev{\cJ}$ is a non-trivial consistency check between the two parametrizations of the same integral.

\subsubsection*{Nonlinear constraint and equation-of-motion insertions}

To determine the first inverse-tension
correction one must retain the nonlinear dependence of 
$\mathcal F$ on the quantum fields. 
At $a=0$  we get 
\begin{equation}
 (\mathsf L^{-1}\mathcal F)^I
 =\zeta^I+\epsilon^2(A_2)^I{}_J\zeta^J
    +\epsilon^3U_3^I
    +O(\epsilon\zeta^2,\zeta^3,\mathrm X^5) \ . 
 \label{n10}
\end{equation}
All collective-coordinate vectors are columns. The entries of the
$12\times12$ matrix $A_2$ are quadratic in quantum fields and have even
Grassmann parity, while $U_3$ is cubic.
The terms of order $\zeta^2$ and $\zeta^3$ in \rf{n10} do not contribute at order $\ep^2$: in the top
component $\mathcal F^1\dots\mathcal F^{12}$ of the localizing Gaussian they would require more than 12 odd
coordinates, and in combination with the shifted root $\zeta_*=-\ep^3U_3+\dots$ (see the footnote below)
their contributions start at order $\ep^4$. The term $U_3$ itself is of order $\ep^3$ and does not contribute
at order $\ep^2$.
The absence of terms of even order in the fluctuation fields in $\cQ=\cQ_1+\cQ_3+\dots$, and of the cubic
vertices in \rf{005}, is a consequence of the $\ZZ_2$ symmetry of the gauge-fixed theory discussed in
Appendix \ref{apf}.

Fermionic integration   then gives a factor\foot{  The inhomogeneous term $\epsilon^3U_3$ in \eqref{n10} shifts the
fermionic constraint root to
$
 \zeta_*=-\epsilon^3U_3+O(\epsilon^5).
$
The quadratic action annihilates the fermionic zero-mode profiles, and
there is no cubic interaction vertex in \rf{005}. A term in the
original quartic action containing $r\geq1$ shifted zero-mode legs
therefore scales as
$
 \epsilon^{-2}\epsilon^{4-r}(\epsilon^3)^r
   =\epsilon^{2+2r},
$
which starts at $\epsilon^4$. In the derivative of the odd constraint,
the terms $O(\epsilon\zeta^2,\zeta^3)$ contribute at the shifted root
only beyond $\epsilon^2$. Hence the odd constraint contributes
at the requested order only through
$\mathcal J_{\mathcal Q}=\operatorname{tr}A_2$.
    } 
\be 
\det(1+\epsilon^2A_2)=1+\epsilon^2\mathcal J_{\mathcal Q}+\dots \ , \qquad \qquad 
\mathcal J_{\mathcal Q}=\operatorname{tr} A_2  \la{m7} 
\ . \ee
The collective-coordinate integrations and the quantum contractions are
two different operations. We first integrate the localizing pairs at
fixed quantum fields, and only then evaluate the quantum expectation value 
with $e^{-S_2}$, i.e.  with quadratic quantum fluctuation  action. 
The zero-mode directions have no quadratic  damping 
 term  in the undeformed action --  their quadratic weight is supplied by
the deformation $\Psi$ term in \rf{n4}. The trace
$\mathcal J_{\mathcal Q}=\operatorname{tr}A_2$ comes from the
Grassmann delta-function determinant. In  contrast,  the trace of  $\mathcal J_{\rm EOM}$ in \rf{n27} below 
comes from an ordinary bosonic
determinant.

In   Appendix~\ref{apf} we compute $\mathcal J_{\mathcal Q}$
using the cubic-order scalar supersymmetry transformation.
We also include the correction $\cJ_{\rm EOM}$ coming from the equation-of-motion terms in the
supersymmetry algebra. On the bosonic fluctuations $\cQ_1^2X^M=\cR X^M$ exactly, but on the
fermions, and at cubic order also on the bosons, applying the gauge-fixed transformation twice gives $\cR$ plus
terms proportional to the fermionic field equation, see \rf{n24},\rf{n25}. The remainder
$\mathcal D^I=(\cQ^2-\cR)\mathcal B^I$ adds $(K\mathcal B)^T\mathcal D$ to $\cQ\Psi_p$ and thus modifies the
quadratic localizing weight. The resulting insertion $\cJ_{\rm EOM}$, obtained after the integration over
the bosonic collective coordinates, is given in \rf{n27}. Its vanishing expectation value shows that the
equation-of-motion terms do not affect the localized coefficient at order $\rT_2^{-1}$; the use of the local
result in the complete integral still rests on the deformation prescription \rf{n4}.

With the same Green
functions and finite projector parts as in section~\ref{sec5}, one
finds
\begin{equation}
 \begin{split}
 \vev{\mathcal J_{\mathcal Q}}_p
     =-\tfrac23(28G_x+13G_y)+\tfrac{5}{3}V^{-1} 
       \ , \ \ \ \qquad \qquad 
 \vev{\mathcal J_{\rm EOM}}=0 .
 \end{split}
 \label{n11}
\end{equation}
In \rf{n11}  and  below, $\vev{\cdots}_p$ denotes the normalized expectation value in the constrained quadratic
quantum theory at the configuration $p$, i.e.\ with the collective directions removed and with the same
projected propagators and analytic prescription as in section \ref{sec5}, i.e. it is not an integral over the
collective coordinates. 

The collective-coordinate integrations and the quantum-field  contractions are two
different operations: the localizing pairs are integrated first at fixed quantum fields, which produces
the insertions $\cJ_\cQ$ and $\cJ_{\rm EOM}$, and only then are the insertions contracted  using  $e^{-S_2}$.
The local result \rf{n11} enters the full integral  due to  \rf{n4},\rf{n41}: the
deformation saturates the fermionic zero modes, the integral concentrates at $p_\pm$, and deformation
independence identifies the result with the prescription-defined integral \rf{n41}. 

\subsubsection*{Vanishing of the integrated coefficient}

The calculation of section \ref{sec5} used the nonlinear collective coordinates \rf{928} and included
the Jacobian insertion $\cJ$ of \rf{939}, computed at the reference point. The calculation of this
section uses the linear variables \rf{n8}, has no Jacobian, and instead produces the insertions
$\cJ_\cQ+\cJ_{\rm EOM}$ from the supersymmetric completion of the deformation that lifts the zero
modes. The two are alternative evaluations of the measure contribution of the zero-mode sector of
one and the same integral, not additional contributions, and their agreement is not automatic.
It 
holds because the lifting is $\mQ$-exact (otherwise the result would depend on $t$) and because both
treatments account for the full measure, by $\Ber\cM$ in section \ref{sec5}   and by the completion terms here.
Equation~\rf{n11} thus establishes that
\be \la{n55}
\vev{\cJ_\cQ+\cJ_{\rm EOM}}_p=\vev{\cJ}_p \ ,
\ee
i.e. that  the two parametrizations agree at order $\Tmem^{-1}$. Conversely, given \rf{n55} the reference-point
computation of section \ref{sec5} is the value of the localized integral at $p$.

Consequently, the localized coefficient $f_{2,p}$  equals the reference-point 
coefficient  $f_2(0; k)$ and  thus also   vanishes: 
combining \rf{n11} with \rf{902},\rf{939}
we get for the $\Tmem^{-1}$  coefficient    for  both configurations  in \rf{n5}
\begin{equation}
 f_{2,p} =\vev{S_4-\mathcal J_{\mathcal Q}-\mathcal J_{\rm EOM}}_p=0\ ,\qquad \ \ 
 \qquad p=p_+,\, p_- .
 \label{n12}
\end{equation}
The computation at $p_-$ is the image of the one at $p_+$ under the unitary transformation of $\CC^4$
exchanging the embedding coordinate planes  $(Y_1,Y_2)$ and $(Y_3,Y_4)$. It is an isometry of $S^7/\ZZ_k$, and it maps the Killing spinor of $\cQ$, which is preserved at both $p_\pm$, to itself up to
the change of the tangent and normal frames. Transporting the frames, the zero-mode profiles and the duals
together, the complement adapted to $\cQ$ at $p_-$ is mapped to the complement $\a=-1$ of \rf{945} at the
reference saddle. Since the results \rf{n23} and \rf{n29} hold for both $\a=\pm1$, and
$f_2(0;k;\a=-1)=0$ by \rf{018}, the local coefficient vanishes also at $p_-$.

From  \rf{n4}   and \rf{n2}  we then   get  
\begin{equation}
 Z=Z_1 \sum_{p=p_+,\, p_-}\, w_p(k)
      \Big[1-\Tmem^{-1}f_{2,p}+O(\Tmem^{-2})\Big] \ , 
 \label{n13}
\end{equation}
where the weights $w_p(k)$ include the oriented paired normalization \rf{m13} and the non-zero-mode
determinants at $p$, and are normalized so that $w_++w_-=1$, in accordance with $\int d\mu_0=1$ in
\rf{n2}.
 Neither their equality nor their individual values are needed for the vanishing of the
$\Tmem^{-1}$ term, since $f_{2,p}=0$ at each point separately. 
The present argument thus fixes the
$\Tmem^{-1}$ coefficient relative to the 1-loop result \rf{02} but not the 1-loop normalization itself:
it determines its structure -- two isolated zeros with Pfaffians of equal magnitude (Appendix \ref{apf};
their relative orientation is not fixed by the local analysis), i.e.\ the
factor 2 of the two configurations $p_\pm$ found in \ci{Gautason:2023igo,Kurlyand:2026yke}.\foot{Note that   
 the labels $p_\pm$ refer to the two
complementary internal planes and not to the two orientations of the  M2 brane: a given supercharge
$\cQ$ is preserved by one orientation of the M2 brane. The opposite orientation (the anti-instanton
sector, with the opposite sign of the Wess--Zumino coupling) is a separate sector, preserving the
supercharges of the opposite chirality, to which the same argument applies with the corresponding
adapted supercharge; we assume that it gives the same result. Together with the factor 2 of the two
planes this is the counting $\gamma=4$ of \ci{Beccaria:2023ujc}, cf. also \ci{Gautason:2023igo,Kurlyand:2026yke}.
 The
absolute normalization is inherited from these references, cf.\ the remark after \rf{n3}.
A possible $\Tmem^{-1}$ correction to a weight would belong to $f_{2,p}$ and cannot be absorbed into the
tension-independent leading normalization.
}

 Consequently,
\begin{equation}
 \ff_2=\sum_p w_p(k)f_{2,p}
          =0 \ .
 \label{n14}
\end{equation}
Let us once again  summarize the relation between the two treatments. 

Section \ref{sec5} and this section compute
the same integral, defined by the paired zero-mode prescription. In section \ref{sec5} the zero-mode
sector is handled by the nonlinear collective coordinates, the integral over which is defined by the
prescription of Appendix \ref{apps}, and the measure contribution appears as the Jacobian insertion
$\cJ$. 

 In this section 
 the zero modes are lifted by the $\mQ$-exact deformation, their integral is defined by  the paired
Gaussian \rf{m13}, and the measure contribution appears as the completion insertions
$\cJ_\cQ+\cJ_{\rm EOM}$. 
The lifting is equivalent, at the quadratic level, to the regularizations used in 
\ci{Gautason:2023igo,Kurlyand:2026yke}. What is new at order $\Tmem^{-1}$ is the completion in  \rf{n54} and
the resulting identity \rf{n55}. The large-$t$ Gaussian produces the factor \rf{m13} or
$\delta^{(12)}(a)\,\zeta^1\cdots\zeta^{12}$, locally in the zero-mode coordinates. This does not mean that
the density in \rf{n2},\rf{n3} had only its highest odd component non-vanishing, and no separate
calculation of the odd components of $f_2(\mathrm a;k)$ is needed.

We conclude that under the stated supersymmetric prescription, the reference-saddle
cancellation extends to the integrated first inverse-tension coefficient $\ff_2$ in \rf{n3}. 
This agrees with the absence of that term in \rf{04},  but does not  yet
  establish the absence  of higher-order  contributions, i.e.
the 1-loop exactness.

\section{Concluding remarks}\label{sec6}

We have studied the  first inverse-tension  correction to the partition function of a Euclidean
M2 brane wrapping $S^3/\mathbb Z_k\subset S^7/\mathbb Z_k$ and located  at a point  in
$\mathrm{AdS}_4$.
 The presence of   bosonic and 
fermionic zero modes requires  using  projected Green's functions,   accounting for a 
collective-coordinate Jacobian  and integrating over the 
collective  coordinate space. 

 For $k>2$, the  $\mathbb Z_k$ quotient produces
anisotropic derivative contractions in $\vev{S_4}$, so that the bosonic, mixed and
four-fermion sectors are not separately proportional to the equation-of-motion
combinations. Their geometric anisotropies cancel in the sum.
 The remaining expression is   nonzero:
the projected Green's  function EOM terms  retain finite zero-mode 
terms even after use of  the $\delta(0)=0$ regularization. Their  contribution is, however,  cancelled by the
 collective-coordinate Jacobian   computed in  a  supersymmetry-adapted  way 
 as shown in section~\ref{sec5}.

The preserved supersymmetry selects the separation of the paired
collective coordinates from the quantum fields. It does not determine
the Jacobian from the quadratic spectrum alone. The latter also
requires the nonlinear broken-symmetry generators, including the
transformations restoring the static and $\kappa$-symmetry  gauges. Within the
$\alpha$-family of  zero-mode complement choices   \rf{018}, the transcendental in $k$  terms cancel independently of
$\alpha$, whereas the rational remainder vanishes  for the two  supersymmetric 
choices $\alpha=\pm1$. The dependence of the local coefficient  $f_2(\mathrm a; k)$ on the zero-mode 
complement is still consistent with independence of the complete integral
when the measure and residual integration are transformed together.

Section~\ref{n1} relates the ${\rm a}=0$  reference-saddle  result $f_2(0;k)=0$ 
 to the one   found after the  integral
over the collective coordinates. The chosen preserved supercharge
has two zeros on the family of saddles, corresponding to the
complementary internal planes in \rf{n5} (their identification is
described  in Appendix~\ref{apf}). We assume that the paired-zero-mode
prescription extends to the interacting path integral and its
regulated measure as specified in \rf{n4}, with the local forms
\rf{n6}. 

Under this prescription the collective-coordinate integral reduces to the contributions of these two
configurations. Near each of them the deformation produces a Gaussian integral over the paired
collective coordinates, and the first inverse-tension correction to it comes from two sources: the
nonlinear dependence of the fermionic part of the deformation on the quantum fields, which gives the
measure insertion $\cJ_\cQ$ whose expectation value agrees with $\vev{\cJ}$ found in section \ref{sec5},
and the equation-of-motion term in the supersymmetry algebra, which gives the insertion $\cJ_{\rm EOM}$
with vanishing expectation value. 
In other words, section \ref{n1} is the computation one would do with ``regularized'' zero-mode
propagators and the interaction terms required by the supersymmetric completion of the regularization,
and its agreement with the collective-coordinate computation of section \ref{sec5} is the statement
that the two parametrizations of the zero-mode sector give the same $\Tmem^{-1}$ coefficient.

Appendix \ref{apf} contains these computations at the reference
configuration; the second configuration is related to it by the symmetry exchanging the two complementary
planes, under which the adapted complement $\a=1$ is mapped to $\a=-1$. Consequently, each local
$\rT_2^{-1}$ coefficient vanishes, and so does their sum in \rf{n14}.

The status of this result should be stated precisely: the reference-saddle coefficient and the local
measure matching are explicit computations with a specified complement and analytic prescription,
while the passage to the full collective-coordinate integral assumes that the paired zero-mode
prescription extends to the interacting theory as a $\mQ$-exact deformation with an invariant
regulated measure, \rf{n4}. Under this assumption, which is of the same type as in any localization
computation and which we have not derived from first principles, the integrated $\Tmem^{-1}$ correction
vanishes.

This  result does not establish or require a point-wise vanishing of
$f_2(\mathrm a;k)$ in an arbitrary quantum-field -- collective-coordinate
split, or  determination of the leading weights. It is a
statement about the integrated coefficient at order $\rT_2^{-1}$
in the common supersymmetric prescription. It agrees with the absence
of this correction in the grand-canonical ensemble  prediction \rf{04} and
thus provides a further test of the ensemble conjecture. It does not by itself 
establish all-order  1-loop exactness of the M2  brane  partition
function.
A direct
construction of the full supersymmetric deformation, including its
gauge completion and integration cycle at finite deformation
parameter would provide an independent justification of the
prescription  based on the assumption \rf{n4}. 

\section*{Acknowledgements}
We   thank  M. Beccaria, S. Kurlyand and J. van Muiden for  useful  discussions.
 This work was supported by  the 
  STFC Consolidated Grant    ST/X000575/1 
 and  STFC Consolidated Award UKRI4230: Abdus Salam Centre for Theoretical Physics at Imperial College London.
 Some of the computations of this paper were done using 
programs written with the help of  ChatGPT models   and checked  using Claude  models.


\appendix

\section{Green's functions and  anisotropy contributions \la{api}}

The cancellation of the $\hh_{33}^2$ terms 
between the three sectors    in section \ref{sec4}
was due to $\Delta_F=\Delta$, i.e. the fact that the 
anisotropy of the fermionic first-derivative term in  \rf{56}  equals the anisotropy of
the bosonic ``induced-metric'' term in \rf{55}, 
\begin{equation}
 \Delta=\Delta_F=2\,G_{x,33}+V^{-1} \ . 
 \label{b1}
\end{equation}
This relation    between  the bosonic and fermionic quantities 
appears to  be   a consequence of  underlying supersymmetry of the quadratic fluctuation spectrum.
Below we   discuss  the  origin of anisotropic contributions 
in derivatives of Green's functions and also explain the relation between $\Delta_F$ and $\Delta$ in \rf{b1}.

\subsubsection*{Bosonic  Green's functions}

For the $\AdS_4$ fluctuations $x^r$  the kinetic operator is the massless 
Laplacian $-\nabla^2$  on $\Sigma_3 = S^3/\ZZ_k$.
To discuss the corresponding Green's function  let us start first with  the Laplacian on $S^3$
  with eigenvalues $\ell(\ell+2)$. 
Functions on $S^3=SU(2)$ decompose into Wigner functions $D^j_{m\,m'}$ with
$j=\ha \ell$. The  index $m'$ corresponds to  the Hopf weight (charge)   $\lambda=2m'$  and
$\ree_3 D^j_{m\,m'}=\ii\lambda D^j_{m\,m'}$ (cf. \rf{3}). 
 On the round sphere
$\int_{S^3}|D^j_{m\,m'}|^2={2\pi^2\ov \ell+1}$, so a mode normalized on $S^3$
has $|u_{S^3}|^2={\ell+1\ov 2 \pi^2} |D^j_{m\,m'}|^2$. 

For a  $\ZZ_k$-invariant
mode $u$ one has  $\int_{\Sigma_3}|u|^2=k^{-1}\int_{S^3}|u|^2$, so normalized on $\Sigma_3$ 
 it is $u=\sqrt k\,u_{S^3}$ and
\begin{equation}
 |u|^2=V^{-1} ({\ell+1}) \,\big|D^j_{m\,m'}\big|^2,
 \qquad
 \sum_m\big|D^j_{m\,m'}\big|^2=1, \qquad \qquad 
 \sum_m|u|^2=V^{-1} ({\ell+1}),
 \qquad V=\tfrac{2\pi^2}{k}.
 \label{p1}
\end{equation}
The coincident  value of the propagator  for a real   scalar  
 is the sum over all modes of
$|u(x)|^2$ divided by the eigenvalue. On $\Sigma_3$ a scalar mode is labelled by
its level  $\ell=2j$, its Hopf weight $\lambda=2m'$ and the left index $m$.
The $\ZZ_k$ identification restricts the weight: for a field  $x^r$ with
trivial holonomy   to $\lambda\in k\ZZ$; for the  scalar 
$\yy^\a$  in \rf{6} to $\lambda\in k\ZZ+1$. 
At fixed $\lambda$ we have 
$\ell\ge|\lambda|$ with $\ell\equiv\lambda$ mod $2$. 

Since the  Laplacian on $S^3$  has  eigenvalues
$\ell(\ell+2)$ 
organizing the mode sum
by weight  $\l$  first and $\ell$  second, and using \rf{p1} for the sum over
$m$ we have 
\begin{equation}
 G_x=\sum_{\rm modes}^{\prime}\tfrac{|u|^2}{\ell(\ell+2)}
 =V^{-1} \sum_{n\in\ZZ}\ \sum_{\substack{\ell\ge|kn|\\ \ell\equiv kn\, \mod\,  2 }}^{\prime}
 \tfrac{\ell+1}{\ell(\ell+2)}
 =\sum_{n\in\ZZ}G^x_{kn}, \qquad \ \ \  G^x_\lambda\equiv V^{-1} \sum_{\substack{\ell\ge|\lambda|\\ \ell\equiv\lambda\, \mod\,  2}}^{\prime}
 \tfrac{\ell+1}{\ell(\ell+2)} .
 \label{p2}
\end{equation}
The prime  means omitting  zero eigenmodes before regularization.
The meaning of $n$ is the Fourier  momentum along the Hopf fibre  (on 
a mode  $e^{\ii n s}$ one has $\ree_3=k\partial_s\to\ii kn$, so that 
$\lambda=kn$).\foot{The inner sum in \rf{p2} 
 is  log UV  divergent  (see below). The outer
one is defined by an analytic continuation, and  the  log UV divergence   is absent in the final result. 
 Equivalently, the sum over $n$ can be Poisson-resummed into the
sum over the $k$ images of the point along the fibre  as in  \rf{63}  which in 3d is manifestly  free of log UV divergence.
}

The same organization applies to  other coincident   Green's   function values.  A fibre
derivative $\ree_3$ at an endpoint inserts $\ii\lambda=\ii k n $, so   we get 
\begin{align}
  G_{x}=&\sum_{n\in\ZZ}\,G^x_{kn}, \qquad  \qquad G_{x,3}=\sum_{n\in\ZZ} kn \,G^x_{kn}=0,  \qquad \qquad 
     G_{x,33}=\sum_{n\in\ZZ}(kn)^2\,G^x_{kn}, \la{pp3} \\
 G_y=&\sum_{n\in\ZZ}G^\yy_{kn+1},
 \qquad \qquad 
 G_{y,3}=\sum_{n\in\ZZ}(kn+1)\,G^\yy_{kn+1},
 \qquad \quad 
 G_{y,33}=\sum_{n\in\ZZ}(kn+1)^2\,G^\yy_{kn+1}.
 \label{p3}
\end{align}
Setting  $\ell=|\lambda|+2l$ with $l\ge0$ and $a=\ha |\lambda|$,  we have 
\begin{equation}
 G^x_\lambda
 =\tfrac1{4V} \sum_{l\ge0}\big(\tfrac1{a+l}+\tfrac1{a+1+l}\big)
 =\tfrac1{4V} \big[\rL-\psi(a)-\psi(a+1)\big],
 \label{q3}
\end{equation}
where $\rL=\log \Lambda^2 $ is the logarithmic UV  divergence  
that goes  away in \rf{p2}  due to $\sum_{n \in \ZZ} 1=0$.\foot{Let us note that the use of projected Green's  functions   before applying  regularization 
and  adopting  the $\delta(0)=0$ regularization for  coincident-point values of the EOM terms 
is  fully consistent  with the  prescription  based   on  sum over images    in \rf{63}.}

 For the internal  $y^m$ fluctuations (organized into $\yy^\a$ as in \rf{6}) 
  the   kinetic  operator in \rf{7}  is
$-\nabla^2-3$ with eigenvalues $\ell(\ell+2)-3$;  since
${(\ell+1)\ov (\ell-1)(\ell+3)}=\frac12[{1\ov \ell-1}+{1\ov \ell+3}]$,  we get   $G_y$ in \rf{p3}  with 
\begin{equation}
 G^\yy_\lambda=\tfrac1{4V}\big[\rL-\psi\big(a-\tfrac12\big)
 -\psi\big(a+\tfrac32\big)\big],\ \ \ \ 
 \qquad a=\tfrac{1}2 |\lambda| \ . 
 \label{aaa4}
\end{equation}
The allowed weights  for $\yy$ are  $\lambda=kn+1$,
 $n\in\ZZ$.
 In the $x$ tower, removal of the $\ell=0$ mode  leaves  
 $G^x_0=\tfrac1{4V}\big[\rL-\psi(1)-\psi(2)\big].$
For  the internal tower, removal of the $\ell=1$ modes  leaves 
$
 G^\yy_1=\tfrac1{4V}\big[\rL-\psi(1)-\psi(3)\big].
$
For $k>2$ these are the only exceptional levels:
$|kn|\ge k$ for $n\ne0$, while $|kn+1|\ge k-1\ge2$ in the internal
tower. 

Using  the expression for $G_{x,33}$ in \rf{pp3} with $G^x_\lambda$ in \rf{q3}
from \rf{44} we get 
\begin{equation}
 \hh_{33}=G_{x,33}+G_{y,33}-G_y .
 \label{b3}
\end{equation}
$ G^\yy_\lambda$   in \rf{aaa4}   may be written as  
\begin{equation}
 G^\yy_\lambda=G^x_{\lambda-1}-\tfrac1{2}V^{-1} \,(\lambda+1)^{-1}  + \tfrac{1}{8} V^{-1}  \delta_{\l,1} \ . 
 \label{b4}
\end{equation}
Inserting \rf{b4} into the sums \rf{p3}  and using that 
 \begin{equation}
 \sum_{n\in\ZZ}1=0,
 \qquad
 \sum_{n\in\ZZ}kn=0,
 \qquad
 \sum_{n\in\ZZ}\te \frac1{kn+2}=\frac\pi k\cot\frac{2\pi}k ,
 \la{cb3}
\end{equation}
one finds 
\al{  
& G_y
 =\te G_x-\frac{\pi}{2kV}\cot\frac{2\pi}k+\frac1{8V} , \qquad \qquad  G_{y,3}=G_x+\frac{\pi}{2kV}\cot\frac{2\pi}k+\frac1{8V} ,\\
& \te  G_{y,3}=2G_x-G_y+\tfrac1{4V} , \qquad  \qquad \qquad G_{y,33}=G_{x,33}+G_x-\frac{\pi}{2kV}\cot\frac{2\pi}k+\frac1{8V}\ . 
 \la{cb6}
}
Thus 
\al{
& G_{y,33}=G_{x,33}+G_y \ , \qquad \ \ \  \ \  \hh_{33}=2G_{x,33}\ . 
 \label{b5}
}
Since according to \rf{79} one has 
$\tfrac13(E_x+E_y)=\tfrac13(-V^{-1}-2V^{-1})=-V^{-1}$, we get for $\Delta$ defined in \rf{55}
\begin{equation}
 \Delta=\hh_{33}-\tfrac13\big(E_x+E_y\big)=2\,G_{x,33}+V^{-1}.
 \label{b6}
\end{equation}
The above relations  can be checked for specific values of $k$. For example, for $k=4$ we find 
\be 
k=4: \ \ \quad  \ G_x= -\tfrac{1}{\pi^2} , \qquad G_y=  -\tfrac{3}{4\pi^2} , \qquad G_{x,33} =  -\tfrac{4}{3\pi^2} , \qquad 
V= \ha \pi^2 \ , \qquad \Delta=  -\tfrac{2}{3\pi^2} \ . \la{a133}  \ee

\subsubsection*{Fermion spectrum and anisotropy}  

The fermionic anisotropy   defined in \rf{56} 
is $\Delta_F=c_F-\tfrac13r$ with $c_F=p_C+p_N$.
Here    $p_C$ and $p_N$  are the
$\PC$ and $\PN$ components of $\rho_3G_{\theta,3}$ in \rf{5222}.
For the orthogonal complement the  inverse of the fermionic  kinetic operator  is
$G_{\theta,\perp}=i\mathrm D(\mathrm D^2)^{-1}$, where the inverse  omits the
zero eigenvalues. Its charged-sector  entries reduce to uncharged scalar levels
at weights $kn$ and $kn\pm2$, and its neutral-sector entries to  the  charged internal scalar levels
at $kn\pm1$. The latter are related to the uncharged scalar resolvents
by \eqref{b4}. 

In more details, 
after the $\ka$-symmetry  gauge the spinor $\theta$ has $16=2\times8$ components. They can be labelled by 
a 2-dimensional   spinor space index, on which the $\rho_a$ act as Pauli
matrices 
and an
8-dimensional   internal index. The internal index splits under
$\Gamma_*$ into the charged block $\PC$ and the uncharged block $\PN$, four
states each (see \rf{9}).
 On $\PC$ the generator $\mathcal T_C =\tfrac\ii2(\Gamma_{45}+\Gamma_{67})$ in \rf{1011}
has eigenvalue $\pm1$; on $\PN$ it vanishes. This eigenvalue is the
internal Hopf charge
$  \HJ\in\{+1,-1,0\},
 $
indicating  how the internal index transforms under the deck rotation of the embedding coordinates 
$Y_m\to e^{2\pi\ii/k}Y_m$, which acts on the normal directions $Y_3,Y_4$ as
well as on the base.


A component of $\theta$ is a Wigner function $D^j_{m\,m'}$ times a tangent
spinor. Its total weight under the  Hopf symmetry action has two parts:
$2m'$ from the orbital action $-\ii \ree_3 $ on the Wigner function,
and $\sigma=\pm1$, the $\rho_3$ eigenvalue, from the tangent spin lift,\foot{The frame $e^1\mp\ii e^2$ rotates along the fibre
and the two tangent-spinor components carry weight $\pm1$.} i.e.
$
 \lambda=2m'+\sigma .
$
Invariance under the deck group requires the total weight  
to vanish mod $k$,  i.e. 
\begin{equation}
 \lambda=kn+\HJ,\qquad n\in\ZZ \ , 
 \label{pw3}
\end{equation}
where $\HJ$ is the internal Hopf  charge. 
 Two states  denoted as  $(\uparrow,\downarrow)$  can have this weight:
$
 (m',\sigma)=(\tfrac{\lambda-1}2,\,+1)
 $ and $
 (m',\sigma)=(\tfrac{\lambda+1}2,\,-1)$
subject to $|m'|\le j$.\foot{Setting  $d=2j+1$,   both states exist when $|\lambda|\le d-2$: this is the two-component
      ``coupled angular space''. 
 Only one exists when $|\lambda|=d$, the edge $m'=\pm j$,  with $\sigma=\operatorname{sgn}\lambda$.
The case of  $|\lambda|=d-1$ never occurs, since $\lambda\equiv d$ mod $2$.}

Let $\rJ_a$ be the generators of the right action of $SU(2)$ on functions on
$S^3=SU(2)$, 
$
 \rJ_a=\tfrac1{2\ii}\,\ree_a,
 \ 
 [\rJ_a,\rJ_b]=\ii\epsilon_{abc}\rJ_c,
\ 
 \rJ_3=\tfrac1{2\ii}\,\ree_3=\tfrac{k}{2\ii}\,\partial_s .
 $
The  above two states are coupled because the  Dirac operator contains (see below) 
\begin{equation}
 \rho\cdot\rJ=\rho_3\rJ_3+\tfrac12\big(\rho_+\rJ_-+\rho_-\rJ_+\big),
 \label{p5}
\end{equation}
and $\rJ_\pm$ shift $m'$ by one while $\rho_\pm$ shift $\sigma$ by two,
preserving $\lambda$.
On a Wigner function $D^j_{m\,m'}$ the right generators act on the second
index:
$
 \rJ^2\to j(j+1),
$    $
 \rJ_3\to m',
 $  and $
 \rJ_\pm $ shifts $ m'\to m'\pm1 $ with 
 coefficient  $ \sqrt{j(j+1)-m'(m'\pm1)} .
$
The left index $m$  supplies the $(2j+1)$-fold degeneracy at
fixed $(j,m')$. 
The scalar Laplacian 
$-\nabla^2=
 - \ree_a \ree_a =4\rJ^2$ has 
 eigenvalues  $ 4j(j+1)=\ell(\ell+2),\  \ell=2j .
 $
 
The Dirac operator on the unit  sphere   in the left-invariant frame is  (cf. \rf{10}) 
\be \rho^a \ree_a+\tfrac{3\ii}2=2\ii\,\rho\cdot\rJ+\tfrac{3\ii}2 \ , \la{pp88}\ee
so   that the Hermitian 
 $\DD=\ii\Dsl$ operator in  \rf{10} is   
 $\DD=-2\,\rho\cdot\rJ-3\PN $  where $\PN$  gives  just a constant  shift. 
  The eigenvalues $j$ and $-(j+1)$ of $\rho\cdot\rJ$ are the standard ones for
coupling spin $\frac12$ to spin $j$: total spin $j+\tfrac12$ gives
$\rho\cdot\rJ=j$, total spin $j-\tfrac12$ gives $-(j+1)$.

The orbital Hopf weight is determined by $\ree_3=2\ii\rJ_3$, so acting on an eigenfunction 
$\ree_3\to2\ii m'$ and the orbital part of the weight is
$
 \lambda_{\rm orbital}=2m' ,
$ which is where the $2m'$ in $\lambda=2m'+\sigma$ comes from.  

In the basis $(\uparrow,\downarrow)$ of the above  two states 
$\rho_3\rJ_3$ is diagonal with entries $\tfrac{\lambda-1}2$ and
$-\tfrac{\lambda+1}2$. The off-diagonal element of $\tfrac12\rho_+\rJ_-$
is the product of  $\langle\uparrow|\rho_+|\downarrow\rangle=2$
and the angular-momentum element
$\langle m'|\rJ_-|m'+1\rangle=\sqrt{j(j+1)-m'(m'+1)}$ at
$m'=\tfrac{\lambda-1}2$, where
\be 
 j(j+1)-m'(m'+1)=\tfrac{1}4\tau^2
 , \qquad \ \ 
 \tau\equiv\sqrt{d^2-\lambda^2},\qquad \ \  d\equiv 2j + 1 \ , \la{pp99}
 \ee
so that 
$
 \rho\cdot\rJ=-\tfrac12+\tfrac\lambda2\rho_3+\tfrac\tau2\rho_1 .
 $
The  Dirac operator  then becomes
\begin{equation}
\DD=-2\rho\cdot\rJ-3\PN \ \ \to \ \  H_{\lambda,j}=\mu_\chi-\lambda\rho_3-\tau\rho_1,
 \qquad
 \mu_\chi=1-3\chi,
 \qquad
 \chi=\begin{cases}0&\text{on }\PC,\\1&\text{on }\PN,\end{cases}\ . 
 \label{aa8}
\end{equation}
 $\chi$ is  the eigenvalue of $\PN$, and $\mu_\chi$ is   1  on the charged block, where
there is no flux term, and $-2$ on the uncharged block, where the
$-3\PN$ flux term  contributes.  One finds that 
\be \la{aa88}
 H_{\lambda,j}^{-1}=\frac{\mu_\chi+\lambda\rho_3+\tau\rho_1}{\mu_\chi^2-d^2},
 \ee
 and the eigenvalues are $\mu_\chi\mp d$:
\begin{equation}
 \PC:\ 1-d=-2j,\quad 1+d=2j+2;
 \qquad
 \PN:\ -2-d=-2j-3,\quad -2+d=2j-1 .
 \label{p10}
\end{equation}
Hence the kernels are  at $j=0$ on the charged block and $j=\tfrac12$ on the
uncharged one, and the other state of the $j=\tfrac12$ uncharged block has
eigenvalue $-4$.
At $|\lambda|=d$ there is no partner, $\rho\cdot\rJ=\rho_3\rJ_3= j$, and
$
 \DD\to \mu_\chi-|\lambda|,
 $ with 
 {multiplicity } $d=|\lambda|$.
Note that 
on the same states,
$
 \ii\rho_3 D_3\to (1-\chi)-\lambda\rho_3.
 $

Using this data one   may analyse  the anisotropy contributions in the fermionic sector. 
After combining the levels $n$ and $-n$, the $P$-parts of the fibre
derivative are  (up to  factors of $\tfrac1{2V}$) 
\begin{equation}
 (p_C)_n=4a_n^2\Psi_n-2,
 \qquad
 (p_N)_n =4a_n^2\Psi_n,
 \qquad
 a_n=\tfrac{1}2kn \ ,
 \qquad
 \Psi_n\equiv \rL -\psi(a_n)-\psi(a_n+1),
 \label{b7}
\end{equation}
where $\Psi_n$ is  the combination that defines $G^x_{\l=kn}$ in
\rf{q3}, so that
$
 \sum_{n>0}a_n^2\Psi_n=\tfrac 12V\,G_{x,33}.
 $
Therefore
\begin{equation}
 p_C=G_{x,33}+\tfrac1{2}{V}^{-1},
 \qquad
 p_N=G_{x,33},
 \qquad
 c_F=2\,G_{x,33}+\tfrac1{2}{V}^{-1}.
 \label{b9}
\end{equation}
Here $\frac1{2}V^{-1} $ in $p_C$ is coming from the constant $-2$ in \rf{b7}
after we use  that   $\sum_{n>0}1=\zeta_R(0)=-\ha $.  
With $r=r_C+r_N=-\tfrac3{2}{V}^{-1}$ we then get 
\begin{equation}
 \Delta_F=c_F-\tfrac13r=2\,G_{x,33}+V^{-1},
 \label{b10}
\end{equation}
which  is the same as $\Delta$ in  \rf{b6}. This  can be  checked also  for particular   values of $k$ like $k=4$ in \rf{a133}.


\subsubsection*{Fermionic coefficients in \rf{88}--\rf{90}}

The per-level inverse \eqref{aa88} determines the coefficients
$s_C,s_N,v_C,v_N$, together with $p_C,p_N,q_C,q_N$,
in \eqref{522}--\eqref{5222}.
With $G_\theta=\ii H^{-1}$ 
the two-component block at fixed $(\lambda,j)$ contributes
\begin{equation}
 \big[G_\theta\big]_{\lambda,j}=\frac{\ii d}{V}\,
 \frac{\mu_\chi+\lambda\rho_3}{\mu_\chi^2-d^2},
 \qquad
 \big[\rho_3G_{\theta,3}\big]_{\lambda,j}=\frac d{V}\,
 \frac{\big[(1-\chi)\mu_\chi-\lambda^2\big]
 +\lambda\big[(1-\chi)-\mu_\chi\big]\rho_3}{\mu_\chi^2-d^2} \ .
 \la{bc8}
\end{equation}
There is no $\rho_1$ term: the two components in
\eqref{aa88} have
different $m'$ and $\sum_mD^j_{m\,m'}\overline{D^j_{m\,m''}}=\delta_{m'm''}$.
The one-component  states at $|\lambda|=d$ contribute to
$\rho_3G_{\theta,3}$ the term $|\lambda|  V^{-1} P_\sigma$ on the charged block,
with $P_\sigma=\ha[1+{\rm sgn}(\lambda)\rho_3]$. 
Summing \eqref{bc8} over $j$ at fixed weight, end states included, gives for
the charged $\upkappa=+1$ tower  (we set  $g_n\equiv G^x_{kn}$ and
$h_n\equiv G^x_{kn+2}$)
\begin{align}
 s_{C,n}&=-\ha\big[(kn+2)g_n-kn\,h_n\big],&
 v_{C,n}&=-\ha\big[(kn+2)g_n+kn\,h_n\big],\notag\\
 p_{C,n}&=\ha kn(kn+2)\big(g_n+h_n\big),&
 q_{C,n}&=\ha kn(kn+2)\big(g_n-h_n\big),
 \la{bc9}
\end{align}
and for the uncharged tower with $\lambda=kn$
(here $w_n^{\pm}\equiv G^\yy_{\pm kn+1}$)
\begin{align}
 s^{(0)}_{N,n}&=\ha\big[(2-\lambda)w_n^-+(2+\lambda)w_n^+\big]
 -\tfrac1{4V}\,\delta_{n0},&
 v^{(0)}_{N,n}&=\ha\big[(2-\lambda)w_n^--(2+\lambda)w_n^+\big],\notag\\
 p_{N,n}&=\ha\big[\lambda(\lambda-2)w_n^-+\lambda(\lambda+2)w_n^+\big],&
 q_{N,n}&=\ha\big[\lambda(\lambda-2)w_n^--\lambda(\lambda+2)w_n^+\big].
 \la{bc10}
\end{align}
The $\HJ=-1$ tower supplies the conjugate internal-charge block.
After both blocks are assembled, the coefficients multiply the full
$P_C,T_C$ matrices used in \eqref{522}.

With $a_n=\ha kn$, \  $\Psi_n$ as in \rf{b7} and $u_n\equiv(a_n^2-1)^{-1}$,
the  functions \rf{q3}, \rf{aaa4} give
(using that $\psi(x+1)=\psi(x)+ x^{-1} $)
\begin{align}
 g_{\pm n}&\te =\tfrac{1}{4} V^{-1} \Psi_n ,&
 h_n&=\tfrac{1}{4} V^{-1} (\Psi_n-a_n^{-1}-(a_n+1)^{-1}),&
 h_{-n}&=\tfrac{1}{4} V^{-1} [ {\Psi_n+a_n^{-1}+(a_n-1)^{-1}}], \notag\\
 &&
 w_n^+&=\tfrac{1}{4} V^{-1} [{\Psi_n-(a_n+1)^{-1}}],&
 w_n^-&=\tfrac{1}{4} V^{-1} [{\Psi_n+(a_n-1)^{-1}}].
 \la{bc11}
\end{align}
Substituting \eqref{bc11} into \eqref{bc9}--\eqref{bc10} and adding the levels
$n$ and $-n$, we get  linear combinations of $\Psi_n$,
$a_n^2\Psi_n$, $u_n$ and $1$, i.e. (in units of $\tfrac1{2V}$)
\begin{center}
\renewcommand{\arraystretch}{1.3}
\begin{tabular}{@{}lcccc@{}}\toprule
 & $s_n$ & $v_n$ & $p_n$ & $q_n$\\\midrule
charged   & $-\Psi_n-u_n-2$ & $-\Psi_n+u_n+2$ & $4a_n^2\Psi_n-2$ & $2$\\
uncharged & $2\Psi_n-2$ & $0$ & $4a_n^2\Psi_n$ & $0$\\\bottomrule
\end{tabular}
\end{center}
At $n=0$  we get  $g_0=h_0=\tfrac{1}{4} V^{-1} (\rL'-1)$ and
$w_0=\tfrac{1}{4} V^{-1}(\rL'-\tfrac32)$ with $\rL'\equiv \rL+2\gamma_E$, so that
\begin{equation}
 s_{C,0}=v_{C,0}=-g_0,
 \qquad
 s^{(0)}_{N,0}=2w_0-\tfrac1{4V},
 \qquad
 p_{C,0}=q_{C,0}=p_{N,0}=q_{N,0}=0 .
 \la{bc12}
\end{equation}
Doing  the sums using 
\begin{align}
 \sum_{n>0}\Psi_n&=2V\big(G_x-g_0\big),\qquad 
 \sum_{n>0}a_n^2\Psi_n =\tfrac 12V \,G_{x,33},\qquad 
 \sum_{n>0}u_n=\ha-\te \frac\pi k\cot\frac{2\pi}k ,
 \la{bc13}
\end{align}
we get 
\begin{align}&
\te  s_C=-G_x+\frac1{4V}+\frac{\pi}{2kV}\cot\frac{2\pi}k,\qquad 
 v_C=-G_x-\frac1{4V}-\frac{\pi}{2kV}\cot\frac{2\pi}k,\qquad 
 p_C=G_{x,33}+\frac1{2V}, \la{bc14}
\\  &
\te  s^{(0)}_N=2G_x,
 \qquad
 v^{(0)}_N=0,
 \qquad
 p_N=G_{x,33},\qquad 
 q_C=-\frac1{2V},
 \qquad
 q_N=0 , \\
  & s_C=-G_y+\tfrac38V^{-1},
 \qquad\qquad 
 v_C=-2G_x+G_y-\tfrac38V^{-1}.
 \la{bc16}\end{align}
The neutral values $s^{(0)}_N=2G_x$, $v^{(0)}_N=0$
correspond to  the orthogonal  choice of the zero-mode complement.  The  expressions in the case of the 
supersymmetry-adapted complement, i.e.
 $s_N=2G_x-\tfrac14V^{-1}$,
$v_N=\tfrac12V^{-1}$, $d_N=-V^{-1}$, is derived in \rf{f1},\rf{f2}.

\section{Zero modes\la{apa}} 

Zero modes in the spectrum of quadratic operators on $S^3/\ZZ_k$ 
 appear as the origin of finite projector  terms
in the Green's function EOM terms   and of  
the  collective-coordinate  Jacobian     so we review them below
  (see also \ci{Gautason:2023igo,Beccaria:2023ujc}).

Every zero mode of the quadratic operator is a  linearization of a symmetry
of the background  which is broken by the  classical solution. 
The membrane surface is localized at a point in $\AdS_4$ 
and wraps $\Sigma_3=S^3/\ZZ_k\subset S^7/\ZZ_k$. Translations in $\AdS_4$ give four real
moduli; rotations of the  complex 2-plane inside $\CC^4$ that commute with
the $\ZZ_k$ action give the rest. For $k>2$ the invariant rotations  form 
 the complex Grassmannian  $\Gr_\CC(2,4)$  of 2-planes through the 
 origin of $\CC^4$  of real dimension eight. The  bosonic moduli space has  therefore   dimension 
$4+8=12$. 

The   $\AdS_4 \times S^7/\ZZ_k $ supercoset  admits   24  supersymmetries, of which the  classical solution 
preserves 12 . The 12  broken ones generate the fermionic zero modes:
4 in the charged spinor  sector  and 8 in the uncharged one. Together
 this leads to   $12$  bosonic plus  12  fermionic   zero modes. 

The same  count  can be found   also  from the  fluctuation spectra.
  For the  $\AdS_4$ scalars,
$-\nabla^2$  has one constant  mode per real component, i.e. 4 modes in total. 
For  the internal scalars $\yy^\a$, $-\nabla^2-3$ at  weight $\lambda=kn+1$, vanishes at
$\ell=1$, $\lambda=1$, with multiplicity  2  per complex field, total of  8 modes.\foot{For $\yy^\a$ operator 
there is no  constant
harmonic for $k>2$, hence no negative mode from the $-3$ term in the fluctuation operator. }
Explicitly, 
in the coordinates $(z,\bar z,s)$ in \rf{3}  
the $\yy^\a$ operator  zero modes  are 
 $\yy^\alpha=\frac{b^\alpha_1+b^\alpha_2\,z}{\sqrt{1+|z|^2}}$
 where $b_1, b_2$   are complex constants, representing  8 
      real parameters.

 For
the  fermions,  the  Dirac operator  $\DD=-2\rho\cdot\rJ-3\PN$  in \rf{aa8}
 has eigenvalues $\{-2j$,
$2(j+1)\}$ on the charged block and $\{-2j-3$, $2j-1\}$ on the uncharged one, so
the kernels sit at $j=0$ (four real modes) and $j=\tfrac12$ (eight). The
partner of the uncharged kernel in its $2\times2$ block has eigenvalue $-4$
and is  not a zero mode.

      In the   charged spinor sector    we get    
constants in the global frame, spanning
      $\tfrac12(\PC+ T_C)$.\foot{The global frame is the left-invariant frame $\ree_a$ of \rf{3} and the spinor
frame that goes with it; ``constants'' means spinors whose 16 components are
constant in that frame -- only those
in the image of $\tfrac12(\PC+ T_C)$.} This projector  selects the
charged block {and} the alignment $\rho_3{\cal T}_C=+1$, i.e.\ tangent-spin
eigenvalue $\sigma$ equal to the internal charge $\HJ$. That is exactly the $j=0$
solution of the weight condition $\lambda=2m'+\sigma=kn+\HJ$: with $m'=0$ it
forces $\sigma=\HJ$ and $n=0$, so the mode is constant along the fibre.  This gives 4 modes.

        In the 
uncharged spinor sector  we get  the zero modes as 
\be \la{pp5}
\aZ_{ht}(u)=U(u)^{-1} \ve_t\otimes n_h \ , \qquad D_a \aZ_{ht}=0 \ , \qquad h=1,\dots,4, \qquad 
      t=1,2  \ , \ee
          which is   annihilated by $D_a$ in \rf{11}. 
      Their block partners are
      \be \nu_{ht}=\rho_3\aZ _{ht}\ , \qquad \ \ \   \DD  \nu_{ht}=-4\nu_{ht} \la{pp6}  \ .\ee
Here $u\in S^3$, $U(u)\in SU(2)$ is the group element 
built from  complex  coordinates $(Y_1,Y_2)$.\foot{$
 U(u)=\begin{pmatrix}Y_1&-\bar Y_2\\Y_2&\bar Y_1\end{pmatrix}\in SU(2)$, where 
 $ |Y_1|^2+|Y_2|^2=1$  so that 
 $\ree_a  U=U\ii\sigma_a$.}
 $\ve_t$  is a basis of the two-dimensional
tangent-spinor space on which the $\rho_a$ act, and $n_h$  is a
basis of the uncharged internal four-space, $\PN n_h=n_h$. $U(u)^{-1}$ acts on
the tangent-spinor index, so the components of $\aZ_{ht}$ are the $j=\tfrac12$
Wigner functions.  This  gives $8=2\times4$  spinor zero modes.

For the Green's function defined using the orthogonal  zero-mode complement 
 the EOM relation  is
$\mathcal O G_\perp=1-P_{0,\perp}$. For a  general   fermionic
complement one must instead use both endpoint equations in \eqref{942}.
Its coincident density \eqref{946} is Hermitian, so
$[P_{0,F}^{\dagger}]=[P_{0,F}]\equiv P_{0,F}(\s,\s) = V^{-1}\mathcal P$ at $\alpha=1$.
 As a  result,   for each  real  field component,  we have for the  EOM   terms in \rf{42},\rf{52},\rf{123}
\begin{equation}
 E_x  =-V^{-1} \ , 
 \qquad
 E_y=-2{V}^{-1},
 \qquad
 E_\theta=- {V}^{-1}\big[\tfrac12\big(\PC+T_C\big)+\PN+T_N\big] \ ,  
 \label{pp33}
\end{equation}
i.e.   in \rf{52} one has 
$r_C=d_C=-{1\ov 2} V^{-1} $, \  $r_N=d_N=- V^{-1}$.
Also, the charged component $q_C$ of $\rho_3G_{\theta,3}$ in \rf{5222}  receives
$V^{-1} $ from every level pair $(n,-n)$ with $n\ne 0$,  so that 
\begin{equation}
 q_C=V^{-1} \sum^\infty_{n=1}1=V^{-1} {\zeta_R(0)}=-\ha V^{-1} \ . 
 \label{pp4}
\end{equation}
In the special  cases of  $k=1$ and $k=2$  the  counting changes. 
The   functions $\hat  \yy^\a= e^{-2\ii s/k}(\bar b^\alpha_1+\bar b^\alpha_2\bar z)/\sqrt{1+|z|^2}$ 
 are formal  zero-modes of the operator $-\nabla^2 -3$ in \rf{7}  for any $k$  but   they are  $2\pi$-periodic in
$s$ only for $k=1,2$. This is why   there are 16 internal zero modes at
$k\le2$ and only eight for $k>2$.  In the $k\leq 2$  case in \rf{pp33} we have 
 $E_y=-4V^{-1}$. For 
$k=1$ the $\ell=0$  constant mode  of $\yy^\a$ operator 
 gives 
four negative modes,  reflecting the instability of the great three-sphere inside $S^7$.  For $k=2$
the weights are odd, the $\ell=0$ mode  is absent, and there are no 
negative modes.

\section{Collective-coordinate  Jacobian\la{apco}}

Here we  shall 
 review  the change of variables in a functional integral when the quadratic fluctuation operator has zero modes
 considering  for simplicity 
 the   bosonic theory case  (see, e.g.,    \cite{Bhattacharya:2024chz}  and references there).  
 The  case with  both  bosons and fermions  is discussed in section \ref{sec5} and Appendix \ref{apd}. 
 
Let us start with   a Euclidean bosonic functional integral
$
 Z=\int [D\phi]\,e^{-S[\phi]} .
$
Under an invertible change of variables
$
 \phi=\phi[\chi]
 $
the measure changes as
\begin{equation}
 [D\phi]=[D\chi]\,J_F[\chi],
 \qquad\quad 
 J_F[\chi]=\det\!\big(\frac{\delta \phi}{\delta\chi}\big).
 \la{Jfield}
\end{equation}
For a  perturbative redefinition
$
 \phi[\chi]=\chi+\epsilon\,\varphi[\chi],
 $
with $\varphi$ depending locally on $\chi$ and finitely many of its derivatives, one may exponentiate the determinant 
 using  auxiliary ghost fields.  The ghost propagator is ultralocal, while the interaction vertices are polynomials in momenta.  In dimensional regularization  (or an analytic regularization with $\delta(0)=0$) 
 the resulting closed ghost loops  vanish, i.e. 
 we get 
$
 J_F=1.
 $\foot{Note that  in general a  field redefinition that removes an equation-of-motion term  at order $\epsilon$ also generates 
 terms at order $\epsilon^2$ and beyond.  These induced terms should be, in general, also  accounted for:
   ``proportional to the equations of motion'' and ``irrelevant in the  path integral''  may not be  equivalent statements beyond the leading  order  (see,  e.g.,  \cite{Criado:2018sdb}).}


Suppose now that a  classical  solution  $\phi\ma(x) $  is a family 
parametrized by   constants   $a^A$ ($A=1, ...,\rq $).   The tangent vectors
\begin{equation}
 \cZ_A(x;a)=\frac{\partial \phi\ma (x)}{\partial a^A}\ , 
 \la{cc22}
\end{equation}
are zero modes of the second variation of the action 
 when this  family is generated by an exact symmetry of the action,
\begin{equation}
 \OO\ma \cZ_A=0,
 \qquad\qquad 
 \OO\ma =\frac{\delta^2 S}{\delta\phi\,\delta\phi}\Big|_{\phi\ma}.
 \la{zeroeq}
\end{equation}
A naive expansion
\begin{equation}
 \phi \ \to  \ \phi+c^A \cZ_A+\eta \ , \
 \la{naiveexp}
\end{equation}
then leads  to  unsuppressed  path integral in the $c^A$ directions   corresponding to 
zero eigenvalues of  $ \OO\ma $.
The standard cure is  to replace $c^A$   by the collective coordinates 
$a^A$  and  to constrain  the remaining fluctuation $\eta$ 
 to lie in a complement to the zero-mode directions.  This can be done  by imposing a    constraint  $C_A=0$ 
   that  removes from $\eta\ma $ the components tangent to the orbit. 
  
  For example,  in the case of a constraint linear in  fluctuation fields we have 
\begin{equation}
 \phi=\phi\ma +\eta, \ \ \ \ 
 \qquad
 C_A(a;\phi)=0, \qquad \qquad  C_A(a;\phi)=\big( \cZ_A(a),\eta \big), 
 \la{cc55}
\end{equation}
where  $(\,  , \, )$ 
is  an inner product on the  field space  and $\cZ_A$ was defined in \rf{cc22}.
To do this   systematically in the path integral  one inserts 
\al{
& 1=\frac{1}{\N_C[\phi]}
 \int d^\rq a\,
 \delta^{(\rq)}\!\big(C(a;\phi)\big)
 \big|\det \M(a;\phi)\big|,
 \qquad \qquad \ \ \   \M_{AB}(a;\phi)=\frac{\partial C_A(a;\phi)}{\partial a^B}\ . 
 \la{Mdef}
}
 $\N_C[\phi]$
 is the number of solutions of $C_A(a;\phi)=0$ for the given field configuration.\foot{In general, 
the factor $\N_C[\phi]$   is  important to account for \cite{Bhattacharya:2024chz}. 
  Its value is  related to an  assumption about  the range of the  collective-coordinate domain.
  For  semiclassical calculations one  has two equivalent choices: either choose a fundamental domain in collective-coordinate space in which each field configuration is represented once, or integrate over a larger domain and divide  by the intersection number. }
  
  Then the  path integral measure may be represented schematically as
\begin{equation}
 [D\phi]
 =d^\rq a\,[D\eta]_\perp\, (\N_C[\phi])^{-1} \, 
 {\CJ[a,\eta]}\ , 
 \qquad\qquad 
 \CJ=\big|\det \M\big| \ , 
 \la{cc8}
\end{equation}
where $[D\eta]_\perp= [D\eta]\,  \delta^{(\rq)}\!\big(C(a;\phi)\big)$. 

In  contrast 
 to  the local field-redefinition Jacobian in \rf{Jfield}  which  is a functional determinant of an operator acting independently at each spacetime point,     the logarithm  of the  collective-coordinate determinant   is 
  not proportional to $\delta(0)$.
Expanding  $\CJ$  generates  interaction vertices in the effective measure that may be important for higher-loop computations as we saw in section \ref{sec5}. 


\section{Instanton zero-mode separation  and  collective-coordinate Jacobian}
\label{apd}

Let us now   specialize to our M2 brane instanton case. 
The zero modes were discussed in Appendix~\ref{apa}. 
Here we define their dual
complements  and the corresponding projected  Green's  functions, and then determine
 the Jacobian in \eqref{933}. The requirement of preserved linear supersymmetry
selects the neutral fermionic complement. 
The broken target-space symmetries, including the transformations
restoring static and $\kappa$-gauge, determine the entries of the 
matrix  $\mathcal M$ in  \rf{928}.  

\subsubsection*{Zero-mode constraints and the projected inverse}

In the normalization of section~\ref{sec5}, the tangent-space split is
\begin{equation}
 \mathrm X=\mathrm a^A\mathcal Z_A+\epsilon\widetilde{\mathrm X},
 \qquad \mathrm a^A=\Lambda^A\mathrm X,
 \qquad \Lambda^A\widetilde{\mathrm X}=0,
 \qquad P_0=\mathcal Z_A\Lambda^A,\quad P_0^2=P_0.
 \label{941}
\end{equation}
Here $\epsilon=\mathrm T_2^{-1/2}$ and
$\Lambda^A\mathcal Z_B=\delta^A{}_B$. The coefficients in this linear
split are tangent  collective coordinates; the nonlinear collective-coordinate map
is \eqref{928}.

For a Hermitian kinetic operator $\mathcal O$, let
$G_\perp$ be its inverse on the orthogonal complement of its
kernel. 
In  general,  dual $\Lambda^A$  need not be orthogonal.\foot{ Two different  choices of  a dual  basis   $\Lambda^A$ obeying
 $\Lambda^A\mathcal Z_B=\delta^A{}_B$  differ by a functional that vanishes on all zero modes, and
correspondingly define different quantum subspaces.  One  natural  choice  is the 
orthogonal  one  when  $\Lambda^A$      is  defined as  
$
 \Lambda^A (...)=\frac{\int d^3\s\sqrt g\;\aZ_A^\dagger\, (...)}
                     {\int d^3\s\sqrt g\;\aZ_A^\dagger \aZ_A}.
 $
This  applies  whenever the kinetic operator
is Hermitian and its zero modes are its own eigenvectors, with every other
eigenvector orthogonal to them. In that case the orthogonal complement of
the kernel is  the space spanned by the non-zero eigenmodes, and
inverting the operator on that space is unambiguous: the constrained
propagator  $G'$ is simply the spectral sum with the zero eigenvalues omitted.
This is the case  for the bosons $x^r$ and $y^m$, whose operators $-\nabla^2$ and
$-\nabla^2-3$ are Hermitian with isolated zero eigenvectors. The same is true   also  for the
charged spinor block, whose zero modes are the constant spinors which are  eigenvectors
of $\DD$  in \rf{10} with eigenvalue zero.
The orthogonal spectral complement is well defined. Below  we  choose a
non-orthogonal complement adapted to the preserved supersymmetry, and  thus  modify both the constrained propagator and the collective-coordinate
measure accordingly.
   }
We have \begin{equation}
\begin{aligned}
 G_\Lambda&=(1-P_0)G_\perp(1-P_0)^\dagger,
 \qquad \Lambda G_\Lambda=0,\qquad 
 G_\Lambda\mathcal O&=1-P_0,
 \qquad \mathcal OG_\Lambda&=1-P_0^\dagger.
\end{aligned}
\label{942}
\end{equation}
For the  fermions one may apply \eqref{942} to
$\mathcal O=\mathrm D=i\slashed D$ in \rf{10}, with 
$G_\theta=iG_\Lambda$ and the corresponding Majorana
conjugation at the second endpoint.

The bosonic duals $\Lambda^A$   are  chosen to be 
orthogonal. 
The four AdS zero modes are constant. To describe the eight internal
ones of $y^m$ kinetic operator, let us use  $\mathbf u\in S^3\subset\mathbb R^4$ of the
covering-space embedding $(Y_1,Y_2)$. Choose eight real complex-linear
matrices $\Omega_I$ with
$\operatorname{tr}(\Omega_I^T\Omega_J)=2\delta_{IJ}$ and set 
$\mathcal Z_I=\Omega_I\mathbf u$. In this covering-space normal frame,
\begin{equation}
\begin{aligned}
&  \Lambda_r x=V^{-1} \int d^3\s\, \sqrt g\,x^r,\qquad 
 \Lambda_I y=2V^{-1}\int d^3\s\, \sqrt g\,\mathcal Z_I\cdot y,\qquad 
 \int d^3\s\, \sqrt g\,\mathcal Z_I\cdot\mathcal Z_J
       =\tfrac12 V \delta_{IJ}.
\end{aligned}
\label{943}
\end{equation}
Writing $t_{Ia}=\nabla_a\mathcal Z_I$, the completeness relations
needed below are
\begin{equation}
 \sum_I\mathcal Z_I^m\mathcal Z_I^n=\delta^{mn},\qquad\qquad 
 \sum_I\mathcal Z_I^m t_{Ia}^n=-(J_y)^{mn}\delta_{a3}.
 \label{944}
\end{equation}
For the four charged fermionic zero modes, the duals are the
orthogonal integrals against constant spinors in
$\tfrac12(P_C+T_C)$.
For the eight neutral modes
$\mathcal Z_A$ of \eqref{pp5}, we choose the family parametrized  by  $\a$
\begin{equation}
\begin{aligned}
 \Lambda^A_\alpha\theta
 =V^{-1} \int d^3\s\, \sqrt g\,
       \mathcal Z_A^\dagger(1+\alpha\, T_N)\theta,\qquad 
 \int d^3\s\, \sqrt g\,\mathcal Z_A^\dagger\mathcal Z_B
 =V\delta_{AB},\qquad
 \int d^3\s\, \sqrt g\,\mathcal Z_A^\dagger T_N\mathcal Z_B=0.
\end{aligned}
\label{945}
\end{equation}
Thus  $\Lambda^A_\alpha\mathcal Z_B=\delta^A{}_B$ for every
$\alpha$.  As a result we get\foot{Only the neutral constraints depend on $\alpha$.
The projector associated with these constraints is
$P_{0,F}^{(\alpha)}=\mathcal Z_A\Lambda_\alpha^A$,
where the charged duals are the orthogonal ones.
The neutral modes  in \eqref{pp5} obey
$\sum_{A\in N}\mathcal Z_A(\sigma)
\mathcal Z_A^\dagger(\sigma)=P_N$.
Using \eqref{945}, and adding the charged contribution, gives
$
 V P_{0,F}^{(\alpha)}(\sigma,\sigma)
 =\tfrac12(P_C+T_C)+P_N+\alpha T_N
 \equiv\mathcal P_\alpha .
 $
At $\alpha=1$ this is the matrix $\mathcal P$ used in
\eqref{932}. } 
\begin{equation}
 [P_{0,F}^{(\alpha)}]=V^{-1} \mathcal P_\alpha,
 \qquad\qquad 
 \mathcal P_\alpha=\tfrac12(P_C+T_C)+P_N+\alpha T_N.
 \label{946}
\end{equation}
Here $[...]$ means   coincident-point value as in \rf{932}. 
 The orthogonal
choice is $\alpha=0$; section~\ref{sec5} used supersymmetric  choice of  $\alpha=1$.

On the lowest neutral block
the inverse (Green's function) value  selected by $\Lambda_\alpha$ is
\begin{equation}
 G_\alpha=-\tfrac18\begin{pmatrix}(1-\alpha)^2&1-\alpha^2\\1-\alpha^2&(1+\alpha)^2\end{pmatrix},
 \qquad
 G_\alpha-G_0=-\tfrac{\alpha^2}4\,|0\rangle\langle0|
 +\tfrac\alpha4\big(|0\rangle\langle r|+|r\rangle\langle0|\big) \ . 
 \label{f1}
\end{equation}
Using  $\sum\aZ\aZ^\dagger=\PN$ and
$|0\rangle\langle r|+|r\rangle\langle0|=\rho_3\to2V^{-1}T_N$, the  the change
relative to the orthogonal choice $\alpha=0$ is
\begin{equation}
 \delta_\alpha G_\theta=\ii V^{-1}\big(-\tfrac{\alpha^2}4\PN+\tfrac\alpha2T_N\big),
 \qquad
 \delta_\alpha E_\theta=-\alpha V^{-1}T_N,
 \qquad
 \delta_\alpha\big[\rho_3G_{\theta,3}\big]=0.
 \label{f2}
\end{equation}
Hence $s_N=2G_x-\tfrac{\alpha^2}4V^{-1}$, $v_N=\tfrac\alpha2V^{-1}$,
$d_N=-\alpha V^{-1}$.

\subsubsection*{Preserved supersymmetry and the neutral spinor  complement}


Let us  now take into account   the   requirement of supersymmetry. 
The linear world-volume supersymmetry of the M2 brane action 
follows from target-space  supersymmetry
combined with the transformation restoring $\kappa$-symmetry  gauge \ci{Claus:1998yf}
(see also \ci{Tseytlin:2025dae} for a review). 
The latter is defined by   $P_+\theta=0$, with $P_\pm=\ha (1\pm\Gamma_\kappa)$ and
$\Gamma_\kappa=\ii\Gamma_{123}$ (see  \rf{y8}).

Here world-volume supersymmetry means the rigid symmetry inherited from a
preserved target-space supersymmetry, supplemented by the transformations
restoring static and $\kappa$-symmetry gauge. Its parameter is a fixed Killing spinor.
The zero-mode functions specify which directions move the classical saddle,
but do not specify how to separate these directions from a general fluctuating
field. The dual $\Lambda$ makes this separation precise: $\Lambda\widetilde{
\mathrm X}=0$ defines the fields to be integrated over while the collective
coordinates are held fixed.  Changing $\alpha$ in \eqref{945} changes this
complement without changing the zero-mode space.

We will  require the  preserved supercharge to respect this separation at the
reference saddle. A variation of the quantum fields should not generate a residual collective
coordinate when all residual coordinates are set to zero.
 In the neutral 
block, \eqref{950}  below states this requirement explicitly: at
$b_0=\zeta_\alpha=0$ the unwanted variations are
$\ha (1-\alpha)\psi$ and $-\ha (1-\alpha)h_0$  and  they vanish for arbitrary quantum
fields only for $\alpha=1$.
One then   needs  an  auxiliary field  formulation  as  the quantum fields are off shell.
Eliminating the auxiliary field $h_0$   from the quadratic action is allowed when computing that
 Gaussian integral, but setting $h_0=0$ inside the transformation rules
is not an off-shell supersymmetry operation: $\mathcal Qh_0=-2\psi$ is generally
nonzero.
i

Let  $\varepsilon$ be a Killing spinor  satisfying 
$P_+\varepsilon=\varepsilon$  for  the  given  M2 brane   instanton  solution.
It satisfies 
$D_r\varepsilon
=\Gamma_r\bar\Gamma\varepsilon$ and
$D_m\varepsilon
=-\ha \Gamma_m\bar\Gamma\varepsilon$, with
$\bar\Gamma=\ii\Gamma_*$  ($r$ and $m$   label 4+4 transverse directions  and $D$ here is the 
target-space spinor derivative). 
In a frame parallel in
the normal directions to the saddle, the transverse expansion is 
\begin{equation}
 \varepsilon(x,y)=(1+ v_1+\dots)\vva,\qquad
 v_1=x^r\Gamma_r\bar\Gamma-\tfrac12y^m\Gamma_m\bar\Gamma.
 \label{971}
\end{equation}
The combined variation  under target space supersymmetry and $\kappa$-symmetry  is
(here $\vartheta = \ha  \theta$   is   standard superspace    odd  coordinate)
\be \delta\vartheta=\varepsilon(x,y)+(1+\Gamma_{\rm B})\kappa  \ . \ee
Expanding  $\Gamma_{\rm B}=\Gamma_\kappa+\Gamma_{(1)}+\dots$ 
 in \rf{kaa}   gives  ($\{\Gamma_\kappa,\rho^aR_a\}=0$, cf. \rf{3399},\rf{903})
\al{
& \Gamma_{(1)}=\Gamma_\kappa\rho^aR_a,\qquad
 R_a=\Gamma_r\nabla_ax^r+\Gamma_m\nabla_ay^m,\qquad
 \Gamma_{(1)}\vva=-\rho^aR_a\vva.
 \label{972}
\\
 & \ \ \ \ \ \ \ \delta_1\vartheta=v_1\vva
                 -\tfrac12\Gamma_{(1)}\vva\ .  
 \label{973}
}
For the bosons, using  $E^M=e^M+\bar\vartheta\Gamma^M D\vartheta+...$ 
and
$\delta_\kappa E^M=0$ the target space supersymmetry and  compensating
$\kappa$-symmetry  transformation give the transverse displacements
$-\bar\vva\Gamma^M\vartheta$ and
$+\bar\vartheta\Gamma^M\vva$, respectively.
Since $C\Gamma^M$ is symmetric and both spinors are odd, their sum is
$-2\bar\vva\Gamma^M\vartheta$.
The leading tangential bilinear vanishes under the  projectors, so restoring
static gauge adds no linear transverse term.

Defining  the supersymmetry transformation as 
$\delta_\nu  =\nu \mathcal Q_1$   and   factorizing the   odd constant parameter $\nu$   by setting 
$\vva\to \nu\varepsilon$ where  $\varepsilon$   is   a  commuting Killing spinor  we get 
\begin{equation}
\begin{aligned}
 \mathcal Q_1x^r=-\bar\varepsilon\Gamma_r\theta,\quad
 \mathcal Q_1y^m=-\bar\varepsilon\Gamma_m\theta,\qquad
 \mathcal Q_1\theta&=\big(\rho^a\Gamma_r\nabla_ax^r
          +\rho^a\Gamma_m\nabla_ay^m
          +2x^r\Gamma_r\bar\Gamma-y^m\Gamma_m\bar\Gamma\big)\varepsilon.
\end{aligned}
\label{947}
\end{equation}
A broken supersymmetry parameter lies  instead  in the kernel  of $P_+$ and  under it the physical
fermion shifts, generating a zero mode. 
For a discussion of the  off-shell auxiliary-field extension  of this quadratic
supersymmetry  multiplet  see 
\cite{Kurlyand:2026yke}.

 Let $b_0$ be an internal $y$-scalar zero-mode
coefficient, $f,\chi$ the two fermionic coefficients in the lowest uncharged 
 block, and $h_0$ an
auxiliary boson.\foot{The uncharged fermionic zero modes  correspond to  
 $j=\tfrac12$, Hopf weight
$\lambda=0$ states  forming a 2-dimensional block. Its two states are the
components $(m',\sigma)=(-\tfrac12,+1)$ and $(+\tfrac12,-1)$  ($\sigma$ is the $\rho_3$ eigenvalue). Let us 
 call the coefficients
of $\theta$ along them $f$ and $\chi$.
 On this block, with $\tau=d=2$  (see \rf{pp99}) the Dirac operator  takes the form 
$
 \DD\ \to \  -2-2\rho_1=-2\begin{pmatrix}1&1\\1&1\end{pmatrix},
 $
which has the zero mode $\aZ\sim (1,-1)$ and a partner $\rho_3\aZ\sim (1,1)$
of eigenvalue $-4$. }
 Then the off-shell  supersymmetry  transformations are
\al{
& \mathcal Q b_0=f,\qquad \mathcal Q f=-2b_0,\qquad 
 \mathcal Q\chi=2b_0+h_0,\qquad 
 \mathcal Qh_0=-2(f+\chi), \ \ \ \   \ \mathcal Q^2=-2 \ . 
\label{948}
}
Let us  set (cf. \rf{945})
\begin{equation}
\begin{aligned}
 \zeta_\alpha&=\tfrac12(1+\alpha)f-\tfrac12(1-\alpha)\chi,
 \qquad \psi=f+\chi,\quad 
 f=\zeta_\alpha+\tfrac12(1-\alpha)\psi,\quad 
 \chi=-\zeta_\alpha+\tfrac12(1+\alpha)\psi.
\end{aligned}
\label{949}
\end{equation}
Then  \eqref{948} becomes
\begin{equation}
\begin{aligned}
 \mathcal Q b_0=\zeta_\alpha+\tfrac12(1-\alpha)\psi,\qquad 
 \mathcal Q\zeta_\alpha=-2b_0-\tfrac12(1-\alpha)h_0,\qquad
 \mathcal Q\psi=h_0,\qquad \mathcal Qh_0=-2\psi.
\end{aligned}
\label{950}
\end{equation}
For the  ``supersymmetric''   choice $\a=1$   the zero-mode  pair of fields $(b_0,\zeta)$ and the quantum pair
$(\psi,h_0)$ transform separately.\foot{For the ``orthogonal'' choice $\a=0$  we have instead 
$\mathcal Qb_0=\zeta_\perp+\ha \psi$ and
$\mathcal Q\zeta_\perp=-2b_0-\ha h_0$.}
With $\mathcal Q\bar\psi=\bar h_0$ and
$\mathcal Q\bar h_0=2\bar\psi$   we have 
\begin{equation}
 \mathcal Q\big(2\bar\psi\psi+\bar h_0h_0\big)
 =2\bar h_0\psi-2\bar\psi h_0
       +2\bar\psi h_0-2\bar h_0\psi=0.
 \label{951}
\end{equation}
 On the conjugate  block the
roles of the components are exchanged; this is 
encoded by $1+T_N$ in \eqref{945}.
Note that  the  supersymmetry  algebra  $\mathcal Q^2=-2$ closes for every $\alpha$.
The value $\alpha=1$ is selected by preservation of the
residual - quantum split under the specified 
supercharge.

\subsubsection*{Broken generators and  matrix  $\mathcal M$}

To find  the  collective coordinate  Jacobian in \rf{929} 
one needs to find  the broken generators, including their
compensator transformations. 
We use the following
fixed-gauge expansion of the target transformations  (the underlying
superisometry construction is  in \cite{Claus:1998yf}).
Given   the original supersymmetry transformation  $\delta_0\theta$, 
the $\kappa$-symmetry gauge restoring   variation is 
\begin{equation}
 \delta\theta=\delta_0\theta
 -(1+\Gamma_{\rm B})P_+
 \big[P_+(1+\Gamma_{\rm B})P_+\big]^{-1}P_+\delta_0\theta,
 \label{953}
\end{equation}
where the inverse is restricted to $P_+$.
An AdS displacement in direction $r$   gives for $\rK$ in  \rf{928} 
\begin{equation}
 \rK_r^{x^s}=(1+x^2)\delta_{rs}-2x^rx^s,
 \qquad \rK_r^y=0,
 \qquad \rK_r^\theta=x^s\Gamma_{rs}\theta+\dots .
 \label{954}
\end{equation}
 For an internal scalar transformation 
we get  a zero-mode shift plus  quadratic terms. 

We 
expand the ``compensated''  generators as
$\mathrm K_B[\epsilon\widetilde{\mathrm X}]
=\mathcal Z_B+\epsilon\mathrm K_B^{(1)}[\widetilde{\mathrm X}]+\dots$,  so that  for 
$\mathcal M$ in \rf{930} we have 
\begin{equation}
 (\mathcal M_1)^A{}_B
       =\Lambda^A\mathrm K_B^{(1)}[\widetilde{\mathrm X}].
 \label{992}
\end{equation}
 Since    $P_-\theta=\theta$,  to the  leading order in the  fluctuating fields we get 
$
 \delta_I\theta\big|_1
       =-\tfrac12 t_{Ia}^m P_-\Gamma_{am}P_-\theta=0
 $ (cf. \rf{943}). 
For a broken spinor parameter, $P_-\varepsilon_0=\varepsilon_0$ and
$\delta_{0}\theta=\varepsilon_0$. 
The term $v_1\varepsilon_0$ in the target
Killing-spinor profile is cancelled in  \eqref{953}, giving
$
 P_+v_1\varepsilon_0=v_1\varepsilon_0,\ \ 
 \delta\theta\big|_1=0.
 $
The four  blocks  in $\mathcal M_1$  are then:
\begin{center}
\renewcommand{\arraystretch}{1.25}
\begin{tabular}{p{0.12\textwidth}p{0.14\textwidth}p{0.60\textwidth}}\hline
row & column & reason the projected first-order  coefficient vanishes\\\hline
 even & even & The scalar isometry expressions  in \eqref{954}, \eqref{958}
 contain a zero-mode shift and quadratic terms, but no linear term.\\
 odd & even & Internal spin rotation gives $P_-\Gamma_{am}P_-=0$;
 the AdS spin rotation  starts with $x\theta$.\\
 even & odd & The possible transverse bilinear has
 $P_-\Gamma_M P_-=0$, with $M$  being any transverse direction.\\
 odd & odd & The first normal Killing-spinor term is cancelled by
 $\kappa$-symmetry  compensating  transformation.\\\hline
\end{tabular}
\end{center}
For the neutral dual the extra matrix $1+\alpha T_N$ commutes with
$P_-$. Hence the overlap of the first-order broken-supersymmetry variation with the
zero-mode duals vanishes
pointwise, separately in the $x$ and $y$ sectors and separately for every
field and derivative coefficient:
\begin{equation}
 \mathcal Z_A^\dagger(1+\alpha T_N)
 P_-\big(v_1-\tfrac12\Gamma_{(1)}\big)P_-\varepsilon_B=0.
 \label{993}
\end{equation}
For example,  its algebraic factors include
$P_-\Gamma_r\bar\Gamma P_-$ and
$P_-\Gamma_m\bar\Gamma P_-$, and its derivative factors include
$P_-\Gamma_\kappa\rho^a\Gamma_M P_-$; each is equal to zero. 
As a result, we  conclude that 
\begin{equation}
 (\mathcal M_1)_{\rm even,even}=(\mathcal M_1)_{\rm odd,even}
 =(\mathcal M_1)_{\rm even,odd}=(\mathcal M_1)_{\rm odd,odd}=0.
 \label{957}
\end{equation}
It follows then  that $\operatorname{STr}\mathcal M_1=0$   and also 
  $\operatorname{STr}\mathcal M_1^2=0$. This is  the
statement required in \eqref{930}. 
 Quadratic
off-diagonal blocks of $\mathcal M_2$ need not vanish, but their
products enter only at a higher order.

\subsubsection*{Computing  $\mathcal J={\rm STr}\, {\cal M}_2$ }

For the diagonal even trace in ${\rm STr}\, {\cal M}_2$   in \rf{931}, only $\rK_r^{x^r}$ and the internal
$y$ response are required. In addition to \eqref{954}, we have 
\begin{equation}
\begin{aligned}
 \rK_I^{y^m}&=\mathcal Z_I^m-\tfrac12y^2\mathcal Z_I^m
       +\xi_I^a\nabla_ay^m
       +\tfrac18t_{Ia}^n\bar\theta\Gamma_{amn}\theta+\dots,\qquad \ \ \ 
 \xi_I^a&=t_{Ia}^n y^n .
\end{aligned}
\label{958}
\end{equation}
Its action on the spinor at the reference bosonic surface,
before restoring the $\kappa$-gauge, is
$-\tfrac12(\nabla_a\mathcal Z_I^m)\Gamma_{am}\theta$,
where $t_{Ia}^m=\nabla_a\mathcal Z_I^m$ as defined above
\eqref{944}. 
Applying  \eqref{953}  gives
\begin{equation}
 \delta_I\theta\big|_{\rm linear}
 =-\tfrac12(\nabla_a\mathcal Z_I^m)
          P_-\Gamma_{am}P_-\theta=0.
 \label{955}
\end{equation}
The bilinear in $\theta$  term is the transverse displacement induced by restoring
the $\kappa$-symmetry gauge. Applying \eqref{943} and \eqref{944}, including this
term, gives the even-coordinate part of the  trace  of ${\cal M}_2$ as  in \rf{931}
\al{& \mathcal J=\operatorname{STr}\mathcal M_2=   \mathcal J_{\rm even}-\mathcal J_{\rm odd}\ , \la{yyy}\\
&
 \mathcal J_{\rm even}=V^{-1}\int d^3\s\, \sqrt g\,
 \Big[2x^2-4y^2+2y^m(J_y)^{mn}\nabla_3y^n
                 -i\bar\theta T_C\theta\Big].
 \label{959}
}
For the odd-coordinate  part of  ${\rm STr}\, {\cal M}_2$, the  quadratic part of the
broken spinor transformation is
\begin{equation}
\begin{aligned}
 \delta\theta\big|_2
  =\big(v_2-\tfrac12\Gamma_{(1)}v_1\big)\varepsilon_0
     +\tfrac1{12}\mathscr M_4\varepsilon_0
     +\tfrac14(\bar\varepsilon_0\rho^a\theta)D_a\theta,\qquad 
 v_2=\tfrac12x^2-\tfrac18y^2
       -\tfrac12(x^r\Gamma_r\bar\Gamma)
                   (y^m\Gamma_m\bar\Gamma).
\end{aligned}
\label{960}
\end{equation}
Here $v_1$ is  given in \eqref{971}, and $\Gamma_{(1)}$ and $\mathscr M_4$ are
 the same matrices as  in 
\eqref{70} and \eqref{72}.  The term
$-\ha \Gamma_{(1)}v_1$ comes from $\kappa$-gauge 
 compensation transformation.\foot{Let us  record the origin of the three coefficients relevant to \eqref{958}
and \eqref{960}. The original superspace fermion is $\vartheta=\ha \theta$, 
and for a broken parameter we write $\varepsilon_\vartheta=\ha \varepsilon_0$, so
that the leading  shift is $\delta\theta=\varepsilon_0$.
For the internal rotation normalized by
$\delta_Iy^m|_0=\mathcal Z_I^m$, the mixed tangent--normal Lorentz
parameter is $\omega_{am}=-t_{Ia}^m$: the displaced tangent has normal
component $\nabla_a\mathcal Z_I^m=t_{Ia}^m$, and the compensating frame
rotation removes that component. The spinor generator
${1\ov 4} \omega_{AB}\Gamma^{AB}$ is therefore
$-\ha t_{Ia}^m\Gamma_{am}$. Then 
$\delta_\kappa\vartheta=\fo t_{Ia}^n\Gamma_{an}\theta$.
Using $\delta_\kappa Z^{\mathcal A}E_{\mathcal A}{}^m(Z)=0$ (here $Z^{\mathcal A}$ is superspace coordinate) 
 and
$E^m=e^m+\bar\vartheta\Gamma^m D\vartheta+...$ gives
$
 \delta_\kappa y^m
 =-\tfrac18t_{Ia}^n\bar\theta\Gamma_m\Gamma_{an}\theta
 =\tfrac18t_{Ia}^n\bar\theta\Gamma_{amn}\theta,
$
as  follows from 
$\Gamma_m\Gamma_{an}=-\Gamma_{amn}-\delta_{mn}\Gamma_a$
and $\bar\theta\Gamma_a\theta=0$, as $C\Gamma_a$ is symmetric.
This derives the  last  term in \eqref{958}.
The non-derivative  term quadratic in the fermions in \rf{960}  comes from the
nonlinear target-space supersymmetry transformation.
In the Killing-spinor coordinates of \cite{Claus:1998yf},  its spinor-matrix factor
$ \mathscr M\coth\mathscr M$        has the
expansion
$\mathbf1+\tfrac13\mathscr M_\vartheta^2+O(\vartheta^4)$.
Using the identification
$\mathscr M_\vartheta^2=\fo \mathscr M_4$  with $\mathscr M_4$
defined in \eqref{72},  this contribution gives 
 $\varepsilon_0+\tfrac1{12}\mathscr M_4\varepsilon_0 +...
$
The bosonic first-order displacement of a broken supersymmetry is
$\delta X^a=-\bar\varepsilon_\vartheta\rho^a\vartheta$.
Restoring static gauge  gives
$
 \xi^a=-\delta X^a=\tfrac14\bar\varepsilon_0\rho^a\theta,
 \ \  \delta_\xi\theta=\xi^aD_a\theta.
$
This is the $\fo $ term of \eqref{960}, with frame transport included in
$D_a$. The $v_2-\ha \Gamma_{(1)}v_1$ term instead comes from the normal
Killing-spinor expansion and the same fixed-gauge compensation. 
}

Taking the trace of  the first term of \eqref{960} with the  zero-mode dual
\eqref{945}  gives the   quadratic in bosons  part of  $\mathcal J_{\rm odd}$
\begin{equation}
\begin{aligned}
 \mathcal J_{\rm odd}^{(b^2)}
 =V^{-1}\int d^3\s\, \sqrt g\,\Big[
 6x^2-\tfrac32y^2+y^m(J_y)^{mn}\nabla_3y^n
       -4\alpha\,  x^r(I_x)^{rs}\nabla_3x^s\Big].
\end{aligned}
\label{961}
\end{equation}
The  algebraic  fermionic part  of  $\mathcal J_{\rm odd}$  is found by writing $\mathscr  M_4$ in  \eqref{72} as
$\mathscr  M_4=\sum_d A_d\theta\, \bar\theta B_d$.
Then on the physical  gauge slice,
$
 \sum_d B_d\mathcal P_\alpha A_d
       =8\ii\big(T_C+\alpha T_N\big),
 $
 and as a  result 
\begin{equation}
 \mathcal J_{\rm odd}^{(\theta^2,\rm alg)}
 =-\tfrac1{12}V^{-1} \int d^3\s\, \sqrt g\,
       \bar\theta\sum_d B_d\mathcal P_\alpha A_d\theta
 =-\tfrac{2\ii}{3}V^{-1} \int d^3\s\, \sqrt g\,
       \bar\theta(T_C+\alpha T_N)\theta.
 \label{963}
\end{equation}
The  last term in  \eqref{960}  leads to the  following    derivative part  of $\mathcal J_{\rm odd}^{(\theta^2)} $ \foot{Note that  $\rho^a\mathcal P_\alpha D_a
 =\big[\tfrac12(P_C-T_C)+P_N-\alpha T_N\big]\slashed D
      +(T_C+2\alpha T_N)\rho_3D_3$    and $\big[\tfrac12(P_C-T_C)+P_N-\alpha T_N\big]\mathcal P_\alpha
            =(1-\alpha^2)P_N.$}
\begin{equation}
 \mathcal J_{\rm odd}^{(\theta^2,\,\mathrm{der})}
       =\tfrac1{4}V^{-1}\int d^3\s\, \sqrt g\,
                 \bar\theta\rho^a\mathcal P_\alpha D_a\theta.
 \label{964}
\end{equation}
Using ~\eqref{959}, \eqref{961}, \eqref{963} and \eqref{964} and setting $\a=1$    we get the expression for  the Jacobian   given in 
 \eqref{933}. 
 
Note that the above  derivation of   \eqref{933}  required, in addition  to the
quadratic fluctuation action,  the  knowledge  of 
 quadratic
coefficients of the gauge-compensating  transformations. The linear supersymmetry  of the 
quadratic action  leading to 
\eqref{950}  fixes $\a=1$   but does not determine those coefficients by itself.


The results for general $\alpha$ quoted in \rf{018} are obtained as follows. The analog of the expectation value of $\mathcal J$ in \rf{936} is 
\begin{equation}
\begin{aligned}
 \langle\mathcal J\rangle_\alpha={}&
 -16G_x-10G_y-4G_{y,3}-\tfrac83v_C+2q_C
       +r_C+2r_N-d_C\\
 &+\alpha\big[-16G_{x,3}+\tfrac{16}{3}v_N
                         +4q_N-2d_N\big].
\end{aligned}
\label{967}
\end{equation}
Here $d_N=-\alpha V^{-1}$, $v_N=\ha\alpha V^{-1}$, $q_N=0$, while $r_N$ and
$p_N$ are $\a$-independent by \rf{f2};  $s_N=2G_x-\tfrac{\a^2}4V^{-1}$ also
depends on $\a$ but drops out of \rf{92} since the $s_F$ terms cancel in the
sum of the three sectors.
As  a  result, 
\begin{equation}
 \langle\mathcal J\rangle_\alpha
    =-\tfrac23(28G_x+13G_y)-3V^{-1}+
                               \tfrac{14}{3}\a^2  V^{-1} .
 \label{968}
\end{equation}
At the   same time, $\vev{L_4}$ in \rf{92}  evaluated with $\a$-dependent   fermionic  zero  mode projector in   \rf{946} gives 
\al{
& \langle \mathcal L_4\rangle_\alpha
    =-\tfrac23V^{-1} (28G_x+13G_y)+ V^{-2} +
                               \tfrac{2}{3}\alpha^2 V^{-2} ,\la{977}\\
 &V\langle\mathcal L_4\rangle_\alpha
              -\langle\mathcal J\rangle_\alpha
    =4(1-\alpha^2)V^{-1}.
\label{969}
}
Note that  the  transcendental  in $k$  functions  $G_x$  and $G_y$
 cancel in \rf{969} for any $\a$. 
 
 The vanishing of the total  2-loop  correction \rf{969} thus   happens  for 
$\alpha=1$ -- the value selected by the supersymmetry in \rf{950} -- and for $\alpha=-1$, which is related to $\alpha=1$ by an $SO(4)$ rotation of $\AdS_4$ (see section~\ref{sec5}).

\section{Paired  bosonic and fermionic  zero modes and  path integral}
\label{apps}

 Integrating over fermionic   collective coordinates    at 1-loop   approximation formally gives   zero. 
 In a   supersymmetric case   with a set of equal numbers  of 
  bosonic and fermionic zero modes  it may be  tempting  to  assign  the path integral a different finite value, 
  assuming  a certain  supersymmetric  IR  regularization.   Below  we review  some  related  
   examples in the string theory and M2-brane   context. 

Suppose that a saddle has $\rq$ real bosonic zero modes $a^I$ and $\rq$ real fermionic zero modes $\zeta^I$.  If they were completely free, their contribution would schematically be
$
 Z_0\sim \int d^\rq a\,d^\rq\zeta\, ...
$ 
If the paired modes admit a common supersymmetry-preserving deformation
that lifts them, their singular Gaussian contribution may be defined
by evaluating the deformed integral before removing the deformation.
Equal numbers of zero eigenvalues alone do not establish the existence
of such a deformation.
Adding to the action  
 $S_\mu =\mu a^* a +\mu \bar\zeta \zeta$  where  $(a,\zeta)$  is  a complex   pair    and using  standard normalizations 
$
 \int {d^2a\over\pi}\,e^{-\mu |a|^2}=\mu^{-1},
 \ 
 \int d\bar\zeta\,d\zeta\,e^{-\mu\bar\zeta\zeta}=\mu,
$   we get 
$
 Z_{\rm pair}(\mu)=1.
$
Then for a  set of  such pairs   we also get 
$
 Z(0)\equiv\lim_{\mu\to0}Z_{\rm pair}(\mu)=1.$
If a supersymmetric deformation localizes the original zero-mode manifold onto several isolated fixed configurations, then, when the local normalizations and orientations agree, one obtains
a sum of the corresponding unit contributions.\foot{
This unit factor concerns the leading Gaussian normalization. It does
not imply that the undeformed collective-coordinate density has only
its highest Grassmann component, or that an interacting insertion can
be replaced by its value at zero odd coordinates without further
argument. The same supersymmetric prescription must be applied to the
leading integral and its inverse-tension correction. In
section~\ref{n1}, the large-deformation evaluation is justified by the
full-field Ward identity \rf{n4}, rather than by the cancellation of
the two Gaussian powers alone.}

A directly relevant example is the $\CP^1$ world-sheet instanton of the
 type IIA string on $\AdS_4\times\CP^3$  discussed   in \cite{Gautason:2023igo}.  The quadratic fluctuation problem has 12   bosonic and 12  fermionic zero modes.  A naive treatment of the fermionic zero modes would make the vacuum partition function vanish.  Instead, ref. \cite{Gautason:2023igo} introduced a supersymmetric deformation of the
 background which lifts the zero modes.  The deformed problem is then well defined, and the undeformed case  is obtained as a limit  of the  regulated path integral.  
The paired bosonic and fermionic  contributions  cancel in the regulated supersymmetric Gaussian integral.   What survives is the contribution from the fixed configurations selected by the deformation.  In the $\CP^1$ instanton problem this procedure produces the additional numerical factor associated with the two localized configurations \cite{Gautason:2023igo}.

Another  precedent is the Euclidean D3-brane calculation in~\cite{Gautason:2024nru}.  There the  quadratic spectrum contains 10 scalar zero modes and 8 fermionic zero modes.  The authors separate these into 8 paired boson--fermion zero modes and two unpaired bosonic zero modes.  For the paired sector they  adopt the prescription that its contribution is one, while the two remaining bosonic modes are treated as genuine collective coordinates and integrated over.  



The M2-brane case is discussed in \cite{Gautason:2025per,Gautason:2025plx}.
The zero-mode moduli spaces are analysed using the preserved world-volume supersymmetry and an $R$-symmetry Killing vector.  Their localization picture gives a geometric interpretation of the regulated zero-mode sector: paired directions are lifted away from the fixed locus, while the path integral localizes onto the supersymmetric  configurations.  In the examples related to the  instanton problem this leads to a finite zero-mode factor that can be interpreted as counting the localized M2 configurations rather than as the value of an unsaturated Grassmann integral.

This prescription  is sharpened in the fluctuation analysis of M2 brane  instantons in $\AdS_4\times Y_7$ in ~\cite{Kurlyand:2026yke}.   Non-cohomological kinetic zero modes can occur in supersymmetry doublets and may be lifted by a regulator that preserves the supersymmetry complex.  
Unpaired fermionic zero modes are to be treated differently  (and may 
 still make  the vacuum path integral  vanish).

This distinction is important for the present case of the  $S^3/\mathbb{Z}_k$ M2 brane  instanton.  Equal numbers of bosonic and fermionic zero eigenvalues are not, by themselves, sufficient to justify a unit factor.  What is required is the stronger statement that the  modes form complete multiplets of the same preserved supercharge and admit a common deformation which lifts them without spoiling the supersymmetry.


No deformation of the 11d supergravity  background is required for
the conditional argument in section~\ref{n1}. There we assume a
supersymmetric deformation on the full fluctuation field space,
including the interactions and measure, with the paired local form
\rf{n6}. Its nonlinear odd constraint and the equation-of-motion
term in the component algebra are retained in the localized measure.
Appendix~\ref{apf} shows that their expectations reproduce the
reference-point  Jacobian contribution, with no additional term at order
$\rT_2^{-1}$. This is the interaction-order comparison needed beyond
the unit Gaussian factor above.

The equality between the regulated integral and its localized
evaluation uses \rf{n4}. 
Once it holds, the  $\rT_2^{-1}$  coefficient vanishes at
each contributing configuration.
 These
statements do not extend the paired prescription to unpaired
fermionic zero modes, and do not establish  vanishing  at higher
orders.


\section{Supersymmetry and  measure in the paired zero-mode prescription}\label{apf}

Here we give the geometric identification of the two configurations in
\rf{n5} and the local measure calculation used in \rf{n11}.
 The  notation follows 
 that of sections~\ref{sec3},
\ref{sec5} and Appendix~\ref{apd}. In particular, $\mathcal P$ in
\rf{932} is the coincident density of an integral projector. 
Traces over spinors below are over the
16  physical components.

\subsubsection*{Preserved and broken supercharges}

The $\mathbb Z_k$-invariant  internal-space  Killing spinors of the $k>2$ background span
the six-dimensional representation
$\mathscr V=\Lambda^2\mathbb C^4$  (the $\mathbf 6$ of $SO(6)\simeq SU(4)$). Relative to a complex 2-plane $\Pi$,
it decomposes as ($\dim_{\mathbb C}E=2,\ \dim_{\mathbb C}F=4$)
\begin{equation}
 \mathscr V =E(\Pi)\oplus F(\Pi),\qquad \qquad
 E(\Pi)=\Lambda^2\Pi\oplus\Lambda^2\Pi^\perp,
       \qquad \qquad
 F(\Pi)=\Pi\wedge\Pi^\perp\simeq\Pi\otimes\Pi^\perp\ .
 \label{n33}
\end{equation}
Let $\mathscr S_C$ and $\mathscr S_N$ be the two AdS Weyl-spinor doublets, distinguished by the
projectors $\PC$ and $\PN$ of \rf{9}, i.e.\ the two chiralities with respect to the $SO(4)=SU(2)\times SU(2)$
of rotations about the instanton position.
The anticommutator of two supercharges
$\psi\otimes v$ and $\psi'\otimes v'$ ($\psi,\psi'\in\mathscr S$, $v,v'\in\mathscr V$) is
$\rU(\psi,\psi')\,(v,v')+\WW(\psi,\psi')\,A(v,v')$, where $(v,v')$ is the invariant inner product on
$\mathscr V$, $\rU(\psi,\psi')\in sp(4)$ is symmetric in $\psi,\psi'$, $\WW(\psi,\psi')$ is the invariant
symplectic form (antisymmetric in $\psi,\psi'$) and $A(v,v')=-A(v',v)\in so(6)$ is the rotation in the
$(v,v')$ plane.
For $\psi,\psi'$ of the same chirality $\rU(\psi,\psi')$ is a rotation of $\AdS_4$ about the instanton
position and $\WW(\psi,\psi')\neq0$ in general; for opposite chiralities $\rU$ is a translation and $\WW=0$.
The rotations $A(v,v')$ with $v,v'\in F(\Pi)$ span $so(4)\subset s(u(2)\oplus u(2))$, and $A(v,v')$
with $v,v'\in E(\Pi)$ generates the relative phase rotation of $\Pi$ and $\Pi^\perp$, while
$A(v,v')$ with $v\in F$, $v'\in E$ does not preserve $\Pi$. 

Requiring the preserved
supercharges to close on the stabilizer of the saddle leaves two possibilities,
$(\mathscr S_C\otimes F)\oplus(\mathscr S_N\otimes E)$ and the one with $C\leftrightarrow N$.
The counting of the broken modes in Appendix \ref{apa} (8 neutral and 4 charged) selects the first.
 The decomposition  is 
\begin{equation}
 \begin{split}
 \text{preserved:}\quad
  (\mathscr S_C\otimes F)\oplus(\mathscr S_N\otimes E)\quad (8+4)  \ ; \qquad 
 \text{broken:}\quad
  (\mathscr S_C\otimes E)\oplus(\mathscr S_N\otimes F) \quad (4+8) \ .
 \end{split}
 \label{n34}
\end{equation}
These representation spaces are complexified; the physical zero-mode
count uses the same reality prescription as in Appendix~\ref{apa}.
In particular, the broken modes are  four charged and eight neutral
ones. The decomposition is unchanged by $\Pi\leftrightarrow\Pi^\perp$.
It describes Killing-spinor solutions and does not assume off-shell
closure on all quantum fluctuations.

The supercharge $\cQ$ of \rf{948} belongs to $\mathscr S_N\otimes E(\Pi_0)$. Writing
$\cQ=\psi_1\otimes \omega_1+\psi_2\otimes \omega_2$ with $\omega_{1,2}$ an orthonormal basis of $E(\Pi_0)$
(the explicit basis is given in \rf{n37} below), one has
\be \la{n42}
\cR=\cQ^2=\ha\big[\rU(\psi_1,\psi_1)+\rU(\psi_2,\psi_2)\big]+\WW(\psi_1,\psi_2)\,A(\omega_1,\omega_2)\ ,
\ee
where the first term is a rotation of $\AdS_4$ in the $SU(2)$ factor acting on $\mathscr S_N$ (with equal
weights in two orthogonal 2-planes) and the second is the relative phase rotation of $\Pi_0$ and $\Pi_0^\perp$,
present provided $\WW(\psi_1,\psi_2)\neq0$, which is the case for \rf{948}. On the tangent directions of the
family of saddles the $\AdS_4$ translations and the broken supercharges in $\mathscr S_N\otimes F$
are rotated by the first term only, while the internal moduli ${\rm Hom}(\Pi_0,\Pi_0^\perp)$ and the broken
supercharges in $\mathscr S_C\otimes E$ are rotated by the second term only. The normalization
$\cQ^2=-2$ on both $b_0$ and $f,\chi$ in \rf{948},\rf{950} fixes the two weights to be equal, so that
$\cR$ has weights of magnitude 2 on all 12 even and 12 odd tangent directions.

In particular, $\ker\cR$ restricted to the tangent space of the family of saddles is trivial: in the
terminology of \ci{Kurlyand:2026yke} the $12+12$ zero modes are non-cohomological $\cQ$-doublets, there are
no BPS moduli of $\cQ$, and $p_\pm$ are isolated.\foot{For comparison with \ci{Kurlyand:2026yke}, use its cone coordinates $z_a$ with $\ZZ_k$ weights
$(1,1,-1,-1)$: in terms of our common-phase coordinates, $z_{1,2}=Y_{1,2}$, $z_{3,4}=\bar Y_{3,4}$, so that
the Reeb $U(1)_r$ of \ci{Kurlyand:2026yke} (common phase of $z_a$) is the relative phase rotation of $\Pi_0$ and
$\Pi_0^\perp$ in \rf{n42}, in agreement with $\cQ^2=-i\mathcal L_\xi+\dots$ there. A plane of $\Gr_\CC(2,4)$
near $\Pi_0$, the graph $Y_{3,4}=B\,Y_{1,2}$, becomes the antiholomorphic graph $z_{3,4}=\bar B\,\bar z_{1,2}$,
so that no nearby plane of the family is a holomorphic cone for the $z$ structure; conversely, a holomorphic
graph $z_{3,4}=C\,z_{1,2}$ is $\ZZ_k$-equivariant only if $(\omega-\omega^{-1})C=0$, i.e.\ $C=0$ for $k>2$.
These are the $n=0$ normal-bundle modes projected out in \ci{Kurlyand:2026yke} for $k>2$, which survive for
$k\le2$ and provide the additional $8$ real bosonic zero modes of the $20+16$ counting of Appendix \ref{apa}.
This is a local statement: the mixed planes, e.g.\ $\{Y_2=Y_4=0\}=\{z_2=z_4=0\}$, are holomorphic for both
structures, but with the opposite orientation on one of their two complex directions. Holomorphicity of the
unoriented cone therefore does not imply preservation of the same oriented supercharge; in the present
construction the mixed planes are fixed by $\cR$ but excluded by $v_{\rm int}=4$ in \rf{n32}.}

Let $\qq$ be the odd vector field generated by $\cQ$ on the family of saddles, as in section~\ref{n1},
and let $\qq_{\rm tan}$ be its linearization at either of the two configurations $p_\pm$.
In terms of the local coordinates $(a^I,\zeta^I)$ of \rf{n8} and of the odd tangent coordinates
\be \la{n46}
\eta^I\equiv \mathsf L^I{}_J\,\zeta^J \ ,
\ee
one has, in the compact convention obtained from the complex normalization $\cQ^2=-2$ of \rf{948} by the
phase rotation $\cQ\to e^{\ii\pi/4}\cQ$, under which $K$ becomes real antisymmetric with $K^2=-4$, while the
combinations $K^{-1}\ML$ and $\mathsf L^{-1}\cQ_3$ entering \rf{n16} and \rf{n27} are unchanged, 
\be \la{n47}
 \qq_{\rm tan}
 =\eta^I\frac{\partial}{\partial a^I}
  +K^I{}_J a^J\frac{\partial^L}{\partial\eta^I},
 \qquad
 \qq_{\rm tan}a^I=\eta^I,\qquad
 \qq_{\rm tan}\eta^I=K^I{}_J a^J,\qquad K^T=-K,\quad K^2=-4\cdot\mathbf 1 .
\ee
Thus $\qq_{\rm tan}^{\,2}$ is the bosonic rotation $\cR$ induced on the tangent collective coordinates,
$\qq_{\rm tan}^{\,2}a=Ka$, in agreement with $\cR\mathcal B=K\mathcal B$ in section~\ref{n1}
(recall that $\mathcal B=a$ and $\mathcal F_1=\mathsf L\zeta=\eta$ by \rf{mm8}). Comparing with
\rf{n44}, the broken component of $\cQ$ near $p_\pm$ is
\be \la{n48}
\phi^I(\mathrm a)=(\mathsf L^{-1}K)^I{}_J\,a^J+\OO(a^2)\ ,
\ee
i.e.\ the body of $\qq\,\zeta$ vanishes linearly at $p_\pm$ with the slope fixed by $\cR$.
Reversing the orientation of one of the six real planes rotated by $\cR$ changes the corresponding
block of $K$ and the sign of the Pfaffian together with the odd integration orientation; it does not
change the local $\rT_2^{-1}$ coefficient. The quadratic localizing form \rf{n7} follows from the pairing
relations \rf{n47}; it is not itself a transformation law.

\subsubsection*{Zero locus of the odd vector field}

Using rotations within $\Pi_0$ and $\Pi_0^\perp$, an orthonormal frame of
$\Pi$ and its complement can be written as
\begin{equation}
 \begin{aligned}
 w_1&=\cos\vartheta_1 e_1+\sin\vartheta_1 e_3,
 &w_2&=\cos\vartheta_2 e_2+\sin\vartheta_2 e_4,\\
 v_1&=-\sin\vartheta_1 e_1+\cos\vartheta_1 e_3,
 &v_2&=-\sin\vartheta_2 e_2+\cos\vartheta_2 e_4 .
 \end{aligned}
 \label{n36}
\end{equation}
The supercharges preserved at $(0,\Pi)$ are given by \rf{n34} with $\Pi_0\to\Pi$; the two-forms
$\omega_{1,2}\in E(\Pi_0)$ of $\cQ$ are preserved at $\Pi$ only through their components in $E(\Pi)$, while
their components in $F(\Pi)=\Pi\wedge\Pi^\perp$ are broken. For the orthonormal basis
$\omega_1=(e_1\wedge e_2+e_3\wedge e_4)/\sqrt2$, $\omega_2=\ii(e_1\wedge e_2-e_3\wedge e_4)/\sqrt2$ of
$E(\Pi_0)$ one finds, expressing $e_1\wedge e_2$ and $e_3\wedge e_4$ in the frame \rf{n36},
\be \la{n37}
 \big|(\omega_1)_{F(\Pi)}\big|^2=\sin^2(\vartheta_1-\vartheta_2)\ ,\qquad\qquad
 \big|(\omega_2)_{F(\Pi)}\big|^2=\sin^2(\vartheta_1+\vartheta_2)\ .
\ee
The internal term of \rf{n32} is twice their sum; the factor 2 is the relative normalization of the internal
and $\AdS_4$ parts of the Killing spinor of $\cQ$, fixed by the equal weights of $\cR$ in \rf{n42}.
The two zeros $\Pi=\Pi_0,\Pi_0^\perp$ do not depend on this factor.

\subsubsection*{Cubic transformations and the insertion $\mathcal J_{\mathcal Q}$}

We now use the component field relations in \rf{947}. 
Let us first note that the gauge-fixed supersymmetry transformations contain only odd powers of the
fluctuation fields, $\cQ=\cQ_1+\cQ_3+\dots$. Indeed, the reflection of all eight transverse directions,
$x^r\to-x^r$ together with $(Y_1,Y_2,Y_3,Y_4)\to(Y_1,Y_2,-Y_3,-Y_4)$, is an isometry of the background
which fixes the instanton point-wise: it is the product of $-1\in SO(4)$ of $\AdS_4$ and of $-1$ on the
normal plane $\Pi_0^\perp=\{Y_1=Y_2=0\}$. It preserves the static gauge, and its spinor lift is proportional to
$\Gamma_{45}\Gamma_{67}\Gamma_{89}\Gamma_{10\,11}\propto\Gk$, so that it preserves also the $\kappa$-symmetry
gauge \rf{y8}. Choosing the sign of the lift to be $\Gk$, the conditions $P_-\theta=\theta$ and
$P_+\varepsilon=\varepsilon$ give $\theta\to-\theta$ and $\varepsilon\to\varepsilon$: all fluctuation fields
change sign while the preserved supercharges are invariant. Covariance of the transformation rules under this $\ZZ_2$
then excludes all terms of even order, $\cQ_{2n}=0$. The same symmetry is responsible for the absence of
the cubic vertices in \rf{005}.

In the language of section \ref{n1}, the computations of this and the next subsection evaluate the
contribution of the completion terms \rf{n54} after the lifted zero modes have been integrated out:
$A_2$ comes from the cubic part $\cQ_3$ of the transformation of the bosonic projections, and the
equation-of-motion term from the cubic closure \rf{n25}.
The coefficient $A_2$  in \rf{n10} is obtained by
inserting a fermionic zero-mode profile in the cubic scalar variation:
\begin{equation}
 (A_2)^J{}_{J'}=(\mathsf L^{-1})^{J}_I\Lambda_B^I
       \big[\mathrm d_\theta(\mathcal Q_3X^M)
                                     [\mathcal Z_{F,J'}]\big],\qquad 
 \qquad  X^M=(x^r,y^m) .
 \label{n16}
\end{equation}
The scalar component selected by the dual is implicit, and
$\mathrm d_\theta$ is a left odd directional derivative. The subscripts
in $\mathcal Q=\mathcal Q_1+\mathcal Q_3+\dots$ specify the degree in
fluctuation fields. The  supersymmetry  transformation is needed through cubic order,  including the  term
with three fermions.

The transformations follow from the superisometries with the same
static- and $\kappa$-gauge restoration as in Appendix~\ref{apd}
(see \ci{deWit:1998yu,Claus:1998yf}). With the commuting Killing parameter
$\varepsilon$, define
\al{
 U_x&=x^r\Gamma_r\bar\Gamma,\qquad
 U_y=y^m\Gamma_m\bar\Gamma,\qquad \bar\Gamma=i\Gamma_*,\qquad 
 \varepsilon_1=(U_x-\tfrac12U_y)\varepsilon,\qquad
 \varepsilon_2=(\tfrac12x^2-\tfrac18y^2-\tfrac12U_xU_y)\varepsilon,\no \\
 \sigma_1&=(\rho^a\Gamma_M\nabla_a  X^M+2U_x-U_y)\varepsilon,\qquad \qquad 
 t^a=\tfrac14\bar\sigma_1\rho^a\theta
                         -\bar\varepsilon_1\rho^a\theta .
 \label{n17}
}
The external odd supersymmetry parameter is moved to the left and
removed, as in \rf{947}. The required scalar terms are
\al{
& (\mathcal Q_3x^r)_{X^2\theta}
     =-\bar\varepsilon_2\Gamma_r\theta
         +x^2\bar\varepsilon\Gamma_r\theta-t^a\nabla_ax^r,\qquad 
 (\mathcal Q_3y^m)_{X^2\theta}
     =-\bar\varepsilon_2\Gamma_m\theta
         +\tfrac12y^my^n\bar\varepsilon\Gamma_n\theta-t^a\nabla_ay^m,\label{n18} \\
& (\mathcal Q_3  X^M)_{\theta^3}
     =-\tfrac1{24}\bar\theta\Gamma_M\mathscr M_\theta^2\varepsilon
                          +\tfrac14\bar\theta\Gamma_M Z_\theta,\qquad 
 Z_\theta=\sum_M(\mathcal Q_1 X^M)\mathcal T_M\theta,
 \qquad \mathcal T_r=-\Gamma_r\bar\Gamma,
 \quad \mathcal T_m=\tfrac12\Gamma_m\bar\Gamma .\no 
}
Here $\mathcal Q_1 X^M=-\bar\varepsilon\Gamma_M\theta$, while
$\sigma_1=\mathcal Q_1\theta$ and 
\begin{equation}
 \mathscr M_\theta^2 \varepsilon =
 \sum_{u=1}^{7}\bar\Gamma\Gamma_{u}\theta
                                      (\bar\theta\Gamma_{u}\varepsilon )
 +2\sum_{r=8}^{11}\bar\Gamma\Gamma_r\theta
                                      (\bar\theta\Gamma_r\varepsilon )-\sum_{1\leq u<v\leq7}\Gamma_{uv}\theta
                      (\bar\theta\Gamma_{uv}\bar\Gamma \varepsilon )
 +2\sum_{8\leq r<s\leq11}\Gamma_{rs}\theta
                      (\bar\theta\Gamma_{rs}\bar\Gamma \varepsilon ) .\ \ 
 \label{n19}
\end{equation}
 Here the  index $u$ runs over the seven internal
tangent and normal axes, whereas $a=1,2,3$ in \rf{n17} is a
3d  frame index.\foot{The physical
spinor satisfies $P_-\theta=\theta$, and the preserved
parameter has the opposite projection, $P_+\varepsilon=\varepsilon$.
The operator $\mathscr M_\theta^2$ acting on $\varepsilon$ must therefore
be formed in the full 32-component space before making the projection. }

Setting  $\mathcal J_{\mathcal Q}=\operatorname{tr} A_2$  
 one finds for the bosonic part\foot{Here and below $\a=\pm1$  labels the two adapted local computations at $p_\pm$, not a change of complement at fixed supercharge.}  
\begin{equation}
 \vev{\mathcal J_{\mathcal Q,B}}
  =E_x+2E_y+8G_x-4G_y-4\alpha G_{x,3}+2G_{y,3} .
 \label{n20}
\end{equation}
The   terms  in $ (\mathcal Q_3  X^M)_{\theta^3}$ in  \rf{n18} give
\al{ \vev{\mathcal J_{\mathcal Q,F}^{(1)}}
      =-6s_C-12s_N-\tfrac{26}{3}v_C+\tfrac{4}{3}\alpha v_N,\qquad \qquad 
 \vev{\mathcal J_{\mathcal Q,F}^{(2)}}
      =18s_C+6s_N+18v_C+6\alpha v_N .
 \label{n21}
}
Differentiating the three-fermion scalar variation gives the quadratic
fermionic insertions in \rf{n21}.

Using the identities \rf{79},\rf{87}--\rf{89} we obtain\foot{Here $E_x,E_y$ are the projected EOM values \rf{pp33}, $G_{x,3}=0$ since $x^r$ is uncharged under the
Hopf $U(1)$, $G_{y,3}$ follows from \rf{87}--\rf{89}, $s_N$ and $v_N$ are the values found in \rf{f2}, and
$s_C$, $v_C$ follow from \rf{79}. The three groups of terms in \rf{n20},\rf{n21} come from the
$X^2\theta$ terms of \rf{n18} (bosonic contractions, giving $E_x$, $E_y$ and the derivative constants) and
from the two structures $\bar\theta\Gamma_M\mathscr M^2_\theta\varepsilon$ and $\bar\theta\Gamma_MZ_\theta$
in the $\theta^3$ terms of \rf{n18} (fermionic contractions, giving $s$ and $v$), respectively.} 
\al{
& E_x=-V^{-1},\qquad E_y=-2V^{-1},\qquad G_{x,3}=0,\qquad
 G_{y,3}=2G_x-G_y+\tfrac14V^{-1},\no \\
 & s_C=-G_y+\tfrac38V^{-1},\qquad s_N=2G_x-\tfrac14V^{-1},\qquad 
 v_C=-2G_x+G_y-\tfrac38V^{-1},\qquad v_N=\tfrac\alpha2V^{-1} , 
 \label{n22}
}
we get 
\al{
 \vev{\mathcal J_{\mathcal Q,B}}&=12G_x-6G_y-\tfrac92V^{-1},\qquad \ \ 
 \vev{\mathcal J_{\mathcal Q,F}^{(1)}}
       =-\tfrac{20}{3}G_x-\tfrac83G_y+\tfrac{14}{3}V^{-1},\qquad \ \ 
 \vev{\mathcal J_{\mathcal Q,F}^{(2)}}=-24G_x+\tfrac32V^{-1} , \no\\
 \vev{\mathcal J_{\mathcal Q}}&=   \vev{\mathcal J_{\mathcal Q,B}}  + \vev{\mathcal J_{\mathcal Q,F}^{(1)}} +  \vev{\mathcal J_{\mathcal Q,F}^{(2)}} = -\tfrac23(28G_x+13G_y)+\tfrac5{3}V^{-1}.
 \label{n23}
}
Thus  $ \vev{\mathcal J_{\mathcal Q}}$ is    precisely the same as $\vev{\mathcal J}$ in 
\rf{939}. This establishes the first equality in \rf{n11} without
specializing the functions $G_x(k)$ and $G_y(k)$.

\subsubsection*{Equation-of-motion term and the insertion $\mathcal J_{\rm EOM}$}

The  square of the linear term in  supersymmetry  generator  acting on physical spinors is
\begin{equation}
 \begin{gathered}
 \mathcal Q_1^2\theta=\mathcal R\theta+N_\alpha\DH\theta,\qquad \ \ 
 \qquad \DH=i\rho^aD_a,\\
 N_\alpha=\alpha\rho_3P_\alpha^\perp,
 \qquad P_\alpha^\perp=\tfrac12(1-T_C-\alpha T_N),
 \qquad \alpha=\pm1 .
 \end{gathered}
 \label{n24}
\end{equation}
The identities
$T_C^2=P_C$, $T_N^2=P_N$ and $T_CT_N=0$ give
$(P_C+T_C)N_\alpha=(P_N+\alpha T_N)N_\alpha=0$.
Hence the adapted duals satisfy $\Lambda_FN_\alpha=0$ pointwise on their
kernels, and $\Lambda_FN_\alpha\DH\theta=0$ for arbitrary fields. 
For completeness, the geometric part of \rf{n24} is, for $\a=1$,
\be \la{n43}
\cR\theta=-\ii\,\ree_3\theta+\rho_3\big(1+T_C-2T_N\big)\theta\ ,
\ee
i.e.\ the Lie derivative along the Hopf Killing vector $\ree_3$ of \rf{044} combined with a constant
rotation matrix with eigenvalues $0,\pm1,\pm2,\pm3$. This fixes the split of the constant terms in
\rf{n24} between $\cR$ and $N_\a\DH$ (the constant part $-3N_\a\PN$ of $N_\a\DH$ belongs to the
equation-of-motion term). On the fermionic zero modes $\DH\theta=0$ and $\cQ_1^2=\cR$, with the
eigenvalues $\pm2$ required by \rf{948}. On the bosonic fluctuations $\cQ_1^2X^M=\cR X^M$ exactly,
the scalar field equation being of second order.

The scalar expressions  \eqref{n18}  determine the odd
constraint insertion, but checking their cubic closure also requires the
cubic fermion variation. With the commuting Killing parameter held fixed,
\be
 (\mathcal Q_1\mathcal Q_3+\mathcal Q_3\mathcal Q_1)X^M
 =\mathcal Q_1(\mathcal Q_3X^M)
  -\bar\varepsilon\Gamma_M\mathcal Q_3\theta .
\ee
The component closure relation used in the remaining local calculation is
\begin{equation}
 \begin{aligned}
 (\mathcal Q_1\mathcal Q_3+\mathcal Q_3\mathcal Q_1)X^M
   =\mathcal V^a\nabla_aX^M,\qquad \qquad 
 \mathcal V^a =\tfrac14\bar\theta\rho^a N_\alpha\DH\theta .
 \end{aligned}
 \label{n25}
\end{equation}
This term is an EOM-dependent world-volume transformation  and it  cannot be
set to zero before  computing the  quantum expectation value.
In particular, an independent derivation of \eqref{n25} must verify
$
 \bar\varepsilon\Gamma_M\mathcal Q_3\theta
 =\mathcal Q_1(\mathcal Q_3X^M)-\mathcal V^a\nabla_aX^M
$
using the fermionic transformation obtained from the same gauge-compensated
superisometry.\foot{The full expression for $\mathcal Q_3\theta$ is not
displayed here. Accordingly, \eqref{n25} is an additional component-closure
input to the local measure calculation; it does not follow from the
above scalar transformations alone. The calculation below verifies the
contribution of this input after projection and contraction, not the full
off-shell identity in \rf{n4}.}

Let us define
$\mathcal D=(\mathcal Q^2-\mathcal R)\mathcal B$, with
$\mathcal R\mathcal B=K\mathcal B$. 
The variation of
\rf{n6} contains $(K\mathcal B)^T\mathcal D$. 
At zero odd collective coordinate,
\begin{equation}
 \mathcal D^I=\epsilon^2\ML^I{}_Ja^J+\epsilon^3d^I+\dots,
 \qquad\qquad 
 \ML^I{}_J=\Lambda_B^I
      \big[\mathcal V^a\nabla_a\mathcal Z_{B,J}\big] .
 \label{n26}
\end{equation}
The functional $\ML$ is quadratic in quantum fermions, and
$d^I=\Lambda_B^I[\mathcal V^a\nabla_a\widetilde{\mathrm X}_B]$ is cubic.
We then get 
\begin{equation}
 \mathcal J_{\rm EOM}
    =-\tfrac12\operatorname{tr}(K^{-1}\ML)
    =-\tfrac{\ii}{2}V^{-1} \int d^3\sigma\sqrt g\,
                    \bar\theta P_\alpha^\perp\DH\theta .
 \label{n27}
\end{equation}
To obtain this expression, we first integrate over
the 12  bosonic collective coordinates $a^I$, keeping the quantum
fields fixed. The matrix $K$ is numerical, whereas $\ML$ in
\eqref{n26} is quadratic in the quantum fermions. Since (according to \rf{mm8}) 
$\mathcal B=a$, the term
$\mathcal D=\epsilon^2\ML a+\dots$ adds
$\epsilon^2(Ka)^T\ML a$ to the quadratic localizing form $|Ka|^2$.
The resulting relative factor from the bosonic integration in \rf{n4}  is
\be 
 \frac{\displaystyle
  \int d^{12}a\,
  \exp\!\big[-t\,a^TK^TKa
             -t\epsilon^2a^TK^T\ML a\big]}
 {\displaystyle
  \int d^{12}a\,\exp\!\big[-t\,a^TK^TKa\big]}
 =
 1-\tfrac{1}{2}\epsilon^2
       \operatorname{tr}(K^{-1}\ML)
 +O(\epsilon^4)
 \equiv
 1+\epsilon^2\mathcal J_{\rm EOM}+\OO(\epsilon^4).
\ee
For the second equality in \rf{n27}, write
$\Lambda_B u=V^{-1}\int d^3\s\, \sqrt g\,R_B(\sigma)u(\sigma)$, where   $R_B$ is the
matrix of dual kernels. The four AdS profiles are constant. The eight
internal profiles and the duals of \rf{943} give, in the component
convention,
\begin{equation}
 \operatorname{tr}
  \big[K^{-1}R_B\nabla_a\mathcal Z_B\big]
       =4i\alpha\,\delta_{a3} .
 \label{n38}
\end{equation}
 Substituting the definition of $\ML$
from \eqref{n26}  and the explicit zero-mode profile identity \eqref{n38}
gives
\be 
 \operatorname{tr}(K^{-1}\ML)
 =
 {4i\alpha}{V}^{-1} \int d^3\sigma\sqrt g\,\mathcal V^3
 =
 {\ii}{V}^{-1} \int d^3\sigma\sqrt g\,
       \bar\theta P_\alpha^\perp\DH\theta ,
\ee
where we used that 
$\mathcal V^3=\tfrac14\bar\theta\rho_3N_\alpha\DH\theta$,
$N_\alpha=\alpha\rho_3P_\alpha^\perp$, and $\alpha^2=1$.\foot{
The other terms from this cubic closure have zero leading Gaussian
expectation or are suppressed at large $t$ at order $\epsilon^2$.
In particular, in local Gaussian variables $a,\zeta=O(t^{-1/2})$,
the term $t\epsilon a^2\zeta\theta$ and its square are suppressed,
whereas $t\epsilon^2a^2\theta^2$ is retained in \rf{n27}.
The term $t\epsilon^2a\zeta X \theta$ has zero leading expectation;
terms starting with $t\epsilon^3a\phi\theta^2$ are beyond this loop order.
}
This yields the second equality in \eqref{n27}.
Thus $\mathcal J_{\rm EOM}$ is a quadratic fermionic insertion
generated by integrating the bosonic collective coordinates.
Its quantum expectation value then produces a fermionic EOM term.

The first-order fermion  Green's function EOM term  is
\begin{equation}
 E_\theta=r_CP_C+r_NP_N+d_CT_C+d_NT_N
       =-V^{-1}\mathcal P_\alpha,
 \qquad
 \mathcal P_\alpha=\tfrac12(P_C+T_C)+P_N+\alpha T_N .
 \label{n28}
\end{equation}
For $\alpha=\pm1$ one has
$P_\alpha^\perp\mathcal P_\alpha=0$. Then taking into account 
\rf{934}  gives
\begin{equation}
 \vev{\mathcal J_{\rm EOM}}
    =-\tfrac12\operatorname{tr}(P_\alpha^\perp E_\theta)
    =-2(r_C+r_N-d_C-\alpha d_N)=0 .
 \label{n29}
\end{equation}
Here the analytic prescription $\delta(0)=0$ removes the local contact term but retains the finite projector
part $E_\theta=-V^{-1}\mathcal P_\alpha$ of \rf{n28}; the vanishing follows from $P^\perp_\alpha\mathcal P_\alpha=0$
for either adapted sign, not from imposing $\DH\theta=0$. This gives the second equality in \rf{n11}.


\subsubsection*{The complementary configuration}

The two local calculations  leading to \rf{n23} and \rf{n29} refer to the same global Killing spinor,
expressed in frames adapted to $p_+$ and $p_-$. The rotation
\begin{equation}
 g_\perp=\begin{pmatrix}0&-\mathbf1_2\\ \mathbf1_2&0\end{pmatrix}
       \in SU(4)
 \label{n30}
\end{equation}
maps $\Pi_0$ to $\Pi_0^\perp$ and commutes with the scalar Hopf
$\mathbb Z_k$ action. 
In the ambient-space  convention
 its spinor action  lift is\footnote{The  choice of directions here is   $(\mathrm{radial},3,-2,1,4,5,6,7)$
 with   the real and imaginary axes of
$(Y_1,Y_2,Y_3,Y_4)$ in order, expressed in the reference frame:
``radial'' is the outward normal to $S^7\subset\mathbb R^8$
(not an AdS direction), and $-2$ denotes $-e_2$.
The rotation $g_\perp$ consists of four quarter-turns, whose
spinor representatives give the four factors in $S_\perp$
and the normalization $(1/\sqrt2)^4=1/4$.
The radial bivector is represented on Killing spinors by
$\gamma_{\rm radial}\gamma_a\mapsto-i\Gamma_a\Gamma_*$.
The inverse in $\varepsilon_-=S_\perp^{-1}\varepsilon_+$
expresses the same global Killing spinor in the rotated frame;
the subscripts label the two configurations, not chirality.}
\begin{equation}
 S_\perp=\tfrac14
 (1+\ii\Gamma_4\Gamma_*)(1-\Gamma_3\Gamma_5)
 (1+\Gamma_2\Gamma_6)(1-\Gamma_1\Gamma_7),\qquad 
 \qquad \varepsilon_-=S_\perp^{-1}\varepsilon_+ .
 \label{n31}
\end{equation}
The subscripts on $\varepsilon_\pm$ label local frame components, not
independently chosen global supersymmetries. Transporting the tangent
and normal frames, zero-mode profiles and duals together gives the
$\alpha=1$ adapted choice at $p_+$ and the $\alpha=-1$ choice at $p_-$.
Thus $v_N$ and $d_N$ reverse sign, while $s_N$ is unchanged.
The factors of $\alpha$ in \eqref{n20}, \eqref{n21} and
\eqref{n38} transform with them. Both signs consequently give the same
three contributions in \eqref{n23}, and \eqref{n29} vanishes at each point.

This comparison fixes the local inverse-tension coefficient, not the
absolute Pfaffian orientation or the individual leading weights.
Those factors are included in the normalized $w_p(k)$ of \eqref{n13}.
No equality of the two weights is needed because their local corrections
vanish separately.

\np

 \small
\baselineskip 11pt
\bibliography{m11}

@article{Schwarz:1995dg,
    author = {Schwarz, Albert S. and Zaboronsky, Oleg},
    title = {Supersymmetry and localization},
    eprint = {hep-th/9511112},
    archivePrefix = {arXiv},
    primaryClass = {hep-th},
    doi = {10.1007/BF02506415},
    journal = {Commun. Math. Phys.},
    volume = {183},
    pages = {463--476},
    year = {1997}
}

@article{Gautason:2024nru,
    author = "Gautason, Fridrik Freyr and van Muiden, Jesse",
    title = "{One-loop quantization of Euclidean D3-branes in holographic backgrounds}",
    eprint = "2402.16779",
    archivePrefix = "arXiv",
    primaryClass = "hep-th",
    doi = "10.1007/JHEP06(2024)073",
    journal = "JHEP",
    volume = "06",
    pages = "073",
    year = "2024"
}

@article{Beccaria:2026zjj,
    author = "Beccaria, M. and Kurlyand, S. A. and Tseytlin, A. A. and van Muiden, J.",
    title = "{M2 brane in AdS$_4\times S^7/\mathbb Z_k$: 2-loop correction to $\frac{1}{2}$-BPS Wilson loop vs localization in ABJM theory}",
    eprint = "2609.14497",
    archivePrefix = "arXiv",
    primaryClass = "hep-th",
    month = "9",
    year = "2026"
}

@article{Criado:2018sdb,
 author = {Criado, Juan Carlos and Perez-Victoria, Manuel},
 title = {{Field redefinitions in effective theories at higher orders}},
 eprint = {1811.09413},
 archivePrefix = {arXiv},
 primaryClass = {hep-ph},
 doi = {10.1007/JHEP03(2019)038},
 journal = {JHEP},
 volume = {03},
 pages = {038},
 year = {2019}
}

@article{Bhattacharya:2024chz,
    author = "Bhattacharya, Arindam and Cotler, Jordan and Dersy, Aur{\'e}lien and Schwartz, Matthew D.",
    title = "{Collective coordinate fix in the path integral}",
    eprint = "2402.18633",
    archivePrefix = "arXiv",
    primaryClass = "hep-th",
    doi = "10.1103/PhysRevD.110.116023",
    journal = "Phys. Rev. D",
    volume = "110",
    number = "11",
    pages = "116023",
    year = "2024"
}

@article{Bergshoeff:1987qx,
    author = "Bergshoeff, E. and Sezgin, E. and Townsend, P. K.",
    title = "{Properties of the Eleven-Dimensional Super Membrane Theory}",
    reportNumber = "IC-87-255",
    doi = "10.1016/0003-4916(88)90050-4",
    journal = "Annals Phys.",
    volume = "185",
    pages = "330",
    year = "1988"
}

@article{Harvey:1999as,
    author = "Harvey, Jeffrey A. and Moore, Gregory W.",
    title = "{Superpotentials and membrane instantons}",
    eprint = "hep-th/9907026",
    archivePrefix = "arXiv",
    reportNumber = "EFI-99-22A, YCTP-P15-99, IASSNS-99-57, EFI-99-22",
    month = "7",
    year = "1999"
}

@article{Beccaria:2023sph,
    author = "Beccaria, Matteo and Giombi, Simone and Tseytlin, Arkady A.",
    title = "{(2,0) theory on S5?$\times $S1 and quantum M2 branes}",
    eprint = "2309.10786",
    archivePrefix = "arXiv",
    primaryClass = "hep-th",
    reportNumber = "PUPT-2648, Imperial-TP-AT-2023-05",
    doi = "10.1016/j.nuclphysb.2023.116400",
    journal = "Nucl. Phys. B",
    volume = "998",
    pages = "116400",
    year = "2024"
}

@article{Drukker:2023bip,
    author = "Drukker, Nadav and Shahpo, Omar",
    title = "{Vortex loop operators and quantum M2-branes}",
    eprint = "2312.17091",
    archivePrefix = "arXiv",
    primaryClass = "hep-th",
    doi = "10.21468/SciPostPhys.17.1.016",
    journal = "SciPost Phys.",
    volume = "17",
    number = "1",
    pages = "016",
    year = "2024"
}

@article{Gautason:2025bft,
    author = "Gautason, Fridrik Freyr and Nix, Alexia",
    title = "{Universal holographic Wilson loops in 3d SCFTs}",
    eprint = "2511.04596",
    archivePrefix = "arXiv",
    primaryClass = "hep-th",
    doi = "10.1007/JHEP06(2026)061",
    journal = "JHEP",
    volume = "06",
    pages = "061",
    year = "2026"
}

@article{Tseytlin:2026rxv,
    author = "Tseytlin, Arkady A. and Wang, Zihan",
    title = "{Wilson loop in AdS3{\texttimes}S3{\texttimes}T4 from quantum M2 brane}",
    eprint = "2603.25590",
    archivePrefix = "arXiv",
    primaryClass = "hep-th",
    doi = "10.1103/wxvx-qrwd",
    journal = "Phys. Rev. D",
    volume = "113",
    number = "12",
    pages = "126006",
    year = "2026"
}

@article{Cagnazzo:2009zh,
    author = "Cagnazzo, Alessandra and Sorokin, Dmitri and Wulff, Linus",
    title = "{String instanton in AdS(4) x CP3}",
    eprint = "0911.5228",
    archivePrefix = "arXiv",
    primaryClass = "hep-th",
    doi = "10.1007/JHEP05(2010)009",
    journal = "JHEP",
    volume = "05",
    pages = "009",
    year = "2010"
}

@article{Gautason:2023igo,
    author = "Gautason, Fridrik Freyr and Puletti, Valentina Giangreco M. and van Muiden, Jesse",
    title = "{Quantized strings and instantons in holography}",
    eprint = "2304.12340",
    archivePrefix = "arXiv",
    primaryClass = "hep-th",
    doi = "10.1007/JHEP08(2023)218",
    journal = "JHEP",
    volume = "08",
    pages = "218",
    year = "2023"
}

@article{Hatsuda:2013gj,
    author = "Hatsuda, Yasuyuki and Moriyama, Sanefumi and Okuyama, Kazumi",
    title = "{Instanton Bound States in ABJM Theory}",
    eprint = "1301.5184",
    archivePrefix = "arXiv",
    primaryClass = "hep-th",
    reportNumber = "DESY-13-010, TIT-HEP-626",
    doi = "10.1007/JHEP05(2013)054",
    journal = "JHEP",
    volume = "05",
    pages = "054",
    year = "2013"
}

@article{Drukker:2011zy,
    author = "Drukker, Nadav and Marino, Marcos and Putrov, Pavel",
    title = "{Nonperturbative aspects of ABJM theory}",
    eprint = "1103.4844",
    archivePrefix = "arXiv",
    primaryClass = "hep-th",
    reportNumber = "IMPERIAL-TP-2011-ND-01",
    doi = "10.1007/JHEP11(2011)141",
    journal = "JHEP",
    volume = "11",
    pages = "141",
    year = "2011"
}

@article{Beccaria:2023ujc,
    author = "Beccaria, Matteo and Giombi, Simone and Tseytlin, Arkady A.",
    title = "{Instanton contributions to the ABJM free energy from quantum M2 branes}",
    eprint = "2307.14112",
    archivePrefix = "arXiv",
    primaryClass = "hep-th",
    reportNumber = "PUPT-2645, Imperial-TP-AT-2023-04",
    doi = "10.1007/JHEP10(2023)029",
    journal = "JHEP",
    volume = "10",
    pages = "029",
    year = "2023"
}

@article{Kurlyand:2026yke,
    author = "Kurlyand, Stefan A.",
    title = "{Membrane instantons and non-perturbative effects in AdS$_{4}$/CFT$_{3}$}",
    eprint = "2606.19467",
    archivePrefix = "arXiv",
    primaryClass = "hep-th",
    doi = "10.1007/JHEP09(2026)222",
    journal = "JHEP",
    volume = "09",
    pages = "222",
    year = "2026"
}

@article{Claus:1998fh,
    author = "Claus, Piet",
    title = "{Super M-brane actions in AdS(4)$\times $ S7 and AdS(7)$\times$S4}",
    eprint = "hep-th/9809045",
    archivePrefix = "arXiv",
    reportNumber = "KUL-TF-98-32",
    doi = "10.1103/PhysRevD.59.066003",
    journal = "Phys. Rev. D",
    volume = "59",
    pages = "066003",
    year = "1999"
}

@article{Giombi:2023vzu,
    author = "Giombi, Simone and Tseytlin, Arkady A.",
    title = "{Wilson Loops at Large N and the Quantum M2-Brane}",
    eprint = "2303.15207",
    archivePrefix = "arXiv",
    primaryClass = "hep-th",
    doi = "10.1103/PhysRevLett.130.201601",
    journal = "Phys. Rev. Lett.",
    volume = "130",
    number = "20",
    pages = "201601",
    year = "2023"
}

@article{Aharony:2008ug,
    author = "Aharony, Ofer and Bergman, Oren and Jafferis, Daniel Louis and Maldacena, Juan",
    title = "{N=6 superconformal Chern-Simons-matter theories, M2-branes and their gravity duals}",
    eprint = "0806.1218",
    archivePrefix = "arXiv",
    primaryClass = "hep-th",
    reportNumber = "WIS-12-08-JUN-DPP",
    doi = "10.1088/1126-6708/2008/10/091",
    journal = "JHEP",
    volume = "10",
    pages = "091",
    year = "2008"
}

@article{Beccaria:2025xry,
    author = "Beccaria, M. and Roiban, R. and Tseytlin, A. A.",
    title = "{2-loop scattering on superstring and supermembrane in flat space}",
    eprint = "2507.09528",
    archivePrefix = "arXiv",
    primaryClass = "hep-th",
    doi = "10.1007/JHEP09(2025)191",
    journal = "JHEP",
    volume = "09",
    pages = "191",
    year = "2025"
}

@article{Tseytlin:2026ctl,
    author = "Tseytlin, Arkady A. and Wang, Zihan",
    title = "{Energy of toroidal M2 brane in flat 11d background}",
    eprint = "2608.23517",
    archivePrefix = "arXiv",
    primaryClass = "hep-th",
    month = "8",
    year = "2026"
}

@article{Tseytlin:2025dae,
    author = "Tseytlin, Arkady A. and Wang, Zihan",
    title = "{On world-volume supersymmetry of supermembrane action in~static gauge}",
    eprint = "2512.04948",
    archivePrefix = "arXiv",
    primaryClass = "hep-th",
    reportNumber = "Imperial-TP{\textendash}2025-AT-02",
    doi = "10.1098/rspa.2025.1058",
    journal = "Proc. Roy. Soc. Lond. A",
    volume = "482",
    number = "2334",
    pages = "20251058",
    year = "2026"
}

@article{Gautason:2025plx,
    author = "Gautason, Fridrik Freyr and van Muiden, Jesse",
    title = "{Ensembles in M-theory and holography}",
    eprint = "2505.21633",
    archivePrefix = "arXiv",
    primaryClass = "hep-th",
    doi = "10.1007/JHEP11(2025)078",
    journal = "JHEP",
    volume = "11",
    pages = "078",
    year = "2025"
}

@article{Gautason:2025per,
    author = "Gautason, Fridrik Freyr and van Muiden, Jesse",
    title = "{Localization of the M2-Brane}",
    eprint = "2503.16597",
    archivePrefix = "arXiv",
    primaryClass = "hep-th",
    doi = "10.1103/67bh-xd42",
    journal = "Phys. Rev. Lett.",
    volume = "135",
    number = "10",
    pages = "101601",
    year = "2025"
}

@article{vanMuiden:2026nsp,
    author = "van Muiden, Jesse",
    title = "{Quantum M2-branes and Holography}",
    eprint = "2603.14544",
    archivePrefix = "arXiv",
    primaryClass = "hep-th",
    doi = "10.22323/1.509.0264",
    journal = "PoS",
    volume = "CORFU2025",
    pages = "264",
    year = "2026"
}

@article{Bobev:2026gir,
    author = "Bobev, Nikolay and Gautason, Fridrik Freyr and van Muiden, Jesse",
    title = "{Holographic Tests of the $\mu$ Ensemble}",
    eprint = "2607.06493",
    archivePrefix = "arXiv",
    primaryClass = "hep-th",
    month = "7",
    year = "2026"
}

@article{Giombi:2024itd,
    author = "Giombi, Simone and Kurlyand, Stefan A. and Tseytlin, Arkady A.",
    title = "{Non-planar corrections in ABJM theory from quantum M2 branes}",
    eprint = "2408.10070",
    archivePrefix = "arXiv",
    primaryClass = "hep-th",
    doi = "10.1007/JHEP11(2024)056",
    journal = "JHEP",
    volume = "11",
    pages = "056",
    year = "2024"
}

@article{Beccaria:2025ahf,
    author = "Beccaria, Matteo and Kurlyand, Stefan A. and Tseytlin, Arkady A.",
    title = "{2-loop free energy of M2 brane in AdS7$\times$S4 and surface defect anomaly in (2,0) theory}",
    eprint = "2511.22306",
    archivePrefix = "arXiv",
    primaryClass = "hep-th",
    doi = "10.1007/JHEP03(2026)257",
    journal = "JHEP",
    volume = "03",
    pages = "257",
    year = "2026"
}

@article{Beccaria:2025vdj,
    author = "Beccaria, Matteo and Tseytlin, Arkady A.",
    title = "{Non-planar corrections to ABJM Bremsstrahlung function from quantum M2 brane}",
    eprint = "2501.06858",
    archivePrefix = "arXiv",
    primaryClass = "hep-th",
    doi = "10.1088/1751-8121/adc9e4",
    journal = "J. Phys. A",
    volume = "58",
    number = "17",
    pages = "175401",
    year = "2025"
}

@article{Beccaria:2025npl,
    author = "Beccaria, Matteo and Kurlyand, Stefan A. and Tseytlin, Arkady A.",
    title = "{On non-planar ABJM anomalous dimensions from M2 branes in AdS$_{4}$ {\texttimes} S$^{7}$/{\ensuremath{\mathbb{Z}}}$_{k}$}",
    eprint = "2503.09360",
    archivePrefix = "arXiv",
    primaryClass = "hep-th",
    doi = "10.1007/JHEP05(2025)187",
    journal = "JHEP",
    volume = "05",
    pages = "187",
    year = "2025"
}

@article{Seibold:2024oyr,
    author = "Seibold, Fiona K. and Tseytlin, Arkady A.",
    title = "{Scattering on the supermembrane}",
    eprint = "2404.09658",
    archivePrefix = "arXiv",
    primaryClass = "hep-th",
    doi = "10.1007/JHEP08(2024)102",
    journal = "JHEP",
    volume = "08",
    pages = "102",
    year = "2024"
}

@article{Duff:1987cs,
    author = "Duff, M. J. and Inami, T. and Pope, C. N. and Sezgin, E. and Stelle, K. S.",
    title = "{Semiclassical Quantization of the Supermembrane}",
    reportNumber = "CERN-TH-4731/87, IC/87/74",
    doi = "10.1016/0550-3213(88)90316-1",
    journal = "Nucl. Phys. B",
    volume = "297",
    pages = "515--538",
    year = "1988"
}

@article{deWit:1998yu,
    author = "de Wit, Bernard and Peeters, Kasper and Plefka, Jan and Sevrin, Alexander",
    title = "{The M theory two-brane in AdS(4)$\times$S7 and AdS(7)$\times$S4}",
    eprint = "hep-th/9808052",
    archivePrefix = "arXiv",
    primaryClass = "hep-th",
    reportNumber = "THU-98-27, NIKHEF-98-024, VUB-TENA-98-4",
    doi = "10.1016/S0370-2693(98)01340-9",
    journal = "Phys. Lett. B",
    volume = "443",
    pages = "153--158",
    year = "1998"
}

@article{Bergshoeff:1987cm,
    author = "Bergshoeff, E. and Sezgin, E. and Townsend, P. K.",
    title = "{Supermembranes and Eleven-Dimensional Supergravity}",
    doi = "10.1016/0370-2693(87)91272-X",
    journal = "Phys. Lett. B",
    volume = "189",
    pages = "75--78",
    year = "1987"
}

@article{Claus:1998yf,
 author = {Claus, Piet and Kallosh, Renata},
 title = {{Superisometries of the AdS x S superspace}},
 eprint = {hep-th/9812087},
 archivePrefix = {arXiv},
 primaryClass = {hep-th},
 reportNumber = {SU-ITP-98-61, KUL-TF-98-56},
 doi = {10.1088/1126-6708/1999/03/014},
 journal = {JHEP},
 volume = {03},
 pages = {014},
 year = {1999}
}
\bibliographystyle{JHEP-v2.9}

\end{document}